\documentclass[aps,showpacs,pre,amsmath,amsfonts,amssymb,superscriptaddress,nofootinbib,notitlepage,a4paper]{revtex4}

\usepackage{cancel}

\usepackage{graphicx}
\usepackage{psfrag}
\usepackage[naturalnames,hypertexnames=true]{hyperref}

\newcommand{\beq}{\begin{equation}} \newcommand{\eeq}{\end{equation}}
\newcommand{\bes}{\begin{split}} \newcommand{\ees}{\end{split}} 
\newcommand{\bea}{\begin{eqnarray}} \newcommand{\eea}{\end{eqnarray}}

\def\s{\sigma}
\def\us{\underline{\sigma}}

\def\hsix{{\widehat{\sigma}_i^x}}

\def\hsiz{{\widehat{\sigma}_i^z}}
\def\hH{\widehat{H}}

\def\la{\langle}
\def\ra{\rangle}

\def\e{\epsilon}

\def\N{\mathcal{N}}
\def\G{\Gamma}
\def\Si{\Sigma}
\def\hT{\widehat{T}}
\def\tr{\textrm{Tr}}
\def\hHp{\widehat{H}_\textrm{P}}
\def\hHq{\widehat{H}_\textrm{Q}}
\def\qcol{q_{\rm{col}}}
\def\Ntraj{\mathcal{N}_{\rm{traj}}}
\def\Npop{\mathcal{N}_{\rm{pop}}}

\def\Tk{T_{\rm{K}}}
\def\Td{T_{\rm{d}}}
\def\Ti{T_{\rm{i}}}
\def\hi{h_{\rm{i}}}
\def\Gi{\Gamma_{\rm{i}}}
\def\Gk{\Gamma_{\rm{K}}}

\def\bs{{\boldsymbol{\sigma}}}
\def\Ncore{\mathcal{N}_{\rm{core}}}
\def\msplq{m^{\rm sp}_{\rm lq}}
\def\msphq{m^{\rm sp}_{\rm hq}}
\def\dd{{\rm d}}

\begin{document}

\title{The effect of quantum fluctuations on the coloring of random graphs}

\author{Victor Bapst}

\author{Guilhem Semerjian}
 
\author{Francesco Zamponi}
\affiliation{LPTENS, Unit\'e Mixte de Recherche (UMR 8549) du CNRS et
  de l'ENS, associ\'ee \`a l'UPMC Univ Paris 06, 24 Rue Lhomond, 75231
  Paris Cedex 05, France.}

\begin{abstract}
We present a study of the coloring problem (antiferromagnetic Potts model) of
random regular graphs, submitted to quantum fluctuations induced by a 
transverse field, using the quantum cavity method and quantum Monte-Carlo 
simulations. We determine the order of the quantum phase transition encountered 
at low temperature as a function of the transverse field and discuss the 
structure of the quantum spin glass phase. In particular, we conclude that 
the quantum adiabatic algorithm would fail to solve efficiently typical 
instances of these problems because of avoided level crossings within the 
quantum spin glass phase, caused by a competition between energetic and
entropic effects.
\end{abstract}

\maketitle

\section{Introduction}

Among the fascinating potential applications of quantum mechanics, the promise of quantum computing is to make use of the laws of quantum mechanics to enhance computers' calculation power. Besides the great effort of research towards the physical realization of quantum computers, a lot of activity has been devoted to develop quantum algorithms and understand their efficiency. A central problem encountered in almost all branches of science is to optimize irregularly shaped cost functions: the Quantum Adiabatic Algorithm (QAA) \cite{qa_first,qa_second,KaNi98,Aeppli99,Fa01} is a generic and universal procedure to tackle such problems. Suppose one wishes to find the ground state of a Hamiltonian $\hHp$ acting on $N$ qubits. The idea of the QAA is to implement an interpolating Hamiltonian $\hH(\G) = \hHp + \G \hHq$ such that the quantum computer can easily be initialized in $\hHq$'s ground state, and to decrease the interpolation parameter $\G$ from a very large value down to zero. If this is done slowly enough the adiabatic theorem \cite{Messiah} ensures that with high probability the system remains at all times in the ground state of the interpolating Hamiltonian. In particular, at the end of the evolution, it is in the ground state of $\hHp$ and the original problem is solved. The crucial question is how slow the evolution should be in the thermodynamic limit $N \rightarrow \infty$, and whether the total evolution time $\mathcal{T}$ has to  grow polynomially or exponentially fast with $N$. This is reminiscent of the classical complexity classes \cite{GareyJohnson,Pa83}, which indeed have quantum counterparts \cite{BeVa97,Wa00,KiShVy02}. Quite generally, the adiabaticity condition requires $\mathcal{T}$ to be larger than the inverse of the squared gap between the ground state and the first excited state of $\hH(\G)$, for all $\G$. Hence, the efficiency of the QAA mainly depends on the rate of closing of the minimal gap along the interpolation, even though more subtle issues such as determining the residual energy after the interpolation \cite{SaMaToCa02,BaSe12} would require a further understanding of the annealing dynamics.

The possibility that the QAA could outperform classical algorithms has triggered a lot of work on optimization problems in quantum fields (see \cite{ST06,qa_book_das_chakrabarti,review_Nishimori,long} for recent reviews), trying in particular to pinpoint the rate of closing of the minimal gap along the interpolation path. As it has now become usual, these studies used random ensembles of constraint satisfaction problems \cite{MitchellSelman92,CheesemanKanefsky91}, focusing on results valid with high probability in the thermodynamic limit. Early results generated excitement by reporting polynomial scaling of the minimal gap for some classically hard problems \cite{Fa01,Hogg03,YKS08}; however, these results where hampered by strong finite-size effects or by the fact that the instances considered were not \textit{typical} of the underlying problem. On the other hand, \cite{JKKM08,YKS10,JKSZ10,FaGoHe12} obtained negative results for the success of the QAA on a certain class of optimization problems: namely, they  found a first order quantum phase transition during the interpolation between the quantum and the classical Hamiltonians. Such a phase transition is not surprising in the context of fully-connected quantum spin glasses \cite{Go90,NR98,BC01bis,CGS01}, and is known to generically lead to a gap vanishing exponentially fast with the system size $N$ \cite{JKKM08,YKS10,JKSZ10} and thus to a blow-up of the time needed by the QAA.  Failure of the QAA because of another mechanism was also discovered for different models in \cite{AC09,AKR10,FSZ10}: because the classical energies have a non-trivial perturbative expansion in $\G$, the energy of an excited state may decrease much faster than the one of the ground state, leading to avoided ``perturbative'' crossings near the classical end of the algorithm. However, these results were obtained either on a toy model, or using well-chosen Hamiltonians $\hHp$ with a peculiar structure of the low-energy spectrum of $\hHp$, that were not typical of the optimization problem considered.

Henceforth, one would like to understand what happens more generically for models with multiple and not necessarily isolated ground states. This is expected to happen for typical optimization problems, which are known to possess complex and intricate configuration spaces that were unveiled by the use of statistical mechanics tools \cite{MoZe,MezardParisi02,KrMoRiSeZd,MM09}. In these more common cases, do the avoided crossings remain finite and isolated in $\G$ (hence leading to singularities in the ground state energy for $N \rightarrow \infty$) or do they proliferate and accumulate, leading to a continuum of level crossings and a gapless phase? Hints for an answer have been obtained on a toy model in \cite{FSZ10,long}; this paper aims to highlight their conclusions on a more realistic optimization problem. This question is also important from a practical point of view as it has been argued that a finite number of level crossings could be eliminated by a suitable redefinition of the quantum Hamiltonian 
\cite{FGGGS10,Ch11,DicAm11,Dic2011}. Also note that the Xorsat problem studied in \cite{JKSZ10} already possessed in some cases an exponential degeneracy of its ground states, but they had a particular structure that smoothened the effect of quantum fluctuations.

The model that we shall study is the coloring one, a famous problem in combinatorics, which is known to be classically very hard to solve - more precisely, NP-complete~\cite{Karp72}. Given a graph and $\qcol$ colors, it consists in coloring the vertices such that no edge connects two vertices of the same color. In physical terms, this corresponds to a $\qcol$-states antiferromagnetic Potts model. This model can be studied on any kind of graph, but we shall focus in the following on random regular graphs, with four colors ($\qcol=4$) and connectivity $c=9$ and $c=13$. Thanks to the quantum cavity method \cite{KRSZ08,LSS08}, we will compute the phase diagram of this model in both cases and unveil the nature of its quantum spin glass phase. Our main results are {\it (i)} the nature of the quantum phase transition that occurs along the interpolation from the quantum phase to the classical one at low temperature, that we find to be continuous in the spin glass language and thus of third order thermodynamically, and {\it (ii)} the presence of a continuum of level crossings within the spin glass phase, that are induced by entropic effects and in particular by the clustered structure of the spin glass phase. 
This last feature should lead the QAA to fail to solve the coloring problem efficiently. Our results will also be corroborated by Monte-Carlo simulations.
Note that result {\it (ii)} confirms
the predictions made in \cite{FSZ10} on a much simpler toy model.

The plan of the paper is as follows. We first briefly recall in Section~\ref{sec_classical_and_expected} the classical features of the model and the definition of its quantum version. Section~\ref{sec_phase_diag} presents the phase diagrams that we obtain for the quantum coloring problems, while Section~\ref{sec_structure_sg} presents in greater details the structure of its quantum spin glass phase.
Technical details are deferred to a series of Appendices. A brief part of our results (mostly Sec.~\ref{sec_q4_c9_clusters}) already appeared on the review paper \cite{long}.

\section{The coloring problem: classical picture and definition of its quantum version}
\label{sec_classical_and_expected}

\subsection{The classical coloring problem}
\label{sec_classical_coloring}

Let us consider a graph $G=(V,L)$ with $V$ a set of $N$ vertices and $L$ a set
of $M$ edges between pairs of vertices. We introduce a Potts variable 
$\s_i \in \{1, \dots, \qcol \}$ on each vertex $i$ of the graph, and denote
$\us=(\s_1,\dots,\s_N)$ the global configuration of the variables. To each of
these configurations we associate an energy, or cost function,
\beq 
\label{eq_def_ham_classic} 
E(\us) = \sum_{(i,j) \in L} \delta_{\sigma_i,\sigma_j} \ ,
\eeq
where the sum runs over the edges of the graph, and $\delta$ denotes the 
Kronecker symbol. Interpreting the Potts variable $\s_i$ as a color given to 
vertex $i$, this cost function counts the number of monochromatic edges in the
configuration $\us$. In physical terms there are antiferromagnetic 
interactions between pairs of variables linked by an edge of the graph.
The Gibbs-Boltzmann probability measure at inverse temperature $\beta$
is then defined as
\beq
\mu(\us) = \frac{1}{Z(\beta)} e^{- \beta E(\us)} \ ,
\label{eq_mu_GB}
\eeq
with the partition function $Z(\beta) = \sum_{\us} e^{- \beta E(\us)}$ ensuring
its normalization.

As mentioned in the introduction the coloring problem is NP-complete (for
$\qcol \ge 3$) in its decision version, namely there does not exist any 
algorithm able to decide the existence of a proper coloring (a configuration
of zero energy) in a number of operations bounded by a polynomial in $N,M$ for
every graph $G$. Computer scientists~\cite{CheesemanKanefsky91} thus turned 
to random ensembles of graphs in an attempt to study the typical difficulty
of the coloring problem, typical meaning here ``with a probability going
to one in the large size (thermodynamic) limit $N \to \infty$''. The 
interesting regime is the one of sparse random graphs, with the number
of edges $M$ of the same order than $N$. The absence of an underlying
Euclidean structure in these random graphs makes the corresponding
statistical model of the mean-field type, hence the methods devised for
the study of mean-field spin glasses models~\cite{Beyond} could be applied
to characterize the typical behavior of the partition function $Z$ and
of the measure $\mu$. For the coloring problem, see~\cite{col_replica} for
the use of the replica method and~\cite{col_sp,col_stab,KrMoRiSeZd,col2,KZ08}
for the application of the cavity method. The main outcome of these studies
is the unveiling of phase transitions in the behavior of the free energy 
density $f=-\frac{1}{N \beta} \ln Z$ (that concentrates around its average
in the thermodynamic limit) and in the properties of $\mu$, as a function
of the inverse temperature $\beta$ and of the parameters of the random graph
ensemble. 

In the following we shall sketch these phase transitions and the cavity
methodology used to derive them; for more details and justifications of the
equations the reader is refered to the above quoted references for the 
coloring problem, and to~\cite{cavity,MM09} for more generic presentations
of the cavity method. We will focus for technical simplicity on the case 
of random $c$-regular graphs: the graphs $G$ are chosen uniformly at random among all the graphs on $N$ 
vertices in which all vertices have the same degree (number of neighbors, or
connectivity) $c$. In the thermodynamic limit ($N \to \infty$), such graphs 
are locally tree-like, meaning that if one selects one of its vertex at random,
the shortest loop around it is larger than any fixed length with a probability 
which goes to one when $N\rightarrow \infty$~\cite{Janson}. The cavity
method exploits this tree-like character, building on the exact solution 
for finite trees that is easily obtained by recursion, and dealing with the
boundary conditions induced by the long loops of the random graphs in a 
self-consistent way. Consider an arbitrary 
variable $i$ in a large random $c$-regular graph, and its marginal probability
$\eta(\s_i)$ that would be obtained from the Gibbs-Boltzmann measure 
(\ref{eq_mu_GB}) if one of the $c$ edges around it were removed.
Forgetting the possible effects of the long loops of the graphs, the 
translational invariance and the recursive structure of the tree implies
that $\eta$ is solution of the following self-consistent equation,
called replica-symmetric (RS) cavity equation:
\beq 
\label{eq_cavity_classical_RS} 
\eta = g(\eta,\dots,\eta), \hspace{1 cm} 
g(\eta_1,\dots,\eta_{c-1})(\sigma) = \frac{1}{z(\eta_1,\dots,\eta_{c-1})} 
\sum_{\s_1,\dots,\s_{c-1}} \prod_{i=1}^{c-1} 
\left( \eta_i(\sigma_i) e^{- \beta \delta_{\sigma, \sigma_i}} \right) \ .  
\eeq
Here $z(\{\eta_i\})$ is a normalization factor ensuring that 
$\sum_\sigma g(\eta)(\sigma)=1$. The cavity method then proposes an expression
for the free energy density $f$ in terms of the solution $\eta$ of this
self-consistent equation.

The equation (\ref{eq_cavity_classical_RS}) and the associated RS prediction
for the free energy density are correct when the spatial decay of correlations 
between variables is fast enough, which is equivalent to the existence of
a single pure state in the Gibbs-Boltzmann measure (or of a finite
number of them, related by simple symmetries). At low enough temperature
and in presence of many constraints between variables (i.e. when $c$ is large
enough) this assumption breaks down. The Gibbs-Boltzmann measure is then
split in an exponentially large number of pure states (or clusters), the 
correlation decay assumption underlying the RS computation is then valid 
only inside one pure state, not in the complete Gibbs-Boltzmann measure.
The 1-step of replica-symmetry breaking (1RSB) cavity method deals with
this situation by making a further ansatz on the structure of the pure states.
One assumes that the typical number of pure states of internal free energy
density $f$ is exponentially large, at the leading order $e^{N\Sigma(f)}$, with
a rate function $\Sigma(f)$ called complexity or configurational entropy. This
function is computed via its Legendre transform~\cite{Mo95}
$\phi(m) =\min_f \{ f - T \Si(f)/m \}$, the parameter $m$ conjugated to $f$
being called the Parisi replica-symmetry breaking parameter. The thermodynamic
potential $\phi(m)$ is obtained from the solution of the 1RSB cavity equation
that generalizes (\ref{eq_cavity_classical_RS}), the order parameter solution
of this self-consistent equation becoming $P(\eta)$, a distribution of the
RS probabilities $\eta$ with respect to the choice of the pure states weighted
proportionally to $e^{-N m \beta f}$:
\beq 
\label{eq_cavity_classical_1RSB}
P(\eta) = \frac{1}{Z}  \int \prod_{i=1}^{c-1} \dd P(\eta_i) \,
\delta \left (\eta-  g(\eta_1,\dots,\eta_{c-1})\right) \,
z(\eta_1,\dots,\eta_{c-1})^m \ ,
\eeq
where $Z$ is a normalization constant and the functions $g$ and $z$ are 
defined in Eq.~(\ref{eq_cavity_classical_RS}).
One can check that the replica-symmetric 
equation (\ref{eq_cavity_classical_RS}) is recovered if $P$ is a 
delta-peaked distribution.

The structure of the Gibbs measure and its decomposition into pure states can be probed by the inter- and intra- state overlaps. The latter are defined, respectively, by:
\beq 
q_0 = \sum_\s \int \dd \widetilde{P}(\eta) \dd \widetilde{P}(\eta') \eta(\s) \eta'(\s) \ ,  \hspace{ 1 cm}  q_1 = \sum_\s \int \dd \widetilde{P}(\eta) \eta(\s) \eta(\s) \ ,
\label{eq_overlap_class} 
\eeq
where $\widetilde{P}$ is defined by the right hand side of (\ref{eq_cavity_classical_1RSB}), but with a product over $c$ terms instead of $c-1$. The energy function (\ref{eq_def_ham_classic}) is symmetric under the exchange of two colors; hence for all $\s$, $\int \dd \widetilde{P}(\eta) \eta(\s)= \frac{1}{\qcol}$, and $q_0$ is always equal to $1/\qcol$. Moreover, on the RS solution, this symmetry enforces $\eta(\s)=1/\qcol$ for all $\s$. The appearance of a non-trivial 1RSB solution to the equation (\ref{eq_cavity_classical_1RSB}) can then be detected by the fact that the intra-state overlap $q_1$ takes a value strictly larger than $1/\qcol$. 

Depending on the values of the temperature and of the degree $c$ the pure state
decomposition of the Gibbs-Boltzmann measure exhibits qualitatively
different properties, that can be read on the type of solutions of the
RS (\ref{eq_cavity_classical_RS}) and 1RSB (\ref{eq_cavity_classical_1RSB}) 
cavity equations:
\begin{itemize}
\item if the 1RSB equation with $m=1$ admits only a RS solution, then almost
all configurations are part of a single pure state, the RS hypothesis is
correct, and its prediction for the free energy density is valid. We shall call
this situation a RS phase, or paramagnetic phase (P) here.
\item if the 1RSB equation with $m=1$ admits a non-trivial solution, and
if the $\min$ in the definition of $\phi(m=1)$ is reached at $f_{\rm int}$ with
$\Sigma(f_{\rm int})>0$, then the system is in a dynamic 1RSB (d1RSB)
phase, or dynamic paramagnetic phase (dP) here. Almost all configurations belong to pure states of internal free energy
density $f_{\rm int}$, and there are $e^{N\Sigma(f_{\rm int})}$ such pure states. It
turns out that the total free energy density $f_{\rm int} - T \Sigma(f_{\rm int})$
coincides with the RS prediction, yet the splitting of the configurations into
pure states has drastic consequences on the dynamics of such a model, that
becomes non-ergodic (on sub-exponential timescales), hence the name of \textit{dynamic} 1RSB phase.
\item if the 1RSB equation with $m=1$ admits a non-trivial solution, but
predicts a negative complexity $\Sigma(f_{\rm int}(m=1))<0$, then the Parisi
breaking parameter $m$ has to be set to the value $m_{\rm s} \in [0,1]$ such
that $\Sigma(f_{\rm int}(m_{\rm s}))=0$. In such a 1RSB phase almost all 
configurations belong to a sub-exponential number of pure states of internal
free energy density $f_{\rm int}(m_{\rm s})$; as the complexity vanishes this value
is also the 1RSB prediction for the total free energy density of the model. Thise phase corresponds to a genuine spin glass (SG) phase.
\end{itemize}

\begin{figure}[t]
\includegraphics[width=10cm]{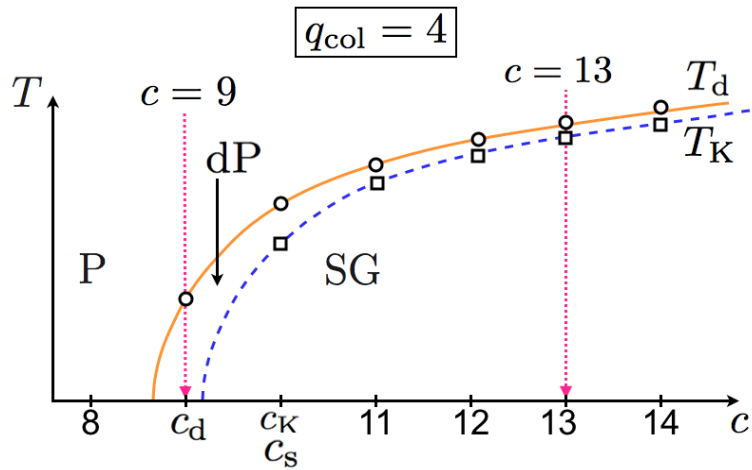}
\caption{Classical phase diagram of the coloring problem for $\qcol=4$ in 
the connectivity-temperature plane, see~\cite{col2,KZ08} for the actual
numerical values. The solid orange line separates the 
paramagnetic phase (P) from the dynamical paramagnet (dP) while the dashed 
blue lines marks the boundary of the genuine spin glass (SG) phase. The model 
is defined only for integer connectivites (indicated by squares and circles); 
lines serve as a guide to the eye. The connectivites $c_{\rm{d}}=9$ and 
$c_{\rm{K}},c_s=10$ are defined in the text. Pink arrows indicates the connectivities that will be studied in the quantum case. }
\label{fig_c_T}
\end{figure}

In general these three phases are encountered in this order upon lowering the
temperature (for large enough degrees) or increasing the degree (for low enough
temperature). The transition between the RS and d1RSB phase occurs at the 
dynamic temperature $T_{\rm d}(c)$ (or at $c_{\rm d}(T)$), and is not accompanied
by any singularity in the free energy. The transition between the d1RSB phase
and the 1RSB phase, at which the Gibbs-Boltzmann measure condenses on the
few largest existing clusters, occurs at $\Tk(c)$. This transition corresponds
indeed to the Kauzmann transition in the Random First Order Theory (RFOT) of
structural glasses~\cite{kirkpatrick:87,kirkpatrick:88}, 
which is a second order transition
from the thermodynamic point of view, even if the order parameter jumps
discontinuously (``first order'') when the transition is crossed. In this
context the dynamic transition $T_{\rm d}$ corresponds to the divergence
of the relaxation time of the Mode Coupling Theory (MCT) of supercooled 
liquids~\cite{mct}. This pattern of transitions has indeed been
found for the coloring of random regular graphs with $\qcol \ge 4$ 
colors~\cite{col2,KZ08}, and is illustrated on Fig.~\ref{fig_c_T}.
The case $\qcol=3$ is singular, there is then no intermediate d1RSB phase,
the transition between the RS and the 1RSB phase occurs continuously, via
a local instability of the RS solution towards the 1RSB one at a temperature
$\Ti(c)$. In the context of optimization problems the zero temperature
limit plays a particularly important role; the transition lines defined above
end in this limit at the critical connectivities $c_{\rm d}$ and $c_{\rm{K}}$. 
A further transition can be defined at zero temperature: the satisfiability
threshold $c_{\rm s}$ is such that the ground state energy vanishes for
$c \le c_{\rm s}$ (in terms of the coloring problem the typical answer to the 
decision question is yes, there are proper colorings of such graphs) and
is strictly positive for $c>c_{\rm s}$.

In the following we shall study the quantum version of this model for 
$\qcol=4$ and $c=9$ or $c=13$; these two cases correspond to the situations 
illustrated by arrows on Fig.~\ref{fig_c_T}. Before turning to this
quantum framework let us make a few remarks. The coloring problem can be
studied for other ensembles of random graphs, notably for Erd\"os-R\'enyi
random graphs in which the degree of vertices are Poisson random variables
of average $c$. The transitions described above are also encountered in
this case~\cite{col2,KZ08}, with the difference that $c$ is now a continuous
parameter. We restricted ourselves to the regular case for technical reasons:
to deal with the fluctuating degrees in the Erd\"os-R\'enyi case one has to
introduce a more complicated order parameter, i.e. in the 1RSB cavity treatment 
a distribution over the distributions $P$ to deal with the fluctuations of 
the probabilities $\eta$ with respect both to the choice of the pure states and
of the local connectivity. Although this complication is affordable in the classical
case, it would become impossible to treat in the quantum setting. We also
neglected in this brief presentation the phenomenon of full replica-symmetry 
breaking~\cite{Beyond}: the 1RSB description of the Gibbs-Boltzmann measure 
is only the first one in a hierarchical construction in which the 
pure states are themselves grouped in clusters of pure states, and so on and
so forth, that become relevant at a so-called Gardner 
transition~\cite{Gardner85}, computed in~\cite{col2,KZ08} for the classical
antiferromagnetic Potts model.
This whole hierarchy can be dealt with in fully-connected models,
most notably the Sherrington-Kirkpatrick one, yet it cannot be handled with
present techniques in finite connectivity models (defined on sparse random 
graphs), even for classical variables, not to speak about quantum models.
The full RSB structure allows to study some important physical properties of these models,
most notably the {\it marginal stability} of glass states; 
however, the 1RSB approximation is expected to give quantitatively good
predictions for thermodynamic properties even when it is unstable towards a full RSB one,
and therefore it will be sufficient for the purposes of this study.

\subsection{Definition of the quantum version of the coloring problem}

In order to define the quantum version of the problem we introduce the 
$\qcol^N$-dimensional Hilbert space spanned by 
$\{| \us \ra , \us \in \{1, \dots, \qcol\}^N\}$. We then define the ``problem''
Hamiltonian $\hH_{\rm P}$ corresponding to the cost function 
(\ref{eq_def_ham_classic}) as the operator diagonal in this (so-called 
computational) basis with diagonal elements equal to the classical energies:
\beq 
\label{eq_def_Hp} 
\hH_{\rm P} = \sum_{\us} E(\us) |\us \ra \la \us | \ .
\eeq
We introduce quantum fluctuations, i.e. off-diagonal matrix elements, by 
defining for each site $i$ an operator $\hT_i$ that flips the color $\s_i$
to any other different color:
\beq
\la \us | \hT_i | \us' \ra = \begin{cases}
1 & \text{if} \ \s_i \neq \s'_i \ \text{and} \ 
\s_j = \s'_j \ \ \forall j \neq i \\
0 & \text{otherwise} \end{cases} \ . 
\label{eq_def_hTi}
\eeq
This is the simplest generalization from the Ising spin case (that corresponds 
to $\qcol=2$) of the Pauli matrix $\hsix$ (taking as the computational basis
the eigenvectors of the $\hsiz$ matrices). The intensity of the quantum
fluctuations is controlled by the ``transverse field'' $\Gamma$, the total
Hamiltonian reading
\beq
\label{eq_def_hH_coloring} \hH(\G) = \hHp + \G \hHq \ , \qquad
\hHq = - \sum_{i=1}^N \hT_i \ .
\eeq
The partition function at inverse temperature $\beta$ is 
$Z(\beta,\G)=\tr[ e^{-\beta \hH(\G)}]$, thermodynamic averages being denoted
$\la \bullet \ra = \tr [ \bullet \ e^{-\beta \hH(\G)}] /Z(\beta,\G)$. 
In particular
we will call $m_x= \langle \hT_i \rangle \in [0, \qcol-1]$ the 
``transverse magnetization'', by analogy with the Ising spin case.
This also explains the name of ``transverse field'' for $\G$.

\subsection{A sketch of the quantum cavity method}

A standard way to compute the partition function of such a quantum model
consists in using the Lie-Suzuki-Trotter formula to disentangle the two
non-commuting parts of the Hamiltonian $\hH(\G)$, introducing copies of the
original degrees of freedom $\s_i$ along an ``imaginary time'' axis of 
length $\beta$. In the limit where the number of such copies goes to infinity 
one thus obtains an exact representation of the quantum partition function as 
a path integral of a classical model, the price to be paid being the
replacement of the spins $\s_i \in \{1, \dots, \qcol \}$ by more complicated degrees of 
freedom $\bs_i$ (we emphasize the use of a bold font here), which are
piecewise constant periodic functions $\s_i(t)$ from $[0,\beta]$ to 
$\{1, \dots, \qcol \}$.

Apart from this replacement the classical model thus obtained from the quantum
one has the same topology of interactions as the classical one 
(\ref{eq_def_ham_classic}); in particular if the latter can be treated with
the classical cavity method (i.e. if it is defined on a locally tree-like 
random graph), then its quantum version can be handled with the quantum 
extension of the cavity method where the spins $\s_i$ are replaced by the
imaginary time trajectories $\bs_i$. This observation was first put to work
in~\cite{LSS08} with a finite number of Suzuki-Trotter slices, the continuous
imaginary time limit being taken in~\cite{KRSZ08}. These two works developed
the quantum cavity method at the RS level, and the quantum 1RSB framework was
then introduced in~\cite{JKSZ10,FSZ11}. Detailed explanations and derivations,
covering the case of the coloring in presence of a transverse field, can
be found in the review~\cite{long}, we shall thus content
ourselves here with a few remarks, some technical details being presented 
in App.~\ref{app_1step_quantum_cavity}, in particular the parallelization 
of the numerical code we used to accelerate the resolution of the cavity
equations.

As explained above the quantum cavity method has exactly the same structure
as the classical one, sketched in Sec.~\ref{sec_classical_coloring}. The
main modification is the fact that the basic object $\eta$ appearing in
the RS (\ref{eq_cavity_classical_RS}) and 1RSB 
(\ref{eq_cavity_classical_1RSB}) cavity equations is now a probability 
distributions over the space of piecewise constant periodic functions from 
$[0,\beta]$ to $\{1,\dots,\qcol\}$. This space is obviously much larger than 
$\{1,\dots,\qcol\}$, and in consequence a single $\eta$, that could be represented by 
$\qcol-1$ real numbers in the classical case, has now to be represented in an 
approximate way for the numerical resolution of Eqs.~(\ref{eq_cavity_classical_RS}), (\ref{eq_cavity_classical_1RSB}). 
A convenient representation is provided by a finite sample of $\Ntraj$ random 
trajectories, $\eta$ being approximated by the empirical distribution of this 
sample.

Finally, the quantum overlaps are defined in analogy with the classical case, taking an extra average over the imaginary time:
\beq \begin{split} 
q_0 &= \sum_{\bs,\bs'} \int \dd \widetilde{P}(\eta) \dd\widetilde{P}(\eta')  
\eta(\bs) \eta'(\bs') 
\frac{1}{\beta} \int_{0}^\beta \delta_{\s(\tau),\s'(\tau)} \, \dd \tau  \ , \\ 
q_1 &= \sum_{\bs,\bs'} \int \dd \widetilde{P}(\eta) \eta(\bs) \eta(\bs' ) 
\frac{1}{\beta} \int_{0}^\beta \delta_{\s(\tau),\s'(\tau)} \, \dd \tau \ .
\label{eq_def_q_quantum}  \end{split} \eeq
As in the classical case, because of the symmetry between colors, $q_0 = 1/\qcol$, and $q_1 \geq 1/\qcol$, with equality on the RS solution.

\subsection{Expected features for the quantum model}

\subsubsection{Quantum phase transitions}
The ground state of the classical coloring problem is diagonal in the
computational basis, while on the other side of the interpolation the ground state of the quantum Hamiltonian $\hHq$ is diagonal in the tensorial product of 
the eigenvectors of the flipping operators $\hT_i$. 
This defines two phases with very different physical properties. Slowly interpolating from high $\G$ down to $\G=0$, the spins have to rotate to go from the quantum paramagnetic phase to the classical spin glass one. If the classical Hamiltonian was simple or resembled the quantum one, for example if it acted as a product of single spin operators, this rotation would be easy and nothing special would happen along the way; but in our case, because of the complicated nature of the interaction (\ref{eq_def_Hp}) and of its associated classical phase, one can expect collective effects to become strongly relevant and a quantum phase transition \cite{sachdev2001} to occur along the way.

It is well established that the gap vanishes at least polynomially fast with $N$ at a quantum second order critical point \cite{sachdev2001,FaGoHe12} (it vanishes exponentially fast in some cases in presence of disorder \cite{Fisher}), while it vanishes exponentially in $N$ at a first order phase transition \cite{JKKM08,YKS10,JKSZ10}. As the scaling of the minimal gap is, according to the adiabatic theorem \cite{Messiah}, critical for the success or failure of the QAA, we have to determine in our case the order of this quantum phase transition. This will be the subject of Sec.~\ref{sec_phase_diag}.

\subsubsection{Level crossings and the role of entropy}
\label{sec_crossings_entropy}

Apart from a quantum phase transition at relatively large values of the transverse field, we also expect the addition of the transverse field to have very important effects also for much smaller $\G$. A first example of such a phenomenon was put forward in \cite{AC09,AKR10}; we sketch it very briefly here. Consider an instance of an optimization problem which has an isolated solution, and another local minimum of its cost function with only one violated clause, and such that these two configurations are far apart in Hamming distance. Computing the continuations of these energies using standard perturbation theories, one can find that, if the model is well chosen, they cross for some value of $\G$. This value shrinks to zero as $N \rightarrow \infty$, hence the name of \textit{perturbative crossings} in the literature. This mechanism is for the moment only understood in perturbation theory, and therefore it holds whenever perturbation theory holds, that is, if at small enough $\G$ the full eigenstates of the quantum problem remain close enough to the classical eigenstates. The other important ingredient is a very particular construction of the instances of the problem, that admit only one solution. But typical instances of generic random optimization problems, even close to the satisfiability threshold, have an exponentially large number of solutions, and so we expect that non-degenerate perturbation theory should not hold~\cite{KnySme10} and that the spectrum should be much more complex. Therefore we would like to understand what happens generically in problems that have multiple and not necessarily isolated solutions.

The quantum random subcubes model \cite{FSZ10,long} 
is built as a quantum extension of the classical random subcubes model~\cite{rcm} 
(note that this model makes use of Ising spins, but these could be changed into Potts spins with any number of states $\qcol$, with only irrelevant changes in numerical prefactors). 
In this model, the classical spin glass phase is represented by a set of random subcubes of the Hilbert space (representing clusters). These subcubes are supposed to be disjoint, of different sizes, and to be associated with random (classical) energies. Adding quantum fluctuations to this model gives rise, under reasonable hypotheses on the distribution of the sizes of the subcubes, to a series of level crossings induced by a combined energetic-entropic effect. More precisely, if one introduces the internal intensive entropy $s$ of a cluster (such that
a cluster contains $e^{N s}$ configurations), the complexity $\Sigma(e,s)$ such that there exist $e^{N \Sigma(e,s)}$ clusters of intensive energy $e$ with such an entropy, and $s_{\rm{max}}(e) = \sup \{s, \Sigma(e,s) \geq 0\}$, then it can be shown that the ground state energy of the model is given by:
\beq e_{\rm{GS}}(\G) = \min_{e} \left[ e - \G (\ln 2) s_{\rm{max}}(e) \right] \eeq
Henceforth, as soon as $\G > 1/((\ln 2) s'_{\rm{max}}(e_{\rm{GS}}(\G=0)))$, the minimum is obtained for a different value of $e$ for each value of $\G$.
In this region the ground state changes abruptly from one cluster to another upon changing $\G$ by an infinitesimal amount, similarly to what is called temperature chaos in classical spin glasses \cite{BrMo87,KrMa02}. Because the clusters are at Hamming distances proportional to $N$, we expect these crossings to be avoided at finite $N$ by producing exponentially small gaps. 

The analysis of the quantum random subcubes model can also be done at finite 
temperature, which reveals the existence of a condensation transition at a 
temperature $T_{\rm{K}}(\G)$ (similar to the one discussed in 
Sec.~\ref{sec_classical_coloring}). The effect of the quantum fluctuations is
particularly drastic when the classical model, at $\G=0$, remains un-condensed
down to zero temperature (i.e. when $T_{\rm{K}}(\G=0)=0$). Indeed in this case, 
for infinitesimal values of $\G>0$ the model suddenly condenses on the largest
clusters of classical groundstates, because the latter undergo the largest
decrease in energy when the transverse field is turned on. One can study more
precisely the limit $\G,T \to 0$, in which the partition function is
equivalent to
\beq
\int \dd s \, \exp \left[ N \left( \Sigma(e_{\rm{GS}}(\G=0),s) 
+ s \frac{\ln (2 \cosh(\beta \G))}{\ln 2} 
\right) \right] \ .
\eeq
The slope of the condensation line $T_{\rm{K}}(\G)$ in the small $\G$ limit
can then be deduced from a saddle-point evaluation of this expression: it
corresponds to the critical value of $\beta \G$ such that the maximum of
the argument of the exponential is reached at the upper limit of integration
$s_{\rm{max}}(e_{\rm{GS}}(\G=0))$, i.e.
\beq \label{eq_Tk_subcubes} 
T_{\rm{K}}(\G) \underset{\G \to 0}{\sim}  \G / \rm{Argcosh[2^{-\partial_s \Sigma(e_{\rm{GS}}(\G=0),s_{\rm{max}}(e_{\rm{GS}}(\G=0)))-1}}] \eeq

As discussed in Sec.~\ref{sec_classical_coloring}, we expect the coloring problem for $\qcol=4$ and $c \geq 9$ to possess a clustered classical phase and thus to exhibit a phenomenology very close to that of the quantum random subcubes model; in particular, crossings within the spin glass phase (see Sec.~\ref{sec_structure_sg}) and a linear condensation temperature $T_{\rm{K}}(\G)$ at small field for $c=9$ (see Sec.~\ref{sec_q4_c9_phase_diagram}). The fact that we shall observe these phenomena in the coloring problem confirms the relevance of this scenario, first proposed in \cite{FSZ10,long}, for generic
random optimization problems.
Let us finally briefly explain why the same properties did not appear when studying the random regular Xorsat problem \cite{JKSZ10}, even when it has a degenerate ground state (for example 4-Xorsat on a regular random graph of connectivity 3): because such a formula is locked~\cite{ZM08}, each solution is isolated and no entropic effects appear when adding quantum fluctuations. Hence the quantum coloring case is the first realistic optimization problem displaying the phenomena outlined above.

\section{The phase diagrams of the coloring model in a transverse field}
\label{sec_phase_diag}

In this section we present our results concerning the phase diagrams 
of the quantum model as a
function of the temperature $T$ and of the transverse field $\Gamma$, for two
different connectivities $c$, emphasizing the modifications of the classical
transitions (at $\Gamma=0$) once the quantum fluctuations are turned on.
Most technical and numerical details of the analysis are deferred to the
Appendices. A further investigation of the structure of the spin glass phase
is presented in Sec.~\ref{sec_structure_sg}.

\subsection{The $\qcol=4, c=13$ case}
\label{sec_q4_c13}

Let us start with the case $c=13$. We recall that in the classical case
($\Gamma=0$) the model exhibits a clustering (dynamical) transition at
$\Td(\G=0)=0.455$, a condensation transition at $\Tk(\G=0)=0.450$, and a
RS instability transition at $\Ti(\G=0)=0.441$~\cite{KZ08}
(we will not discuss the Gardner transition towards full RSB at
$T_{\rm G} =0.185$~\cite{KZ08}).
From a thermodynamic point of view the only relevant transition in this
classical limit is $\Tk$, the free energy has a discontinuity in its
second derivatives at this temperature, while it is non-singular
at $\Td$ and $\Ti$. The latter marks the limit of local stability of the RS
solution, but it is here irrelevant because of the discontinuous transition
towards the 1RSB phase that occurs at higher temperature.

\begin{figure}[t]
\includegraphics[width = 10cm] {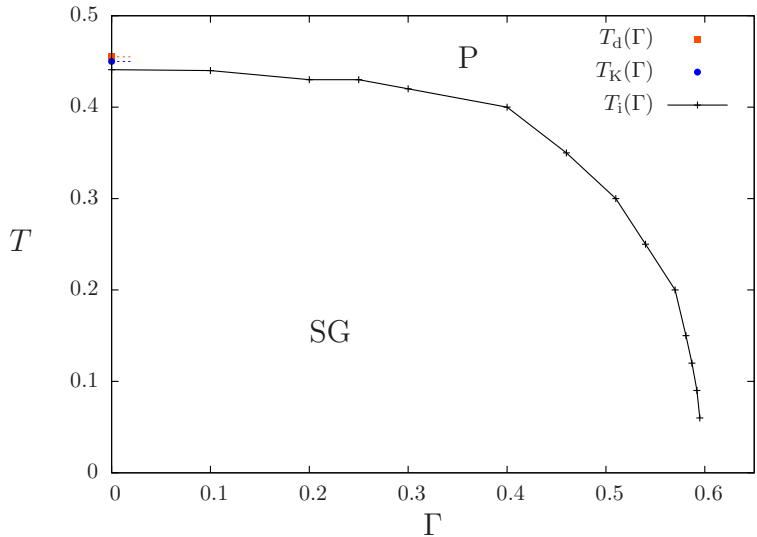}
\caption{Phase diagram of the coloring problem for $\qcol=4$ and $c=13$. 
The solid black line is the continuous transition line between the 1RSB spin glass 
phase and the RS (quantum) paramagnetic phase. For numerical accuracy reasons
we were not able to determine the continuation of the dynamical $\Td$ and 
Kauzmann $\Tk$ transitions when $\G>0$. 
}
\label{fig_phas_diagram_q4_c12}
\end{figure}

\begin{figure}[h]
\includegraphics[width=8.2cm]{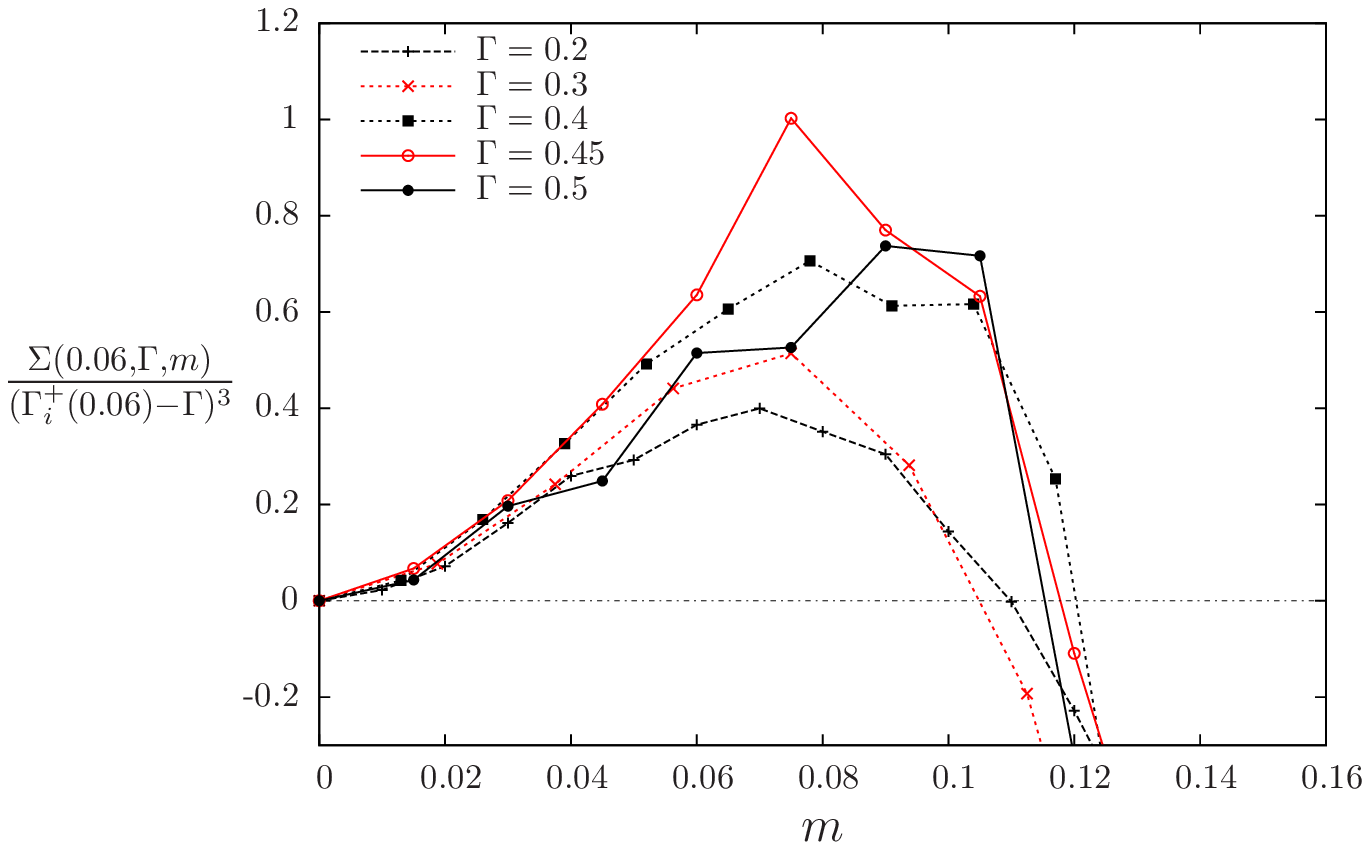} \hspace{0.5 cm}
\includegraphics[width=7cm]{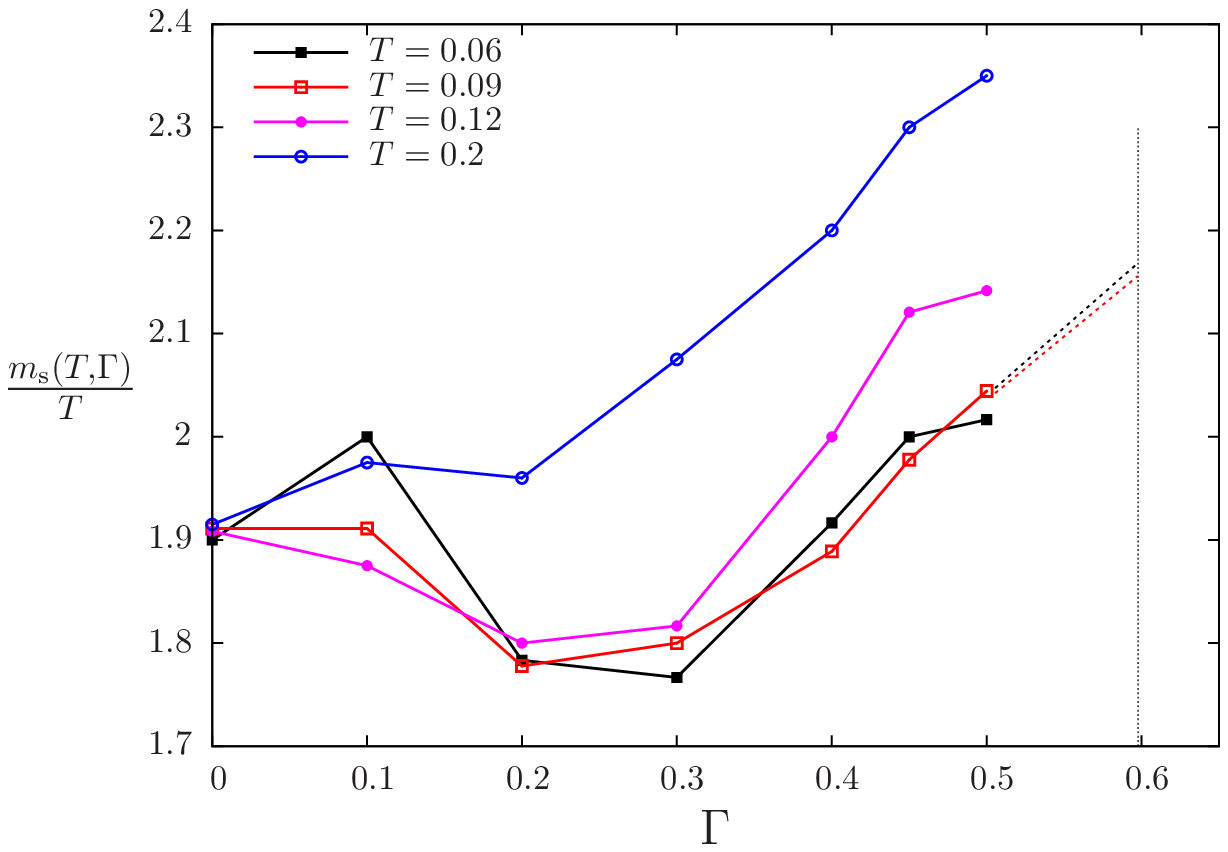}
\caption{
Results for the coloring problem for $\qcol=4$ and $c=13$.
Left panel: scaling form for the complexity upon approaching the third order transition. The temperature is fixed to $T=0.06$, while $\G$ is varied. Upon approaching the critical field $\Gi(T=0.06)=0.595$, the curves collapse. Note that the static value of $m$, $m_{\rm{s}}(T,\G)$, at which the complexity $\Sigma(T,\G,m)$ vanishes, is not singular in $\Gamma$ close to the transition. \\
Right panel: static value of the Parisi parameter divided by $T$, for various values of $T$. The dashed vertical line is the zero temperature limit for the critical field, $\Gi(T=0) = 0.598$. In the low temperature limit, curves collapse. The black (resp. red) dashed lines are fit to $m_{\rm{s}}(T,\Gamma)/T$ for $T=0.06$ (resp. $T=0.09$) near the critical field.}
\label{fig_static_parisi_q4_c12}
\end{figure}

We report on Fig.~\ref{fig_phas_diagram_q4_c12} the phase diagram that we 
obtain with the quantum 1RSB cavity method. We do find, for $\G$ large 
enough, only one transition line, $\Ti(\G)$, that corresponds to a local
instability of the RS solution towards a non-trivial solution of the 1RSB
equation. This transition is continuous, in the sense that the overlap $q_1$
defined in Eq.~(\ref{eq_def_q_quantum}) grows continuously from its replica-symmetric value $1/\qcol$ when
one enters the spin glass 1RSB phase. In order to locate it as precisely
as possible we resorted to a finite population size scaling analysis, as
explained in more details in Appendix~\ref{sec_c9_linear_instabilities}.
On this line the free energy of the model is singular, and this
corresponds to a thermodynamically third order transition. This pattern
of a second order RFOT like condensation transition (at $\Tk$ for $\G=0$) 
that becomes a 
thermodynamically third order transition upon changing one parameter was
actually met in earlier studies of mean-field disordered models, namely
in the fully-connected spherical $p$-spin model in presence of a magnetic
field~\cite{CrSo92,CrHoSo93}. For the convenience of the reader we gathered
in Appendix~\ref{app_pspin} the main aspects of the analysis of this
(classical) model; the third order character of the thermodynamic transition
derives, in both cases, from the linear growth of the overlap with respect
to the distance to the continuous transition. A further evidence for the
order of the transition is presented on Fig.~\ref{fig_static_parisi_q4_c12}:
on the left panel one observes a rather good collapse of the complexity
curves $\Sigma(m)$, for various values of the transverse field $\G$ at a fixed
temperature $T$, rescaled by $(\Gi(T)-\G)^3$. These curves vanish at the
static Parisi parameter $m_{\rm s}(\G,T)$, that admits a finite limit (smaller
than 1) when $\G \to \Gi(T)$. As shown on the right panel of 
Fig.~\ref{fig_static_parisi_q4_c12} this value is proportional to $T$ when
$T$ is reduced for a fixed value of $\G$.

In the classical limit $\G\to 0$, the line $\Ti(\G)$ falls below the classical transitions $\Td$ and $\Tk$. This means that these two classical transitions
should extend into two lines at finite but small $\G$, and at some finite $\G$ the transition must change nature from discontinuous to continuous.
However, in this model all this happens in a range of $T$ and $\G$ too small 
to be resolved within the accuracy of the numerical resolution of the quantum
1RSB equation.

We should emphasize here that the incorporation of the replica-symmetry 
breaking effects was crucial to unveil the phase diagram of this model.
As a matter of fact the RS computation predicts, incorrectly, a first order 
phase transition as a function of the transverse field at low enough
temperature. We present in Appendix~\ref{sec_c13_RS_spurious} the detailed 
numerical results and arguments we used to rule out this spurious prediction
of the RS computation.
A further confirmation on the continuous nature of the RSB
transition will be provided by the
Monte-Carlo simulations presented in Sec.~\ref{sec_structure_sg}.

\subsection{The $\qcol=4, c=9$ case}
\label{sec_q4_c9}
\label{sec_q4_c9_phase_diagram}

We now turn to the case $c=9$.
Before discussing the quantum case, we recall~\cite{KZ08} that for $\qcol=4$ and $c=9$, the model is classically satisfiable, meaning that the ground states' energy is zero and that graphs are typically colorable. These ground states are exponentially numerous and are arranged in an exponentially large number of clusters; this corresponds to the region $c_{\rm{d}} \leq c < c_{\rm{K}}$, meaning that $\Tk(\G=0)=0$, while $\Td(\G=0)=0.153$ is finite. The model is thus described, for $\G=0$ and $ T \leq \Td$, by the 1RSB equation with Parisi parameter $m=1$.

\subsubsection{Solutions of the cavity equations and spinodal lines}
\label{sec_spinodales_q4_c9}

As discussed in details in~\cite{col2,long}, the central quantity that is computed by the cavity method is the free energy function $\phi(m)$, which is the Legendre transform of the
complexity function, $\phi(m) =\min_f \{ f - T \Si(f)/m \}$. The thermodynamic free energy of the system is obtained by {\it maximizing} $\phi(m)$ in the interval $m\in[0,1]$, therefore
in the following we restrict ourselves to these values of $m$.

When solving the cavity equation, we found that in addition to the trivial RS solution, in some regions of $(T,\G)$ there exist two different non-trivial 1RSB solutions.
The first one develops continuously from the RS solution through a linear instability: hence the overlap $q_1$ grows continuously from its minimal value
$1/\qcol$, and we refer to this solution
as the ``low overlap'' solution ${\rm 1RSB_{lq}}$. The second solution appears discontinuously, as in the classical coloring problem: the overlap $q_1$ is typically
larger, hence we refer to this solution as the ``high overlap" solution ${\rm 1RSB_{hq}}$. The coexistence of two different 1RSB solutions is quite unusual, but
had been observed before: a good pedagogical example is discussed in details in Appendix~\ref{app_pspin}.

\begin{figure}[t]
\includegraphics[width = 8 cm]{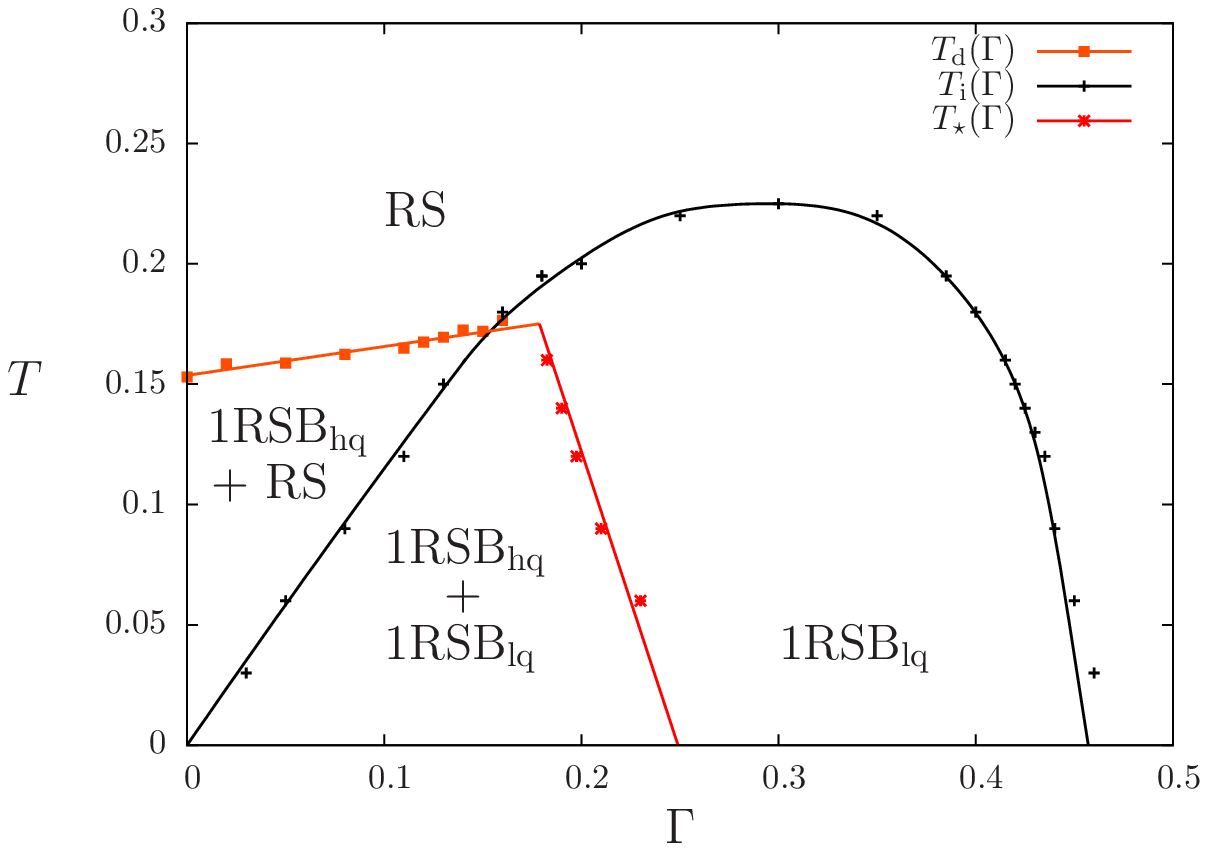}
\includegraphics[width = 7 cm]{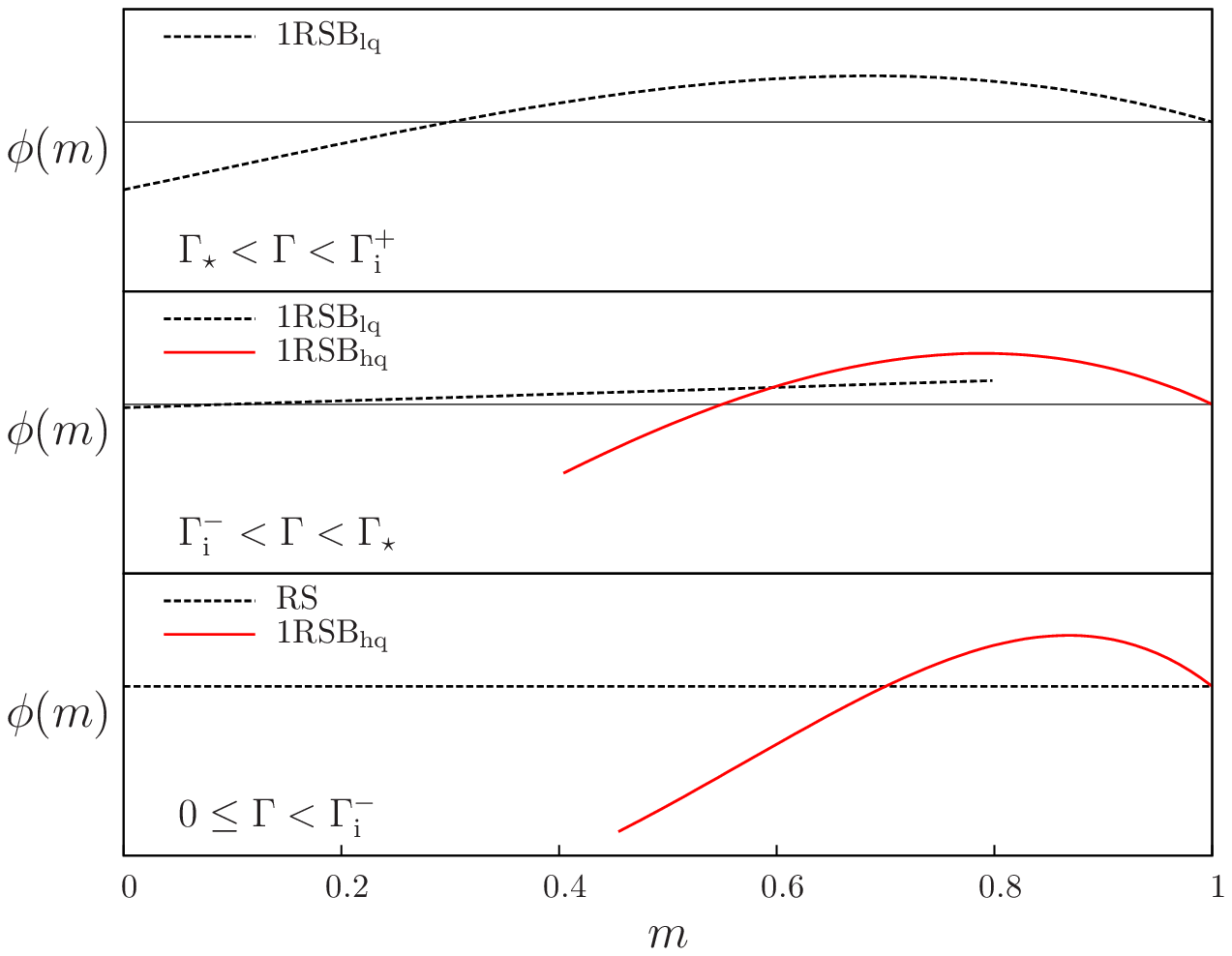}
\caption{({\it Left}) Spinodal lines of the coloring problem for $\qcol=4$ and $c=9$. 
The orange line with squares is the dynamic transition line $\Td(\G)$, the black line with ticks is the continuous transition line $\Ti(\G)$, the red line with crosses is the line $T_\star(\G)$. 
Lines are fits to the data which are shown as guides to the eye.  
({\it Right}) Sketches of the function $\phi(m)$ corresponding to the RS and the two 1RSB solutions in the three regions defined by the intersections of the lines in the left panel. For instance, this would correspond to fix $T=0.1$ and reduce $\Gamma$ from top to bottom.
}
\label{fig_phase_diagram_q4_c9_sp}
\end{figure}

We find that the ${\rm 1RSB_{lq}}$ solution exists in an interval $m \in [0, \msplq ]$, while the ${\rm 1RSB_{hq}}$ solution exists in an interval $m \in [ \msphq ,1]$,
as illustrated in Fig.~\ref{fig_phase_diagram_q4_c9_sp}. Both $\msplq$ and $\msphq$ depend on $T$ and $\Gamma$, to lighten the notations we keep implicit these dependencies.
To delimit the region of existence of these solutions in the $(T,\G)$ phase diagram, we define the following
transition lines:
\begin{itemize}
\item The clustering (or dynamic) transition line $\Td(\G)$ is defined by the condition $\msphq = 1$. It is the continuation of the classical clustering transition $\Td(\G=0)$
to the quantum regime. Above this line, the ${\rm 1RSB_{hq}}$ solution only exists outside the interval $m\in [0,1]$ and is therefore irrelevant for the thermodynamics of the model.
\item The transition line $\Ti(\G)$ is defined as the point where the RS solution becomes linearly unstable. 
Note that the instability of the RS solution is independent of $m$ and leads to the ${\rm 1RSB_{lq}}$ solution, hence we cannot have coexistence of 
the RS and ${\rm 1RSB_{lq}}$ solutions.
Above the line $\Ti(\G)$, the ${\rm 1RSB_{lq}}$ solution does not exist as it coincides
with the RS one. Below this line, the RS solution does not exist as it becomes the ${\rm 1RSB_{lq}}$ one, at least in the interval $m \in [0,\msplq]$. 
The temperature $\Ti(\G)$ is non-monotonous, hence its inverse function has two branches that we denote by $\Gamma^+_{\rm i}(T)$ and $\Gamma^-_{\rm i}(T)$.
\item A third line $T_\star(\G)$ is defined by the condition $\msplq = \msphq$. When this happens, the two distinct 1RSB solutions merge into a unique 1RSB solution.
Note that $\phi(m)$ is a convex function of $m$. When $\msplq = \msphq = m_\star$ the two solutions merge continuously
and with continuous first derivative, $\phi_{\rm lq}(m_\star) = \phi_{\rm hq}(m_\star)$ and $\phi_{\rm lq}'(m_\star) = \phi_{\rm hq}'(m_\star)$. In this sense the line $T_\star(\G)$ 
corresponds to a ``critical point'': below this line, there is a first order transition in $m$ between the two non-trivial solutions; on the line, the first order transition disappears, a
transition still exists and is of second order; above the line, there is a unique analytic 1RSB solution that we still call ${\rm 1RSB_{lq}}$ because it connects continuously to the RS one.
\end{itemize}
As illustrated in Fig.~\ref{fig_phase_diagram_q4_c9_sp}, these three lines divide the $(T,\G)$ plane in four different regions.
At high enough temperature, $T > \max(\Ti, \Td)$, only the RS solution exists. 
In the region below $\Td$ and above $\Ti$, the RS solution remains stable but it coexists with the ${\rm 1RSB_{hq}}$ one (right, lower panel).
In the region below $\min(\Td, \Ti, T_\star)$, the two non-trivial 1RSB solution coexist (right, middle panel). 
In the region below $\Ti$ and above $\max(\Td,T_\star)$, only the ${\rm 1RSB_{lq}}$ exists, either because the ${\rm 1RSB_{hq}}$ has moved outside the interval
$m\in[0,1]$ at $\Td$, or because the two solutions have merged at $T_\star$ (right, top panel). 
Note that in the numerical computation each solution, for the same point of the phase diagram, can be reached by following different routes: starting from high $\G$ and decreasing it to reach the ${\rm 1RSB_{lq}}$ solution, or starting from $\G=0$ and increasing it to reach the ${\rm 1RSB_{hq}}$ solution.

\subsubsection{Thermodynamic phase diagram}

\begin{figure}[t]
\includegraphics[width = 8 cm]{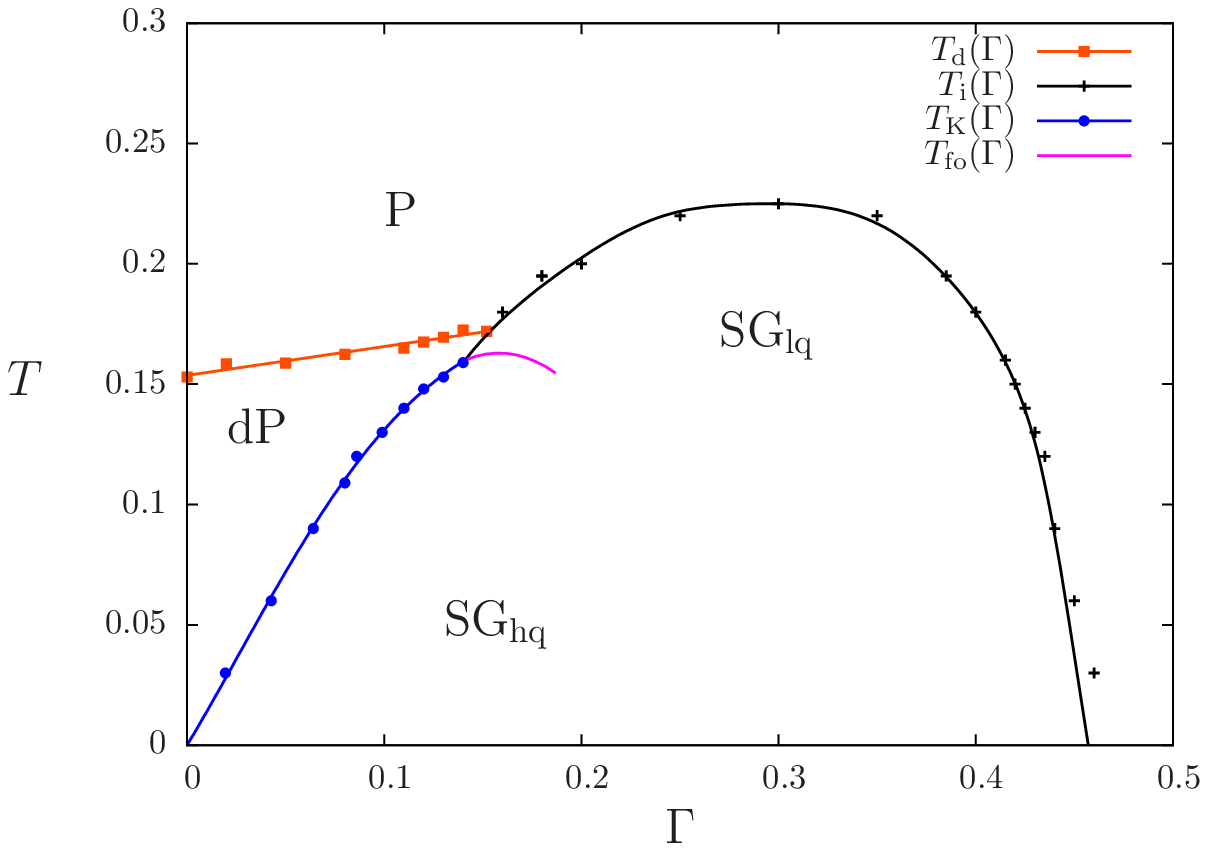}
\includegraphics[width = 8 cm]{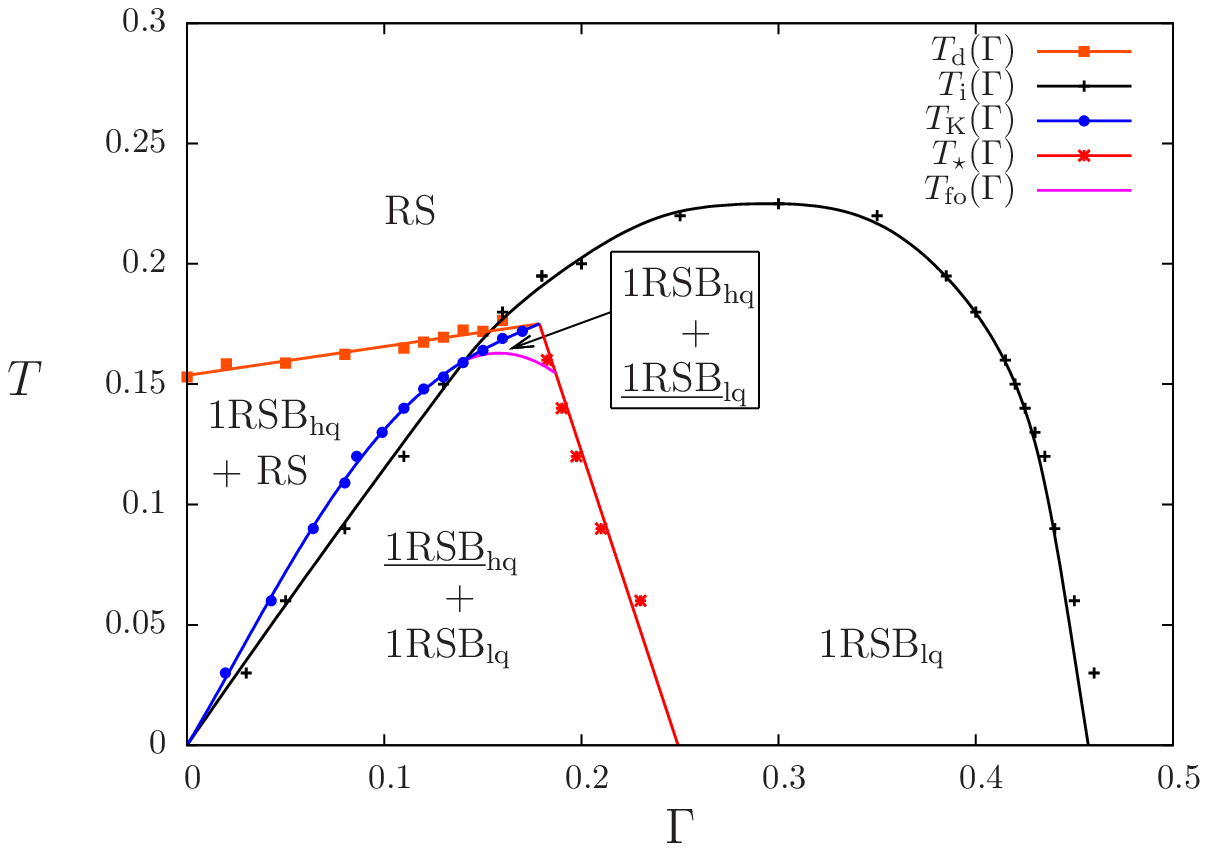}
\caption{Phase diagram of the coloring problem for $\qcol=4$ and $c=9$. In the left panel only the thermodynamic lines and the thermodynamic phases are reported. 
The right panel contains all the lines reported
in the left panel as well as in Fig.~\ref{fig_phase_diagram_q4_c9_sp}: when two solutions of the cavity equations coexist, the stable one is underlined.
The orange line with squares is the dynamic transition line $\Td(\G)$, the black line with ticks is the continuous transition line $\Ti(\G)$, the red line with crosses is the line $T_\star(\G)$, 
the blue line with circles is the condensation transition $\Tk(\G)$. Lines are fits to the data which are shown as guides to the eye. In addition, we report a purely conjectural shape
of the first order transition $T_{\rm fo}(\G)$ as a purple full line.
At large $\Gamma$, $\Ti(\G)$ ends at $\Gamma_{\rm i}^+(T=0) \simeq 0.47$.}
\label{fig_phase_diagram_q4_c9}
\end{figure}

The thermodynamic free energy of the problem corresponds to the maximum of $\phi(m)$, which is reached in $m=m_{\rm{s}}(T,\G)$.
We have seen that in some regions of the phase diagram several solutions for $\phi(m)$ can coexist, leading to different branches of this function.
This leads to several thermodynamic phase transitions that we now describe. For clarity, in Fig.~\ref{fig_phase_diagram_q4_c9} we report
in the left panel the thermodynamic lines only, while in the right panel we report the complete phase diagram including the spinodal lines
that were already discussed in Fig.~\ref{fig_phase_diagram_q4_c9_sp}.

Let us recall the following properties of the 1RSB solution. First of all, $\phi(m=1)$ is always equal to the replica-symmetric free energy. Second,
the complexity $\Si(m) = \beta m^2 \phi'(m)$ is related to the first derivative of $\phi(m)$.
We find the following phase transitions in the model:
\begin{itemize}
\item At high temperature the system is in a RS paramagnetic (P) phase. Upon lowering the temperature at low enough $\G$, the dynamical transition $\Td(\G)$ is met and
below it the ${\rm 1RSB_{hq}}$ appears.
As in the classical limit $\G=0$, for $\G$ small enough the new solution is such that $\Si(m=1) \propto \phi'(m=1) > 0$. By convexity, for any $m<1$ the free energy of the
${\rm 1RSB_{hq}}$ solution is smaller than the one of the RS solution and the system remains in the paramagnetic phase. However, as in the classical case the paramagnetic
phase is not ergodic because it is a superposition of many metastable states, and we refer to it as ``dynamical paramagnet'' or dP.
\item Upon lowering further the temperature, the complexity $\Si(m=1)$ decreases and reaches zero at the
condensation temperature $\Tk(\G)$. At this point the slope of $\phi_{\rm hq}(m)$ in 1 becomes negative, in such a way that $\phi_{\rm hq}(m)$ has a maximum in the interval
$m\in [0,1]$, and its value at the maximum is necessarily larger than the RS solution (Fig.~\ref{fig_phase_diagram_q4_c9_sp}, right, lower panel). Hence at $\Tk$ a thermodynamic
transition happens between the RS and ${\rm 1RSB_{hq}}$ phase. This transition is a standard RFOT transition: it is of second order thermodynamically, because the maximum of
$\phi_{\rm hq}(m)$ moves smoothly enough away from $m=1$, but it is of first order from the point of view of the order parameter because the system is jumping discontinuously
from the RS to the ${\rm 1RSB_{hq}}$ phase. We call the spin glass phase below $\Tk$ the ${\rm SG_{hq}}$ phase.
Note that we find numerically that the line $\Tk(\G)$ grows linearly with $\G$ at small $\G$: this 
is a consequence of the clustered structure of the classical spin glass phase, as will be discussed in Sec.~\ref{sec_q4_c9_clusters}.
\item If the temperature is lowered below $\Ti(\G)$ at high enough $\G$, the RS phase becomes linearly unstable before the line $\Tk$ is met. In this case, the possible presence
of the ${\rm 1RSB_{hq}}$ solution is irrelevant as we already discussed. On the other hand, the solution ${\rm 1RSB_{lq}}$ appears at $\Ti(\G)$ and we always find that it appears
with a negative complexity at $m=1$, $\Si(m=1)<0$, which implies that the slope of $\phi_{\rm lq}(m)$ in $m=1$ is negative. The maximum of $\phi(m)$ is thus assumed on the ${\rm 1RSB_{lq}}$ in this
case (Fig.~\ref{fig_phase_diagram_q4_c9_sp}, right, upper panel). The transition at $\Ti(\G)$ is a thermodynamical third order transition (see Appendix~\ref{app_pspin}), 
and the overlap grows continuously from its replica-symmetric value.
We call the phase below $\Ti$ the ${\rm SG_{lq}}$ phase.
\item Finally we should discuss the behavior in the spin glass region, for $T< \min(\Tk,\Ti)$. The line $\Tk(\G)$ can be continued in this region, and above it $\phi'_{\rm hq}(m=1)>0$, hence
the ${\rm 1RSB_{hq}}$ is surely metastable with respect to the ${\rm 1RSB_{lq}}$ solution. However, below this line, $\phi'_{\rm hq}(m=1)<0$, 
hence both the ${\rm 1RSB_{lq}}$ and ${\rm 1RSB_{hq}}$ solutions have a maximum for $m\in [0,1]$. This can lead to a thermodynamical first order transition between these two solutions
if the values of $\phi(m)$ at the two maxima cross. This must happen at some temperature $T_{\rm fo}(\G) < \Tk(\G)$, because we know that the ${\rm 1RSB_{lq}}$ solution is stable
at high temperature while the ${\rm 1RSB_{hq}}$ is stable at low temperatures. Unfortunately, the free energy differences are so small in this region that our numerical accuracy does not
allow us to determine the first order transition line. The line drawn in Fig.~\ref{fig_phase_diagram_q4_c9} is therefore schematic. We can only say that the variation of the complexity at $m=1$
is found to be much higher in the ${\rm 1RSB_{hq}}$ solution than in the ${\rm 1RSB_{lq}}$ one.
This is actually related to the fact that $\Tk$ is a second order transition while $\Ti$ is a third order transition. It implies in particular that $T_{\rm fo}$ must be tangent to $\Tk$ when
they separate, and it also suggests that $T_{\rm fo}(\G)$ must be always very close to $\Tk(\G)$.
Note also that the line $T_{\rm fo}(\G)$ must necessarily end on the line $T_\star(\G)$ where the distinction between the two 1RSB phases disappears.
\end{itemize}
As a final remark, we note that within our numerical accuracy it seems that the three lines $\Td(\G)$, $\Tk(\G)$ and $T_\star(\G)$ cross at a single point. This finding is perfectly compatible
with all of our data. To better understand this point let us call $(T_0,\G_0)$ the point where the lines $\Td(\G)$ and $T_\star(\G)$ cross.
If $\Tk(\G)$ does not cross the lines in the same point, there are two possibilities: {\it (i)} $\Tk$ crosses $T_\star$ at some $T<T_0$. However, this can be excluded
because at $T_\star$ the two 1RSB solutions must merge and convexity implies that the slope of $\phi_{\rm hq}$ in $m=1$ must be negative. 
{\it (ii)} $\Tk$ crosses $\Td$ at $\G < \G_0$. This scenario seems to us logically possible. However one must stress that in most cases the lines $\Tk$ and $\Td$ do not cross,
but merge into some kind of critical point. Therefore the scenario of a triple crossing at $(T_0,\G_0)$ seems likely. Unfortunately, we could not devise a more solid argument.

In summary, the thermodynamic phase diagram contains a paramagnetic (P) phase, a dynamical paramagnet (dP), a ``high overlap'' spin glass (${\rm SG_{hq}}$) and
a ``low overlap" spin glass (${\rm SG_{lq}}$). The transition at $\Td$ from the P to the dP phase is a standard dynamical (clustering) transition, and has no thermodynamic consequences.
The transition $\Tk$ from dP to ${\rm SG_{hq}}$ is also a standard RFOT transition, i.e. a second order transition. The transition $\Ti$ from P to ${\rm SG_{lq}}$ is a third order thermodynamic
transition. Finally, a first order thermodynamical transition between ${\rm SG_{lq}}$ and ${\rm SG_{hq}}$ must exist, even if our numerical accuracy is not sufficient to determine it precisely.

In particular we find that in the limit of zero temperature, the system becomes a spin glass under the action of an infinitesimal transverse field. At very low temperatures the two spin glass phases transform smoothly into each other, and the only phase transition is the third order transition at $\G_{\rm i}^+(T\to 0) \simeq 0.47$.

The artifact of the spurious first order transition predicted by the RS 
computation, discussed for $c=13$ at the end of Sec.~\ref{sec_q4_c13} and 
in Appendix~\ref{sec_c13_RS_spurious}, occurs also in the case $c=9$, 
even if at even lower temperatures with respect to $c=13$.

\section{Structure of the spin glass phase}
\label{sec_structure_sg}

\subsection{Clusters crossings in the spin glass phase}
\label{sec_q4_c9_clusters}

As sketched in Sec.~\ref{sec_crossings_entropy}, apart from the various transition lines that appear for this model when the quantum field is turned on, we also expect the spin glass phase to exhibit an interesting quantum behaviour due to its rich classical structure. The intuition based on the study of the quantum
random subcubes model~\cite{FSZ10,long} recalled in 
Sec.~\ref{sec_crossings_entropy} is that a continuum of (avoided)
level crossings occurs in the spin-glass phase, because the classical model
admits a pure state decomposition with clusters of various energies and 
entropies. As clusters with a larger classical energy have also a larger
entropy, their total (quantum) energy is reduced faster when $\G$ increases,
hence the crossings which lead to exponentially small gaps because of the
extensive Hamming distance between clusters. The linear behavior of the
line $\Tk(\G)$ reported in Sec.~\ref{sec_q4_c9} was a first confirmation of 
this mechanism, we report here a further evidence in its favour (the following 
discussion was already partially undertaken in \cite{long}).

The coloring problem for $\qcol=4, c=9$ is in its clustered phase at zero temperature and zero field. Its clusters have internal entropies that are distributed according to a large deviation function (the complexity), 
$\N(s) = \exp[N \Si(s)]$.  Typical configurations are found in clusters that 
have zero energy and value of the entropy $s^*$ such that $\Si'(s^*)=-1$, 
and the corresponding complexity $\Si^*=\Si(s^*)$ is strictly positive.
In particular, for this model $\Sigma^* \simeq 0.012$ and $s^* \simeq 0.08$.
However, many other clusters with larger and smaller entropies exist, as well as clusters with positive energies. 
When $\G\gtrsim 0$, each cluster $A$ of degenerate classical states transforms continuously into a set of quantum states, 
the lowest of which (the ``ground state of cluster $A$'' $|GS(A)\ra$) has an energy per spin
\beq\label{eq:e_COL}
\begin{split}
e(A,\G) &= e_{\rm cl}(A,\G) - \G m_x(A,\G) \ , \\
e_{\rm cl}(A,\G) &= \langle GS(A) |  \widehat{H}_P | GS(A) \rangle/N \ , \\
m_x(A,\G) &=  \langle GS(A) |  \sum_i \hT_i | GS(A) \rangle/N \ .
\end{split}\eeq
The key observation is that, as in the quantum random subcubes model \cite{FSZ10}, we expect $m_x(A,\G)$ to be finite when $\G\to 0$,
$\lim_{\G\to 0} m_x(A,\G) = m_x^0(A)$, because there exist degenerate classical 
ground states at Hamming distance 1 inside $A$, and we expect $m_x^0(A)$ to be positively
correlated with the classical entropy of the cluster.
At the same time, $e_{\rm cl}(A,\G) \propto \G^2$ at small $\G$.

Therefore, the energy of the ground state of a cluster is linear at small $\G$, $e(\G) \sim -\G m_x^0(A)$.
Largest clusters yield the greatest decrease in energy when quantum fluctuations are switched on, 
and they dominate at zero temperature as soon as $\G>0$.
Because these are the states with maximal entropy, they correspond to $\Si(s_{\rm max})=0$.
Hence as soon as $\G>0$, the zero temperature complexity abruptly drops to zero. 
In other words, we expect the system to condense into the largest clusters under an infinitesimal amount of quantum fluctuations. 
This in particular implies that, as in the quantum random subcubes model,
a non-zero $\Tk(\G)$ should emerge, and that $\Tk(\Gamma) \propto \G$ for 
small $\G$, the effect of the transverse field entering the computation of
the free energy under the combination $\beta \G$.
This feature is confirmed by the phase diagram presented on 
Fig.~\ref{fig_phase_diagram_q4_c9}.

\begin{figure}[t]
\includegraphics[width=9 cm]{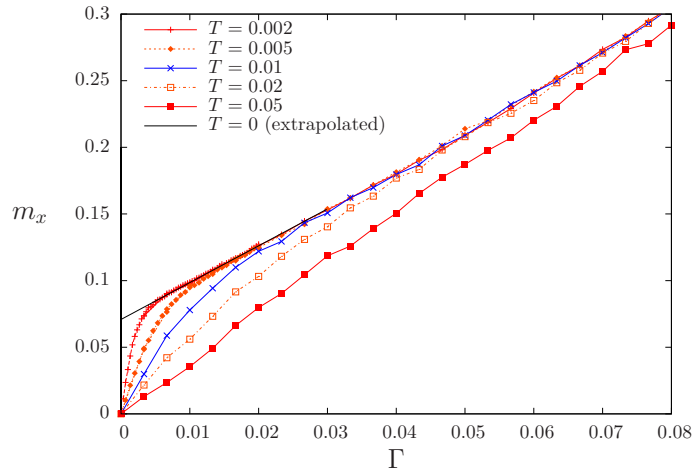}
\caption{Transverse magnetization obtained by path-integral Monte-Carlo simulations with $N=1000$ spins. All the simulations were started from the same planted state at $T=0$, and then runned increasing $\G$ at fixed $T$. The black line is the extrapolation at $T=0$, showing that $m_x^0(A) = \lim_{\Gamma \rightarrow 0} \lim_{T \rightarrow 0} m_x(A,T, \Gamma) > 0$.}
\label{fig_mx_col}
\end{figure}

Let us now probe the hypothesis $m_x^0(A)>0$, first numerically. Our path integral formalism does not allow to work directly at zero temperature, hence we had to run several Monte-Carlo simulations at low temperature and extrapolate the results to the limit $T \rightarrow 0$. In this case, because the system is in a dynamical 1RSB phase at $\G=0$, it is possible to use ``quiet planting'' \cite{QuietPlanting} to construct configurations equilibrated at $T=0$. We started all the simulations in the same quietly planted configuration at $\G=0$, for the same instance of the coloring problem; in this way we assumed that the simulations all follow the evolution of the same cluster. Extrapolating the results to $T=0$ gives the ground state properties of the cluster as $m_x^0(A)=\lim_{\Gamma \rightarrow 0} \lim_{T \rightarrow 0} m_x(A,T, \Gamma)$, as shown on Fig.~\ref{fig_mx_col}. Note that at the very low temperatures we investigated, and given
the size of the graph, the planted solution at $T=0$ is actually an equilibrium
configuration of the model at $\G=0$ for all temperatures of Fig.~\ref{fig_mx_col}.

It is also possible to get a theoretical understanding of the existence of this finite limit for $\lim_{\Gamma \rightarrow 0} \lim_{T \rightarrow 0} m_x(A,T, \Gamma)$. First of all in the quantum random subcubes model, one can derive the exact expression $m_x(A,T,\Gamma) = \frac{s(A)}{\ln 2} \tanh( \Gamma/T)$. Hence $m_x^0(A)=\frac{s(A)}{\ln 2}$, and a perturbation theory in $\Gamma$ to compute $m_x(A,T,\Gamma)$ will be divergent when $T \rightarrow 0$, the coefficient of the term linear in $\Gamma$ being proportional to $m_x(A) /T$. For the coloring problem, no closed expression can be derived for $m_x(T,\Gamma)$, but it is still possible to compute it perturbatively in $\Gamma$ within the path-integral representation of the partition function. The details of the computation are given in Appendix~\ref{app_small_field}. The important conclusion is that in the limit of small $\G$ and $T$ (with the limit $\G \to 0$ taken before $T\to 0$) one gets the asymptotic 
\beq
m_x(A,T,\G) \sim  \frac{\G}{T} \frac{1}{N} \sum_{i=1}^N
\left \langle q_{\rm auth}(\us,i) -1  \right \rangle_A \ , 
\label{eq_dev_mx}
\eeq
where $\us$ is
an uniformly chosen groundstate of the cluster $A$, and $q_{\rm auth}(\us,i)$ 
denotes the number of colors the site $i$ can take without creating a 
monochromatic edge if the rest of $\us$ is kept fixed (in other words the
number of colors that do not appear in the neighbors of $i$ in $\us$).
The expansion with $\G \to 0$ before $T \to 0$ is thus the same as for the quantum random subcubes model. Altough we cannot prove that $m_x^0$ (i.e. with the order of the limits reversed) is again proportional to $\la q_{\rm auth}-1\ra_A$ in this case, we expect that it will still be positively correlated with it, and with the internal entropy of the cluster. This could be checked numerically by repeating the above computation for many different clusters of different entropies. In any case the demonstration given above of a non-zero $m_x^0(A)$ is an additional piece of evidence in favour of the mechanism of crossings induced by the competition between the energy and the entropy of the classical pure states.

\subsection{Quantum Monte Carlo annealing}

\begin{figure}[t]
\includegraphics[width = 7cm]{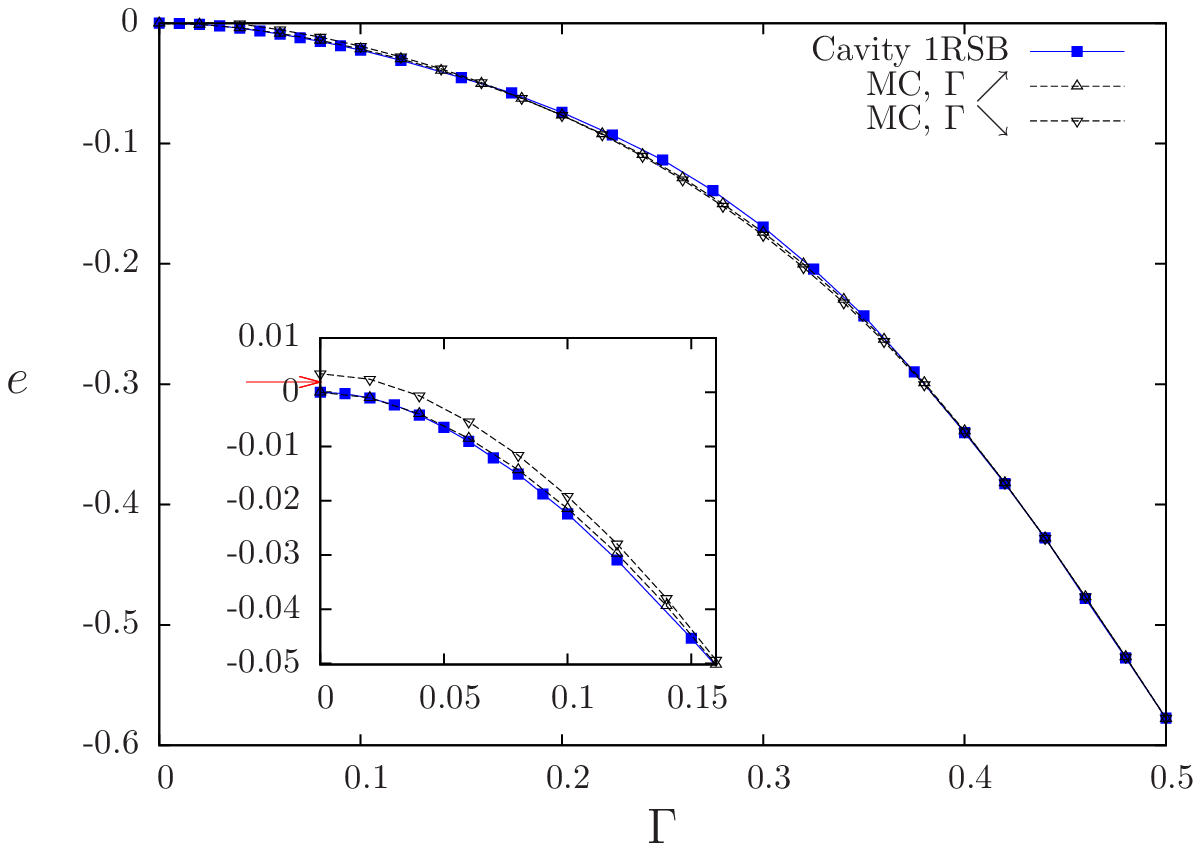} \hspace{0.5 cm}
\includegraphics[width = 7cm]{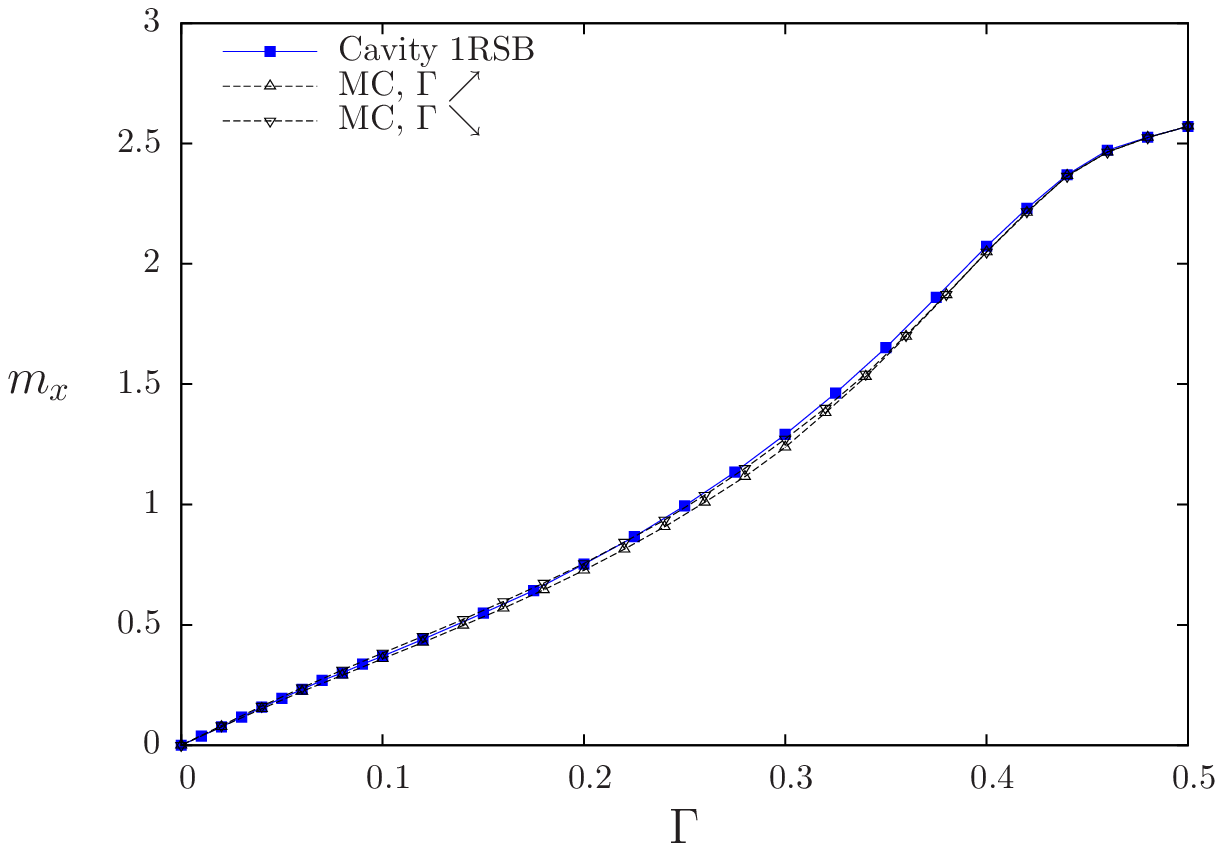}
\caption{ Energy (left panel) and transverse magnetization (right panel) as a function of $\G$ at fixed temperature $T=0.06$ for $\qcol=4$ and $c=9$.
Quantum cavity computations are reported as blue solid lines, path-integral Monte-Carlo simulations (for a single system of size $N=10000$) as dashed black lines with triangles, both for increasing and decreasing $\G$. The inset of the left panel shows a zoom on the region of low $\G$ where hysteresis is observed in the Monte-Carlo simulation. The red arrow indicates the value of the classical energy (0.0019) that corresponds to a classical annealing starting from $T_{\rm d}$.}
\label{fig_mc_q4_c9}
\end{figure}

Let us conclude this section by pointing out the consequences of this clustered structure of the spin glass phase for a quantum annealing of the problem. One expects that, because of this complicated structure, a quantum annealing of the problem starting from high $\G$ down to $\G=0$ will remain stuck in large clusters of higher energy when $\G$ gets small, and will not be able to find the smaller clusters that contain the ground states. However, it is not possible to simulate the real-time Schr\"odinger quantum dynamics of such a diluted model for reasonable values of $N$ (the size of the Hilbert space is equal to $4^N$). Still, the effects of clustering can be seen on a \textit{classical} alternative to quantum annealing, namely the annealing of a path integral Monte-Carlo computation (PIMC), whose results are shown on Fig.~\ref{fig_mc_q4_c9}. These simulations were performed in two ways: in the first run, we prepared a typical graph together with one of its typical configurations at a very small temperature $T=0.06$ via the quiet planting technique~\cite{QuietPlanting}; we initialized the PIMC at $\G=0$ in this configuration (which, given the size of the graph, is actually at zero energy), and then slowly increased $\G$. Note that the results of this run perfectly match the results of the 1RSB quantum cavity computation, confirming the validity of our analysis. In the second run, we initialized and equilibrated the PIMC at $\G=2$ (in the paramagnetic phase), and then decreased $\G$ down to $\G=0$. On the scale of the figure, no difference between the two PIMC runs is observed, however a closer look (inset of left panel in Fig.~\ref{fig_mc_q4_c9}) reveals that decreasing $\G$ one obtains a positive residual energy at $\G=0$.  The latter is found to be larger than the residual energy after an infinitely slow thermal annealing, which is obtained by preparing a quietly planted configuration at $T_{\rm d}$ and performing a slow classical annealing down to $T=0$~\cite{QuietPlanting,0295-5075-90-6-66002}.
Although a direct comparison is not possible because the PIMC annealing has not been extrapolated to the infinitely slow limit (we used steps of transverse field $d\Gamma=0.02$ with $1 000$ sweeps per step), this result suggests that an annealing of a PIMC simulation is not more efficient than a thermal annealing for this model. We believe that this is once again related to entropic level crossings inside the spin glass phase, as in the quantum random subcubes model. Note that the residual energy of the PIMC annealing is \textit{a priori} not directly related to the one of a true quantum annealing.

\begin{figure}[t]
\includegraphics[width=7cm]{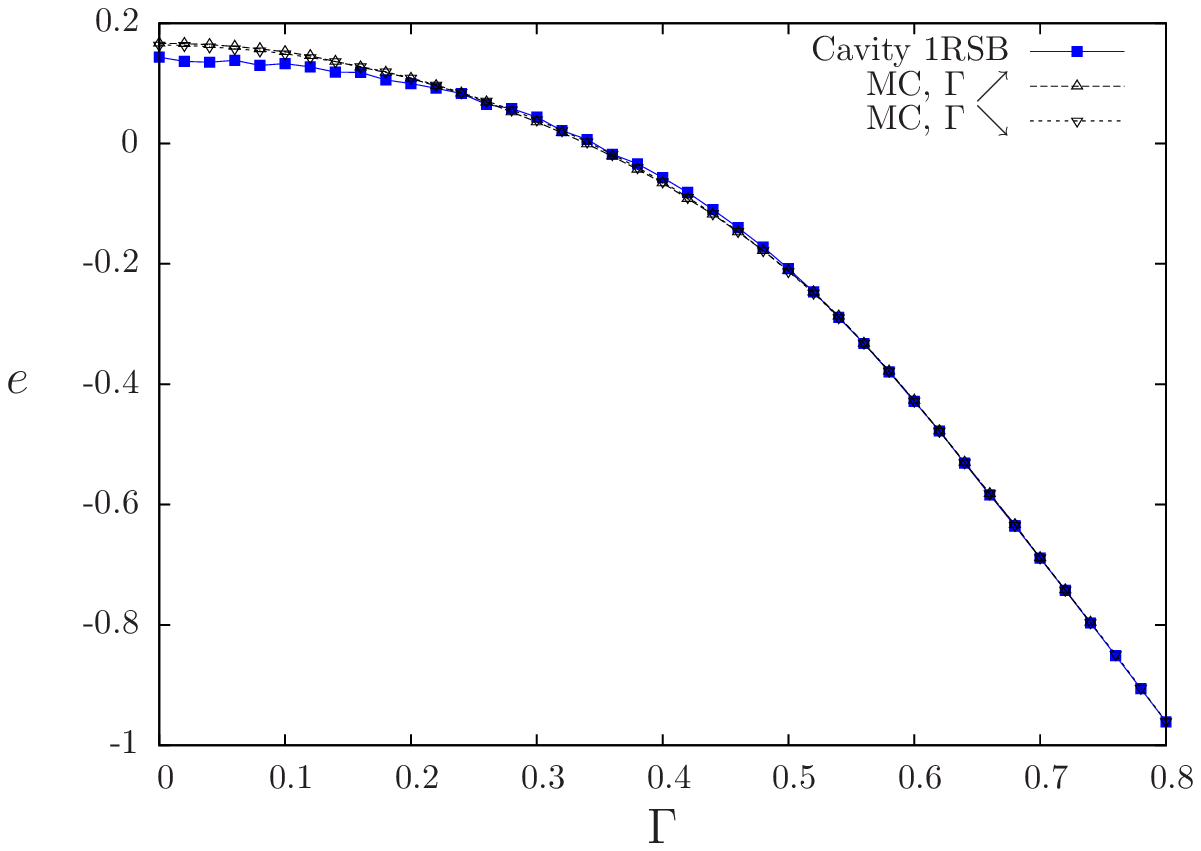} \hspace{ 0.5 cm}
\includegraphics[width=7cm]{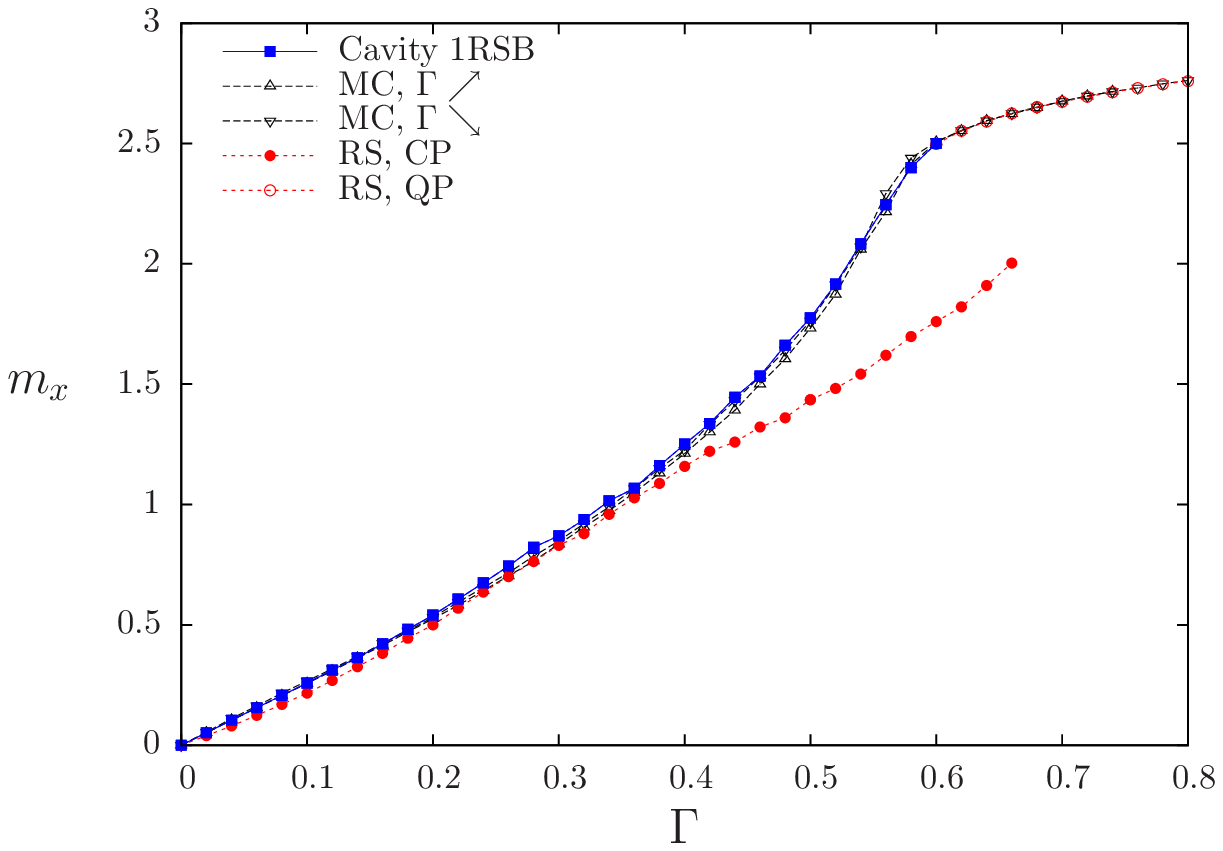}
\caption{Energy (left panel) and transverse magnetization (right panel) as a function of $\G$ at fixed temperature $T=0.06$ for $\qcol=4$ and $c=13$. Quantum 1RSB cavity computations are reported as blue solid lines, path-integral Monte-Carlo simulations (for a single system of size $N=10000$) as dashed black lines with triangles, both for increasing and decreasing $\G$; those obtained starting from $\G=0$ were obtained after a slow thermal annealing of the system from $\Td(\G=0)$ (in this case, because $\Tk(\G=0)$ is finite, it is not possible to use quiet planting at low temperature), hence the small discrepancy in the energy for $\G=0$ between the cavity computation and the Monte Carlo results. The (erroneous) replica-symmetric results (for the magnetization only) are shown in red with dash dotted lines and circles; the absence of hysteresis around the transition ($\G \simeq 0.56$) confirms that the first-order transition predicted by the RS computation is spurious.}
\label{fig_mc_q4_c12}
\end{figure}

Similar results are obtained for connectivity $c=13$, however in this case the quiet planting technique cannot be used at $T=0$ because $\Tk>0$. 
Therefore we used as a starting point for the increasing $\G$ run a classical 
configuration obtained through a slow thermal annealing from $\Td$. These results are reported on Fig.~\ref{fig_mc_q4_c12}.

\section{Conclusions}
\label{sec_conclusion}

In this paper we presented a study of the classical coloring problem, to which a quantum transverse field is added to represent the action of a quantum computer performing a quantum annealing with the
aim of finding solutions to the classical problem.
The study of this particular problem was motivated by the fact that among classical random optimization problems, it is the simplest one 
that shows an exponential degeneracy of solutions {\it inside clusters} (in the jargon of~\cite{ZM08},
it is a non-locked problem).
Therefore, this is the simplest non-trivial model where the predictions of~\cite{FSZ10}, that were obtained on a toy model, could be tested. We believe that similar conclusions would be reached
on similar problems, such as random $K$-SAT.

Our main results are the characterization of the phase diagram in the $(T,\G)$
plane of parameters, and a further description of the spin-glass phase
at low temperature and transverse field. Concerning the latter issue, we
found evidences that the mechanism described in~\cite{FSZ10} is at work
also in the case of the coloring, more precisely that {\it (i)} at low $\G$, the line $\Tk(\G)$ is linear in $\G$ for $c=9$, with $\Tk(\G=0)=0$ and $\Td(\G=0)>0$ (see Fig.~\ref{fig_phase_diagram_q4_c9}); 
and that {\it (ii)} the transverse magnetization of
the ground state of a cluster goes to a positive constant in the limit $\G\to 0$ (see Fig.~\ref{fig_mx_col}).
Both these results are direct consequences of the exponential degeneracy
of solutions in the clusters~\cite{FSZ10}, and they show that the general scenario proposed in~\cite{FSZ10} using a toy model is at work in realistic random optimization problems.

A new feature of the present study, which could not be expected based on the analogy with the random subcubes model, 
is that the quantum phase transition between the spin glass and quantum paramagnetic phases at low temperature is of third order
thermodynamically. Although we do not know the scaling of the gap at such a transition, the analogy with the results of~\cite{FaGoHe12} leads us to believe that the gap will be
polynomial right at the quantum phase transition. On the other hand, based on the results of~\cite{FSZ10}, we expect that the spin glass phase is characterized by a continuum 
of level crossings with an everywhere exponentially small gap. It would be extremely interesting to study numerically the gap in given instances of this model; exact diagonalization
is impossible due to the rapid growth of the Hilbert space with $N$ (as $4^N$), however this might be doable through Quantum Monte Carlo following~\cite{HY11,FaGoHe12}.

\acknowledgments
We warmly thank Laura Foini and Florent Krzakala for useful discussions related to this work.
Numerical computations were performed in part at the MesoPSL computing center, with support from R\'egion Ile de France and ANR, in part using HPC resources from GENCI-CCRT/TGCC (Grant 2012056924), and in part using local computational resources that were bought thanks to the PIR grant of ENS ``Optimization in complex system''.

\appendix

\section{A reminder on the fully-connected $p$-spin spherical model in a longitudinal field}
\label{sec_spherical_pspin}
\label{app_pspin}

\subsection{Definition and phase diagram}

During the study of the quantum version of the coloring problem we have
encountered several phenomena (in particular a third order continuous 
transition towards a spin glass phase, and multiple RSB solutions) that
also appear in a much simpler, classical disordered model, namely
the fully-connected $p$-spin spherical model in a field~\cite{CrSo92}. 
Analytical computations, and in particular expansions around the transition 
line, can be performed explicitly in this model and have constituted a very 
useful guideline for the analysis of the quantum coloring model. For these 
reasons we briefly collect in this Appendix the main results on this 
well-known model that are enlightening with this application in mind and
refer the reader to~\cite{CrSo92,CrHoSo93,CuKu93,CaGaGi99,CC05,FrTr06} 
for more details on this model.

Its Hamiltonian is
\beq 
\label{eq_def_ham_pspin} 
H(\s_1,\dots,\s_N) = - \sum_{1 \le i_1 < \dots < i_p \le N} J_{i_1, \dots, i_p} 
\s_{i_1} \dots \s_{i_p} - h \sum_{i=1}^N \sigma_i \ , 
\eeq
where the quenched couplings $J_{i_1, \dots, i_p}$ are independent Gaussian
random variables of zero mean and variance $p!/2 N^{p-1}$,
and the degrees of freedom $\s_i$ are continuous real variables subject to the
spherical constraint $\sum_{i=1}^N \s_i^2 = N$. The partition function is 
defined as the integration of the Gibbs-Boltzmann factor $e^{-\beta H}$ over the
$N$-dimensional sphere of radius $\sqrt{N}$, and the associated free energy 
density concentrates in the thermodynamic limit around its quenched average.
The latter can be computed with the help of the replica method and reads
\beq
\phi(\beta,h)= \sup_{\substack{0 \leq m \leq 1 \\ 0 \leq q_0 \leq q_1 \leq 1}} 
\phi_{\rm 1rsb}(\beta,h;m,q_1,q_0) \ ,
\label{eq_phi_pspin}
\eeq
where the variational function is
\beq 
\label{phi_1RSB_classic}\begin{split} 
\phi_{\rm 1rsb}(\beta,h;m,q_1,q_0) &= 
- \frac{1}{2 \beta} \left\{1 + \log( 2 \pi) +
\frac{\beta^2}{2} \left[ (1 -q_1^p) +m(q_1^p-q_0^p) \right]  
+ (\beta h)^2 (1-q_1+m(q_1-q_0))\right. \\ & 
\hspace{2 cm} \left. + \frac{q_0}{1-q_1+m(q_1-q_0)}
+ \frac{m-1}{m}  \log \left[1-q_1 \right] 
+ \frac{1}{m} \log \left[ 1-q_1+m(q_1-q_0) \right]  \right \} \ .
\end{split} 
\eeq
In this expression $m$ is the Parisi parameter discussed in the main text,
while $q_0$ (resp. $q_1$) is the typical overlap between two configurations in
different (resp. in the same) pure states.
Note that the 1RSB potential $\phi(m)$ discussed in the main text
corresponds to the maximization of this function with respect to
$q_1$ and $q_0$, with $m$ fixed.

\begin{figure}[h]
\includegraphics[width=8 cm]{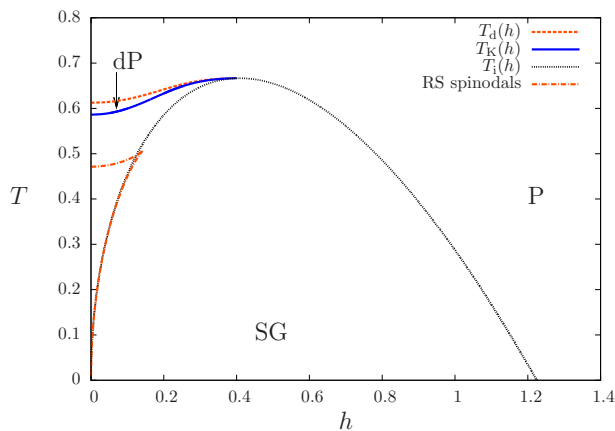}
\caption{Phase diagram of the spherical $p$-spin model for $p=3$ (all values
of $p \ge 3$ are qualitatively similar). The dashed 
orange line corresponds to the dynamic transition, 
the solid blue one to the Kauzmann (condensation) transition line, 
and the dotted black one to the continuous 
transition line. 
SG refers to the 1RSB spin glass phase, dP to the 
\textit{dynamical} paramagnet (1RSB phase with $m=1$), and $P$ to the 
paramagnetic (RS) phase. The three lines meet at the point 
$(h_{\rm c},T_{\rm c})$. The orange 
dash-dotted line is the limit of coexistence of the two RS solutions,
whose free energies cross at a spurious first order phase transition
not shown on this figure.}
\label{fig_phase_diagram_pspin}
\end{figure}

The phase diagram of this model, shown on Fig.~\ref{fig_phase_diagram_pspin},
is obtained by solving the maximization
problem defined in (\ref{eq_phi_pspin},\ref{phi_1RSB_classic}); depending
on the values of $\beta,h$ the supremum of $\phi_{\rm 1rsb}$ is either reached
in the subspace of parameters $q_0=q_1$ (this corresponds to a replica 
symmetric situation), or on a non-trivial 1RSB solution with
$q_0 < q_1$. The phase transition that separates these two regimes (and that
reveals itself as a singularity in $\phi(\beta,h)$) as a function of the 
temperature, changes qualitatively depending on the value of $h$. For $h$
smaller than $h_{\rm c}$ the transition is of
the RFOT type, as described in Sec.~\ref{sec_classical_coloring}: there
exists a dynamical transition temperature $T_{\rm d}(h)$ and a condensation
(Kauzmann) transition temperature $T_{\rm K}(h) < T_{\rm d}(h)$, the free energy
undergoing a thermodynamic phase transition at the 
latter. On this line $T_{\rm K}(h)$ the overlap order parameter is discontinuous,
yet the value of the static Parisi parameter $m_{\rm s}$, i.e. the one
maximizing (\ref{phi_1RSB_classic}), is $m_{\rm s}(h,T_{\rm K}(h))=1$, 
which makes this
discontinuous transition second order from a thermodynamic point of view. 
On the contrary for $h>h_{\rm c}$ (but of course not too large) there is
a single transition temperature $T_{\rm i}(h)$, at which the order parameter
of the RSB phase grows continuously with $m_{\rm s}(h,T_{\rm i}(h))<1$, and which 
is thermodynamically of third order. The three lines $T_{\rm d},T_{\rm K}$ and
$T_{\rm i}$ meet at $h=h_{\rm c}$.
In the following we give some details on the derivation of these properties
of the phase diagram, and also on some features that are not directly
relevant for its thermodynamic properties (a spurious first order RS 
transition and the coexistence of different 1RSB solutions 
at $m \neq m_{\rm s}$).

\subsection{The RS solution(s) and its stability limit}
\label{sec_pspin_continuous}

Let us define the replica-symmetric variational free energy, function of
a single overlap $q$, by substituting $q_0=q_1=q$ in (\ref{phi_1RSB_classic});
one then sees that this expression is independent of $m$, and is equal to
\beq
\phi_{\rm rs}(\beta,h;q) = - \frac{1}{2 \beta} \left\{1 + \log( 2 \pi) +
\frac{\beta^2}{2} (1 -q^p) + (\beta h)^2 (1-q) + \frac{q}{1-q} + \log(1-q)
\right \} \ .
\eeq
The stationary points of this function of $q$ are solutions of the
following equation,
\beq 
\label{eq_pspin_rs} 
\frac{\beta^2}{2} p q^{p-1}+(\beta h)^2 = \frac{q}{(1-q)^2} \ .
\eeq
Depending on the values of the parameters $\beta,h$ there exists either
one or three solutions to the RS equation (\ref{eq_pspin_rs}). In the latter
case the intermediate one corresponds to a minimum of $\phi_{\rm rs}$ and can
be discarded, while the two extreme ones compete to maximize $\phi_{\rm rs}$ and
thus cross at a first order transition line. The boundary of the domain of
existence of multiple RS solutions in the $(h,T)$ plane is more easily
expressed parametrically, as a function of $q$; on this boundary one
has the relation (\ref{eq_pspin_rs}) and in addition
\beq 
\label{eq_pspin_spinodal_rs} 
\frac{\beta^2}{2} p(p-1) q^{p-2}=\frac{1+q}{(1-q)^3} \ .
\eeq
The orange dash-dotted line in Fig.~\ref{fig_phase_diagram_pspin} represents
this spinodal limit of existence of multiple RS solutions. The first order
transition between these two solutions occurs on a line, not shown on the
figure, that starts at some $T>0,h=0$ and joins the spinodals at their cusp.
We shall see that this transition is irrelevant thermodynamically, as it 
occurs inside the 1RSB phase.

It is easy to check that if $q$ is a stationay point of $\phi_{\rm rs}$, then
the first derivatives of $\phi_{\rm 1rsb}$ with respect to $q_0,q_1$ and $m$ vanish
in $(q_0,q_1,m)=(q,q,m)$, for all values of $m$, i.e. that $(q,q,m)$ is a
stationary point of $\phi_{\rm 1rsb}$. Let us discuss more precisely its
nature, assuming $q$ is a local maximum of $\phi_{\rm rs}$.
All second derivatives of $\phi_{\rm 1rsb}$ with respect to $q_0,q_1,m$
which involve at least one derivative with respect to $m$ vanish, hence one 
can concentrate on the Hessian of $\phi_{\rm 1rsb}$ with respect to $q_0$ and 
$q_1$, and in particular on its determinant. A short computation leads to
\beq 
\label{eq_det_hess_pspin} \det \textrm{Hess}(m,q) 
= \left . \begin{vmatrix} \frac{\partial^2 \phi_{\rm 1rsb}}{\partial q_0^2} & 
\frac{\partial^2 \phi_{\rm 1rsb}}{\partial q_0 \partial q_1} \\ 
\frac{\partial^2 \phi_{\rm 1rsb}}{\partial q_0 \partial q_1}  & 
\frac{\partial^2 \phi_{\rm 1rsb}}{\partial q_1^2} \end{vmatrix}
\right._{\substack{q_0=q\\ q_1=q}} 
= \frac{m(1-m)}{2 \beta}  \frac{\partial^2 \phi_{\rm rs}}{\partial q^2}
\left[ \frac{\beta^2}{2}p(p-1)q^{p-2}-\frac{1}{(1-q)^2} \right] 
\ .
\eeq
As $m \in [0,1]$ and $\partial^2 \phi_{\rm rs} / \partial q^2 \le 0$ on a maximum
of the RS potential, the sign of the determinant of the Hessian can only
change when the last factor of this equation vanishes, i.e. when
\beq 
\label{eq_continuous_pspin} 
\frac{\beta^2}{2}p(p-1)q^{p-2}=\frac{1}{(1-q)^2} \ .
\eeq
The black line $T_{\rm i}(h)$ (or, alternatively, the two branches 
$h_{\rm i}^{\pm}(T)$) in Fig.~\ref{fig_phase_diagram_pspin} corresponds
to the parametric representation, as a function of $q$, of the conjoint
solution of (\ref{eq_continuous_pspin}) and of the RS equation 
(\ref{eq_pspin_rs}). It starts from $(T=0, \hi^-(T=0)=0)$ (corresponding to 
$q=0$), reaches a maximum in 
$\left (T_{\rm c}= \sqrt{2 \frac{(p-1)(p-2)^{p-2}}{p^{p-1}}},
h_{\rm c} = \sqrt{\frac{(p-2)^p}{2p^{p-2}}} \right )$ (when $q=\frac{p-2}{p}$), 
before hitting the zero temperature line again at 
$\hi^+(T=0)=\sqrt{\frac{p(p-2)}{2}}$ (for $q=1$).

Outside the region enclosed by $T_{\rm i}(h)$, i.e. at high temperature or
field, the determinant of the Hessian is positive, corresponding to two
negative eigenvalues. A local maximum of $\phi_{\rm rs}$ then corresponds to
a local maximum of $\phi_{\rm 1rsb}$ in the larger RSB subspace of parameters, 
and such a RS solution is at least locally correct. On the contrary once
$T_{\rm i}(h)$ is crossed one of the eigenvalues of the Hessian becomes
positive, there is one direction that increases $\phi_{\rm 1rsb}$ starting from
the RS local maximum, which is thus locally unstable (for all values of the
Parisi parameter in $[0,1]$). We will come back in more details
in Appendix~\ref{sec_app_linear_instability} on the behavior of the model 
around the instability line $T_{\rm i}(h)$.
Let us also mention as a technical detail the issue of the stability of the
RS solutions in their domain of coexistence. The latter is traversed by the
small field branch of the instability line, $h_{\rm i}^-(T)$; in the 
intersection of the domain of coexistence of two RS solutions with the
interior of the instability domain $T<T_{\rm i}(h)$ both RS solutions are
locally unstable. When two RS solutions coexist with $T>T_{\rm i}(h)$ the one
continuously connected to the high temperature regime is locally stable, while
the other one is locally unstable. In any case even the locally stable 
solution is irrelevant as the non-trivial 1RSB solution will have a larger
free energy, as discussed below.

\subsection{The RFOT-like transitions at small $h$}
\label{app_pspin_dyn_tr}

We have seen above one mechanism for the appearance of replica-symmetry
breaking, namely the transformation of a local maximum $q$ of $\phi_{\rm rs}$ from
a local maximum of $\phi_{\rm 1rsb}$ with $q_0=q_1=q$ to a saddle. 
Another possibility is the appearance of a local maximum of $\phi_{\rm 1rsb}$
with $q_0 \neq q_1$, i.e. far away from the replica-symmetric subspace 
(hence the ``discontinuous'' character of such a transition). At variance
with the local instability, which is independent of the value of $m$, the
occurence of the discontinuous transition is $m$-dependent. We shall start
by considering the most important case $m=1$, and come back on this 
$m$-dependence in Appendix~\ref{sec_app_coexistence}.

At $m=1$ the stationary conditions of $\phi_{\rm 1rsb}$, 
$\frac{\partial \phi_{\rm 1rsb}}{\partial q_0} = 
\frac{\partial \phi_{\rm 1rsb}}{\partial q_1} = 0 $ imply that
$q_0$ is solution of the RS equation (\ref{eq_pspin_rs}), 
while $q_1\ge q_0$ verifies:
\beq 
f(q_1)=f(q_0) \ , \qquad \text{with} \ \ 
f(x) = \frac{\beta^2}{2}p x^{p-1} - \frac{1}{1-x} \ .
\eeq
Only for some range of parameters the equation $f(x)=f(q_0)$ admits
a non-trivial solution $q_1>q_0$. On its limit of existence one has
the additional condition $f'(q_1)=0$, i.e.:
\beq 
\label{eq_pspin_q1d} 
\frac{\beta^2}{2}p(p-1)q_1^{p-2} = \frac{1}{(1-q_1)^2} \ . 
\eeq
The line $T_{\rm d}(h)$, corresponding to this limit of existence and
drawn in dashed orange on Fig.~\ref{fig_phase_diagram_pspin}, can be obtained
parametrically: from (\ref{eq_pspin_q1d}) one obtains $\beta$ as a function
of $q_1$, then $q_0$ is expressed in terms of $q_1$ as a solution of 
$f(q_0)=f(q_1)$, and finally the field is obtained from (\ref{eq_pspin_rs}),
$q_0$ being solution of this RS equation. The relevant interval for this parametrization is 
$q_1 \in [\frac{p-2}{p},\frac{p-2}{p-1}]$. At the lower limit
the line $T_{\rm d}$ merges with the local instability $T_{\rm i}$ 
at $h_{\rm c}$; in this limit the two extrema of $f$ merge, 
$q_1 \rightarrow q_0$ and (\ref{eq_pspin_q1d}) indeed reduces to 
(\ref{eq_continuous_pspin}). The upper limit $q_1=\frac{p-2}{p-1}$
corresponds to $h = 0$, $q_0=0$, and one can find the explicit expression 
$T_{\rm d}(h=0) = \sqrt{\frac{p(p-2)^{p-2}}{2 (p-1)^{p-1}}}$. 

The non-trivial 1RSB solution just found at $m=1$ might
be the global maximum of $\phi_{\rm 1rsb}$, or not. To test this one has to
compute the derivative of $\phi_{\rm 1rsb}$ with respect to $m$, in this
point. As discussed in Sec.~\ref{sec_classical_coloring} this is proportional
to the complexity; when the latter is positive $m=1$ is indeed a maximum
of $\phi_{\rm 1rsb}$, otherwise there will be a maximum at a value $m_{\rm s}<1$.
The Kauzmann, or condensation, transition separates these two regimes, and
corresponds to the line drawn in solid blue on 
Fig.~\ref{fig_phase_diagram_pspin}.

\subsection{The third order transition at large $h$}
\label{sec_app_linear_instability}

\begin{figure}[t]
\includegraphics[width=8 cm]{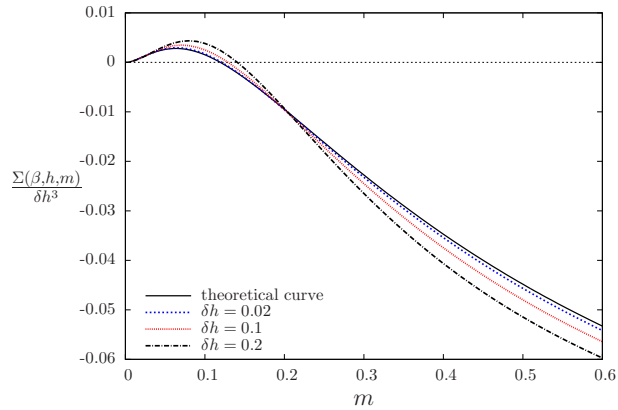}
\caption{Scaling form of the complexity $\Sigma(\beta,h,m)/\delta h^3$ of 
the spherical $3$-spin model for $\beta = 1/0.3$ and 
$h = \hi^+(\beta)-\delta h$. The solid black line was obtained from the
analytical expansion given in Eq.~(\ref{eq_dev_phi_instab}).}
\label{fig_scaling_complexity_pspin}
\end{figure}

We now come back in more details on the behavior of the free energy
around the RS local instability. We want to justify that the 
perturbations in the overlaps $q_0$ and $q_1$ with respect to their common
RS value are linear in the distance from the instability line, and
that the first discontinuities appear in the 
third derivatives of the free energy $\phi(\beta,h)$ (hence the transition 
is thermodynamically of third order). In particular, the complexity 
takes a scaling form in $(\delta T+\delta h)^3$ 
when one approaches the instability line 
(see Fig.~\ref{fig_scaling_complexity_pspin}). These last two facts have been
used in the analysis of the quantum coloring problem.

As explained in Appendix~\ref{sec_pspin_continuous}, 
when crossing the instability line $\Ti(h)$ the RS extremum goes
from a maximum to a saddle (in the larger 1RSB space of parameters).
By definition the determinant of the Hessian matrix written in 
Eq.~(\ref{eq_det_hess_pspin}) vanishes in $(h,\Ti(h))$. More precisely,
one can diagonalize the matrix in this point and find that it has one
eigenvalue 0 with eigenvector $\begin{pmatrix}-(1-m) \\ m \end{pmatrix}$
and one strictly negative eigenvalue 
associated to the eigenvector $\begin{pmatrix}m \\ 1-m \end{pmatrix}$.

Consider now some parameters $(\beta,h)$ inside the unstable region,
denote $q_{\rm rs}(\beta,h)$ the associated solution of the RS
equation, and expand the 1RSB potential around this point in powers of $\delta_+,\delta_-$ as
\beq
\delta \phi = \phi_{\rm 1rsb}(\beta,h;
q_0=q_{\rm rs}(\beta,h) - (1-m) \delta_+ + m \delta_-,
q_1=q_{\rm rs}(\beta,h) + m \delta_+ + (1-m) \delta_-, m )
- \phi_{\rm rs}(\beta,h; q_{\rm rs}(\beta,h) ) \ .
\eeq
As $q_{\rm rs}$ is a RS solution, the
expansion begins with second order terms. In the following we shall only need the
coefficients of the terms of order $\delta_+^2$ and $\delta_+^3$, denoted
$a/2$ and $b/3$ respectively, with
\bea
a &=& \frac{1}{2 \beta} m (1-m)
\left[ \frac{\beta^2}{2}p(p-1)q_{\rm rs}(\beta,h)^{p-2}
-\frac{1}{(1-q_{\rm rs}(\beta,h))^2} \right] \ , \\
b&=&-\frac{1}{2 \beta} m (1-m) \left[ 
(1-2m)\frac{\beta^2}{4}p(p-1)(p-2) q_{\rm rs}(\beta,h)^{p-3}
+ \frac{m}{(1-q_{\rm rs}(\beta,h))^3} \right] \ .
\eea
Consider now that $(\beta,h)$ is at a small distance, call it $\epsilon$,
of the instability line; a short reasoning reveals that the coefficients
of the terms in $\delta_+^2$ and $\delta_+ \delta_-$ are of order $\epsilon$,
while all others are of order 1 in this limit (in fact, note that the 
perturbation $\delta_+$ is in the direction of the vanishing eigenvalue of the
Hessian). One then sees that the maximization of $\delta \phi$ over
$\delta_+$ and $\delta_-$ will lead to $\delta_+=O(\epsilon)$, 
$\delta_-=O(\epsilon^2)$, and that at the lowest order $\delta \phi$ is
simply given by the maximization of $a \delta_+^2 /2 + b \delta_+^3 /3$.
We thus obtain in this limit
\beq
\phi(\beta,h;m) - \phi_{\rm rs}(\beta,h) \approx
\frac{(1-q)^6}{12 \beta} 
\frac{m(1-m)}{\left(m + (1-2m) \frac{(p-2)(1-q)}{2 q} \right)^2}
\left[\frac{\beta^2}{2}p(p-1)q_{\rm rs}(\beta,h)^{p-2}
-\frac{1}{(1-q_{\rm rs}(\beta,h))^2}  \right]^3 \ ,
\label{eq_dev_phi_instab}
\eeq
where we made some simplifications using Eq.~(\ref{eq_continuous_pspin})
verified by $q$, the value of the overlap at the point of the instability
line approached in this limit. The term in square brackets is positive in
the RS unstable region, and vanishes linearly on the instability line. 
As $\phi_{\rm rs}(\beta,h)$ is regular across this line this demonstrates the
third order character of the transition. The dependence on $m$ of the
1RSB potential can then be easily studied, in particular its maximum is
reached in $\widehat{m}_{\rm s}(q) = \frac{(p-2)(1-q)}{2 q}$. This behaviour is checked numerically by the scaling plots of
the complexity, see Fig.~\ref{fig_scaling_complexity_pspin}.
We emphasize that as $\delta_+ = O(\epsilon)$ on the maximum of $\phi$,
the overlap order parameter $q_1 - q_0$ grows linearly away from the
instability.

It could seem at this point that the above study is valid along the two 
branches $h_{\rm i}^\pm(T)$; this is however not the case, as further 
considerations reveal. The maximization over the 1RSB overlaps must
enforce the condition $q_1 \ge q_0$, which translates here into 
$\delta_+ \ge 0$. For the local maximum of $a \delta_+^2 /2 + b \delta_+^3 /3$
to happen on this side of the origin one must have $b<0$ (we always have $a>0$
in the unstable regime). Assuming $m \in [0,1]$, and after some simplifications,
one sees this condition to be equivalent to 
$\widehat{m}_{\rm s}(q) + m (1-2\widehat{m}_{\rm s}(q))>0$.
Two cases can now be distinguished:
\begin{itemize}
\item The high-field branch $\hi^+(T)$ corresponds, in the $q$-parametrized
representation, to $q \in \left[\frac{p-2}{p},1 \right]$. Then 
$\widehat{m}_{\rm s}(q) \in [0,1]$, thus the condition $q_1 \ge q_0$ at the
local maximum is respected for all $m \in [0,1]$, and the maximum of $\phi(m)$
is reached in an acceptable value $\widehat{m}_{\rm s}(q) \in [0,1]$ of the 
Parisi parameter.

\item On the contrary for the low-field part of the instability line $\hi^-(T)$,
$\widehat{m}_{\rm s}(q)>1$. The condition $q_1 \ge q_0$ is only fulfilled
for $m \in [0,\msplq]$, where 
$\msplq = \frac{\widehat{m}_{\rm s}(q)}{2 \widehat{m}_{\rm s}(q) -1}$. For these
values of $m$ there exists a non-trivial solution of the 1RSB equations that
grows continuously from the RS one. But for $m \in [\msplq,1]$ one sees that
the maximum of the expansion would correspond to $\delta_+ \to \infty$, in
other words the maximum of the 1RSB potential is far from the RS solution,
and corresponds to the discontinuous solution.

\end{itemize}

\subsection{Coexistence of RSB solutions}
\label{sec_app_coexistence}

\begin{figure}[t]
\includegraphics[width = 8 cm]{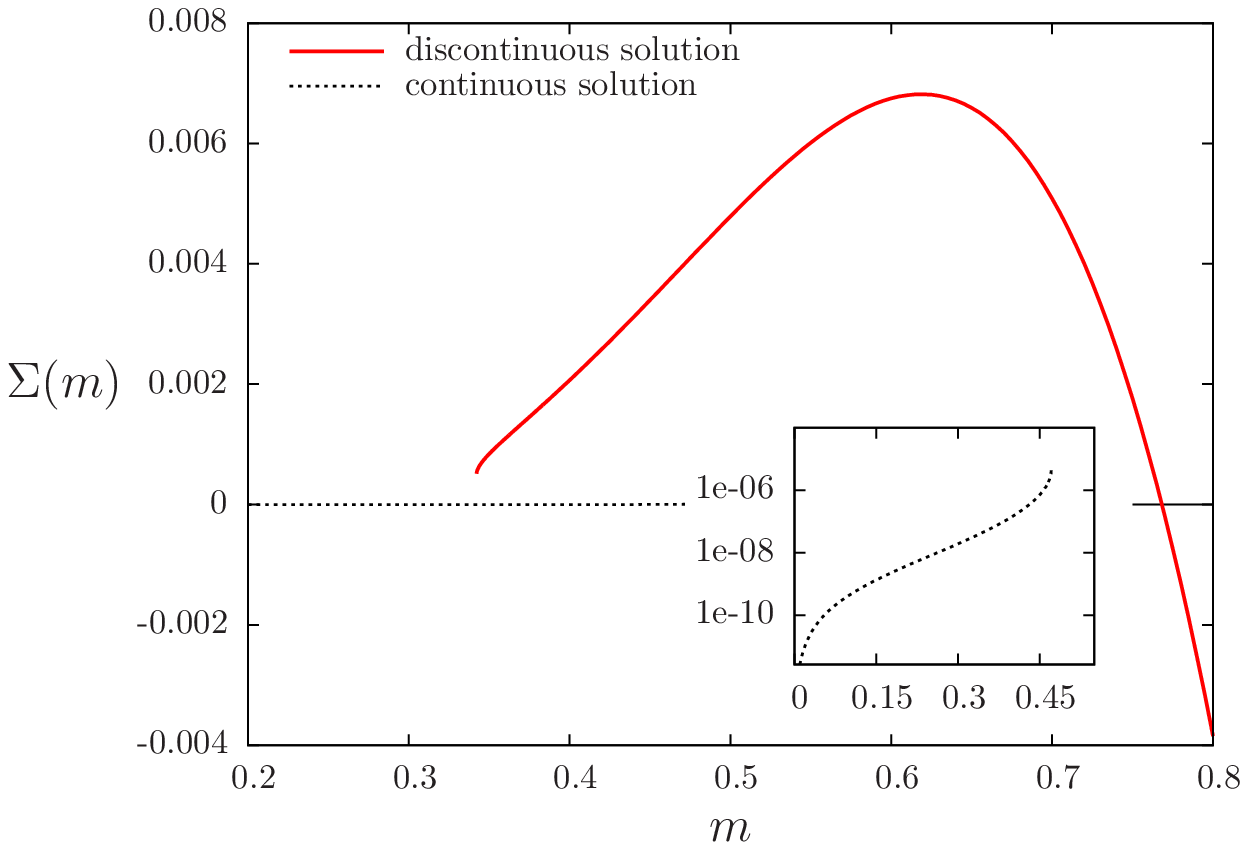} \hspace{6mm}
\includegraphics[width = 8 cm]{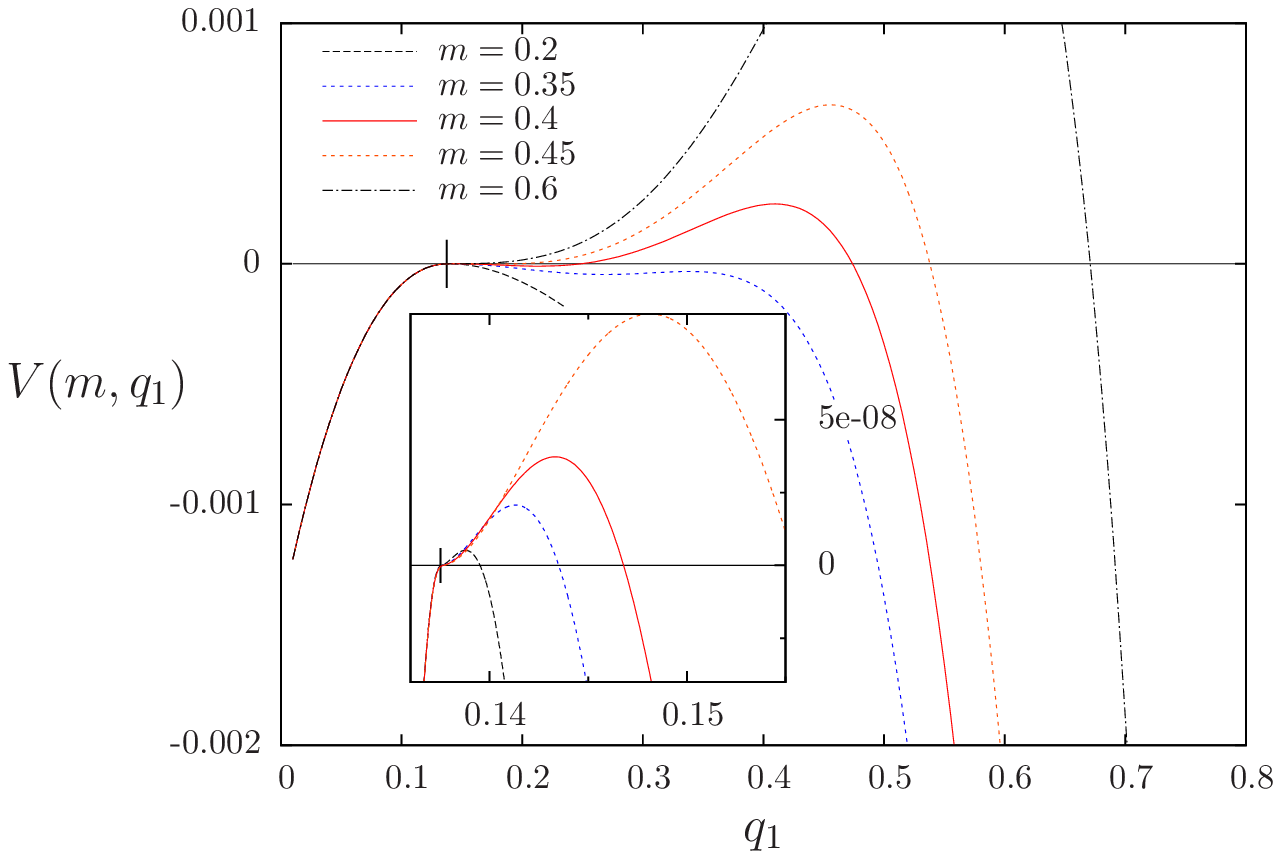}
\caption{Study of the coexistence of two RSB solutions, for $T=0.55$ and 
a field $h=0.166$ in the narrow range 
$[\hi^-(T) \simeq 0.1649, h_\star(T) \simeq 0.167]$ where the two RSB solutions
have a common domain of existence $m \in [\msphq,\msplq]$. 
For $T=0.55$ and $h=0.166$
the continuous solution exists for $m\in[0,\msplq]$, with 
$\msplq \simeq 0.48$, while the discontinuous one is defined for 
$m \in [\msphq,1]$, with $\msphq \simeq 0.34$.\\
Left panel: Complexity $\Sigma(\beta,h,m)$ as a function of $m$. The 
discontinuous solution has the largest complexity in the domain of coexistence
$[\msphq,\msplq]$. Note that the complexity of 
the continuous solution is very small, but non-zero. The relevant
value of the parameter $m$ for the thermodynamics, $m_{\rm s}(T,h) \simeq 0.77$
where the complexity vanishes, is outside the domain of existence of the
continuous solution.
\\
Right panel: The potential $V(m,q_1)$ defined in (\ref{eq_def_V}) as a
function of $q_1$ for several values of $m$. The inset is a zoom near 
$q_1=q_0$, and the ticks indicate the replica-symmetric solution. 
}
\label{fig_coexistence}
\end{figure}

We have just seen that close to the low-field branch $h_{\rm i}^-(T)<h_{\rm c}$ 
of the instability line the locally unstable RS solution gives birth
continuously to a non-trivial solution for $m\in [0,\msplq(T,h)]$. On the
other hand we have also shown that for $m=1$ a discontinuous 1RSB solution
exists for $T<T_{\rm d}(h)$; actually the latter exists for $m\in [\msphq(T,h),1]$,
hence two 1RSB solutions coexist for $m\in[\msphq(T,h),\msplq(T,h)]$
(whenever $\msphq(T,h)<\msplq(T,h)$). 
This leads the physical observables to have two 
branches when plotted as a function of $m$ 
(see left panel of Fig.~\ref{fig_coexistence}). 
However, once one also extremizes over $m$, no sign of the continuous 
solution to the 1RSB equation remains, and the physical solution is always 
the discontinuous one. 
To justify this point it is convenient to consider the reduced potential: 
\beq 
\label{eq_def_V} 
V(m,q_1) = \sup_{0 \leq q_0 \leq q_1} \phi_{\rm 1rsb}(\beta,h;m,q_1,q_0) 
- \sup_{q \in [0,1]} \phi_{\rm 1rsb}(\beta,h;m,q,q) \ ,
\eeq
where the dependencies on $T$ and $h$ are kept understood, and the 
normalization has been chosen such that $V$ vanishes on the replica-symmetric 
solution. A plot of $V$ as a function of $q_1$ for $h$ slightly larger than 
$h_{\rm{i}}^-(T)$ and several values of $m$  is shown on the right
panel of Fig.~\ref{fig_coexistence}. It can be seen that, as expected from the 
previous discussion, $V$ has for $m$ small enough ($m < \msplq(T,h)$) a 
local maximum near the replica-symmetric point $q_1=q_0$ (indicated by a tick 
on the figure), corresponding to the continuous solution. On the other hand, 
there exists for $m$ sufficiently large ($m \geq \msphq(T,h)$) another local 
maximum of $V$ at a value of $q_1$ which is further away from $q_0$. This 
second maximum corresponds to the discontinuous solution. When 
$m \rightarrow \msplq(T,h)$, the continuous solution merges onto the 
discontinuous one. Increasing $h$, $\msplq(T,h)$ shrinks and the limit of 
coexistence of both solutions (at fixed $m$) is given by the critical 
field $h_\star(T)$ such that $\msplq(T,h_\star(T))=\msphq(T,h_\star(T))$. In 
the region $h \in [h_{\rm{i}}^-(T), h_\star(T)]$ thermodynamical observables 
acquire 
two branches when computed at fixed $m$, as shown for the complexity on
the left panel of Fig.~\ref{fig_coexistence}. Finally, the relevant solution 
from a thermodynamical point of view is the one which maximizes $V(m,q_1)$ 
over both $m$ and $q_1$; because the continuous maximum is always very small 
and merges with the discontinuous one upon increasing $m$, one finds that the 
relevant solution is always the discontinuous one.


\section{$1$-step quantum cavity equations and their numerical resolution}
\label{app_1step_quantum_cavity}

We briefly recalled in Sec.~\ref{sec_classical_coloring} how the classical 
coloring problem on a diluted random graph could be solved analytically in 
the thermodynamic limit by the cavity method, summarized by the RS
(\ref{eq_cavity_classical_RS}) and 1RSB 
(\ref{eq_cavity_classical_1RSB}) cavity equations.
The same route can be undertaken in the quantum case, with the important 
complication that the probability distributions $\eta$ appearing in these
two equations are now over ``trajectories'' instead of colors, a trajectory 
$\bs$ being a piecewise constant periodic function from $[0,\beta]$ to 
$\{1,\dots,\qcol\}$. The only modification to
Eqs.~(\ref{eq_cavity_classical_RS},\ref{eq_cavity_classical_1RSB})
implied by this change is the replacement of the functions $g,z$ by 
(see~\cite{long} for more details)
\beq 
\label{eq_1rsb_b} 
g(\eta_1,\dots,\eta_{c-1}) (\bs) = 
\frac{1}{z(\eta_1,\dots,\eta_{c-1})} \Gamma^{n(\bs)} 
\sum_{\{\bs_i\}} \prod_{i=1}^{c-1} \eta_i(\bs_i) 
\exp \left[ \int_0^\beta \sum_{i=1}^{c-1} \delta_{\s(t),\s_i(t)} dt \right] 
\; , \eeq
$z(\eta_1,\dots,\eta_{c-1})$ being defined by normalization, and $n(\bs)$ 
counting the number of discontinuities in the trajectory $\bs$. 
Note that for $\G=0$, only constant trajectories remain and one recovers the
definition given in (\ref{eq_cavity_classical_RS}). 

The population dynamics method~\cite{abou1973,cavity} is a convenient way
to solve these equations numerically. Probability distributions are 
approximated by weighted samples of representative elements:
\beq
P(\eta) = \sum_{i=1}^{\Npop} a_i \delta( \eta- \eta_i) \ , \qquad
\eta_i(\bs) = \sum_{j=1}^{\Ntraj} b_{i,j} \delta( \bs- \bs_{i,j}) \ .
\eeq
We therefore have two deal with two level of populations: 
one of $\Npop$ messages $\eta$, 
each of these messages being represented by $\Ntraj$ trajectories $\bs$.
Each trajectory is encoded by the imaginary times in $[0,\beta]$ where it
changes value (there are of order $\beta \G$ such times), and by its constant
value between these jumps.
Then the 1RSB equation (\ref{eq_cavity_classical_1RSB},\ref{eq_1rsb_b})
can be solved by 
iteration on these samples and the associated weights $a_i, b_{i,j}$, the
current estimation of $P$ being inserted in the r.h.s. of 
Eq.~(\ref{eq_cavity_classical_1RSB}), and the l.h.s. is approximated by
a new discrete representation.
This procedure would be exact only if $\Npop$ and $\Ntraj$ were infinite, 
while the memory available on present days computers limits the values of 
$\Npop$ and $\Ntraj$ to rather small values 
(see below and App.~\ref{app:numerics_cavity} for concrete examples). 
This induces both systematic deviations of the empirical mean from the exact 
value and noise in its estimation; extrapolations to 
$\Npop, \Ntraj \rightarrow \infty$ via finite size analysis can in principle 
be performed to reduce these effects, we show one example of such treatement
in App.~\ref{sec_c9_linear_instabilities}. Further difficulties arise
when the weights $a_i$ (or $b_{i,j}$) become very heterogeneous: then the
population is effectively supported only on the representants with the 
largest weights, hence the effective size of the population can be much 
smaller than the number of samples. These population representations of
probability distributions are known in statistics as particle 
approximations~\cite{particle_filters}; many resampling techniques are known
to fight this impoverishment of the sample representativity, but there does
not seem to be an universal way to avoid it.

To increase the speed of our numerical code we made use of the parallelization
opportunities offered by multi-core computers. Let us sketch how the
1RSB equation resolution can be distributed on several processing units.
The update procedure that we 
alluded to above can simply be thought of as a way to generate a sample at 
step $\tau+1$, $\{ (\eta_i(\tau+1),a_i(\tau+1))\}$, from a sample at step 
$\tau$, $\{ (\eta_i(\tau),a_i(\tau))\}$. The important point is that the 
new representants $(\eta_i(\tau+1),a_i(\tau+1))$ are generated independently 
one from the other (apart from the normalization condition $\sum_i a_i =1$
that can easily be enforced). 
The update procedure can then be parallelized as follows: 
each core is first sent the whole sample of messages and weights 
$\{\eta_i(\tau),a_i(\tau)\}_{1 \leq i \leq \Npop}$. Then each core $c$, with 
$1 \leq c \leq \Ncore$, generates independently a new sample of 
$\Npop / \Ncore$ messages: 
$\{(\eta_{c,i}(\tau+1),a_{c,i}(\tau+1))\}_{1 \leq i \leq \Npop/\Ncore}$. 
These new messages are then gathered together to form the full sample at step 
$\tau+1$: $\{(\eta_i(\tau),a_i(\tau))\}_{1 \leq i \leq \Npop}$.
This method does not change the memory limit of the procedure, 
because each core is sent the whole population in the first step, but 
allows for a gain in time roughly proportional to $\Ncore$ (the communication
between the processors usually takes much less time than the generation
of the samples). Typically we used $\Ncore$ between 12 and 64; values of 
$\Npop$ and $\Ntraj$ are limited by the amount of memory available on a 
single core, leading in our case to the constraint 
$\Npop \times \Ntraj \lesssim 3.10^7 / (\beta \G) $. A moment of thought 
reveals that it is also possible to avoid the step of population gathering, 
and to keep the population of messages 
$\{\eta_i(\tau),a_i(\tau)\}_{1 \leq i \leq \Npop}$ spread over the $\Ncore$ cores.
This allows for larger populations, but strongly increases the time 
needed to update the population, as much more information has to be 
exchanged between cores; therefore, we dit not use this second procedure in 
this work.
 
\section{Details on the numerical results and fitting procedures}
\label{app:numerics_cavity}

\subsection{Finite population scaling for the instability of the RS solution}
\label{sec_c9_linear_instabilities}
 
\begin{figure}[t]
\includegraphics[width= 8 cm]{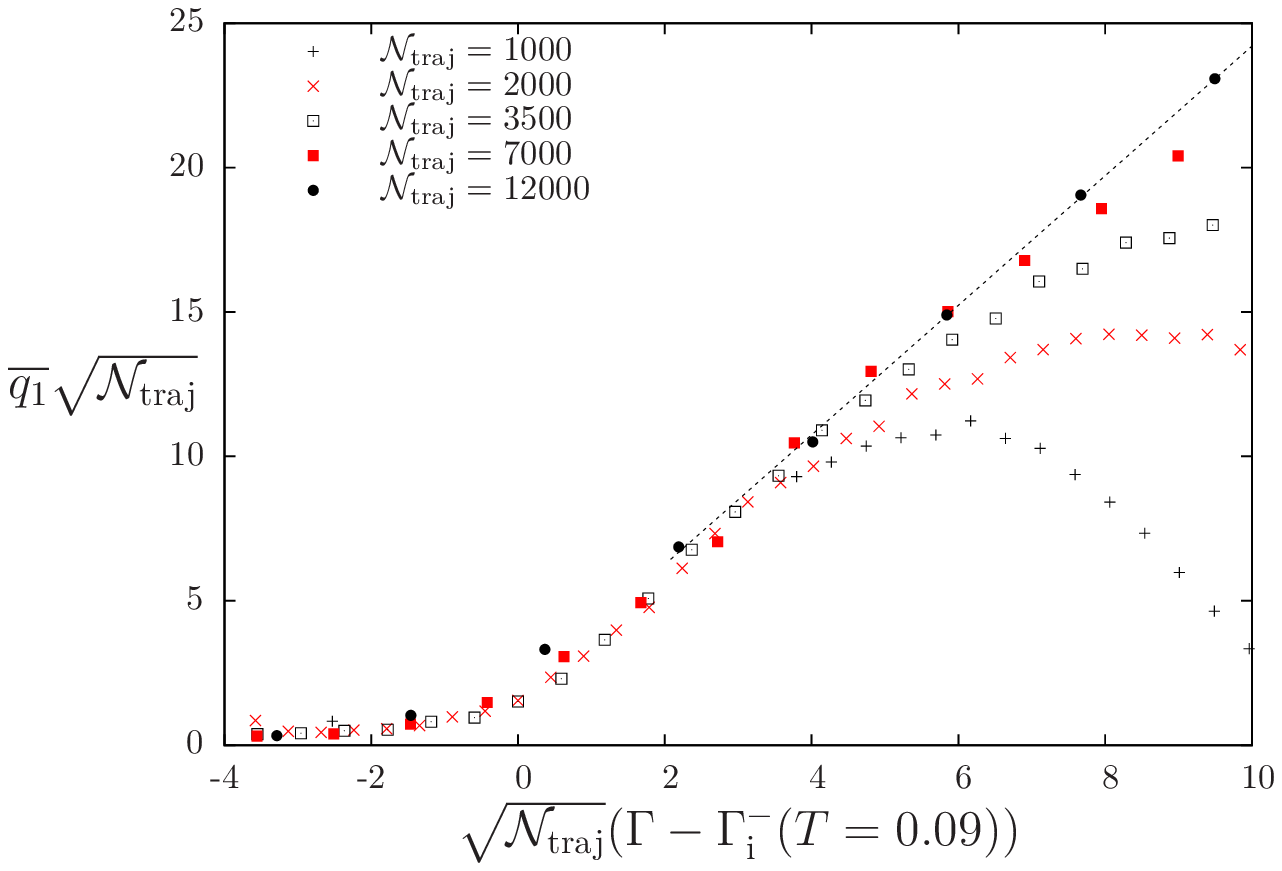} \hspace{0.5 cm}
\includegraphics[width= 8 cm]{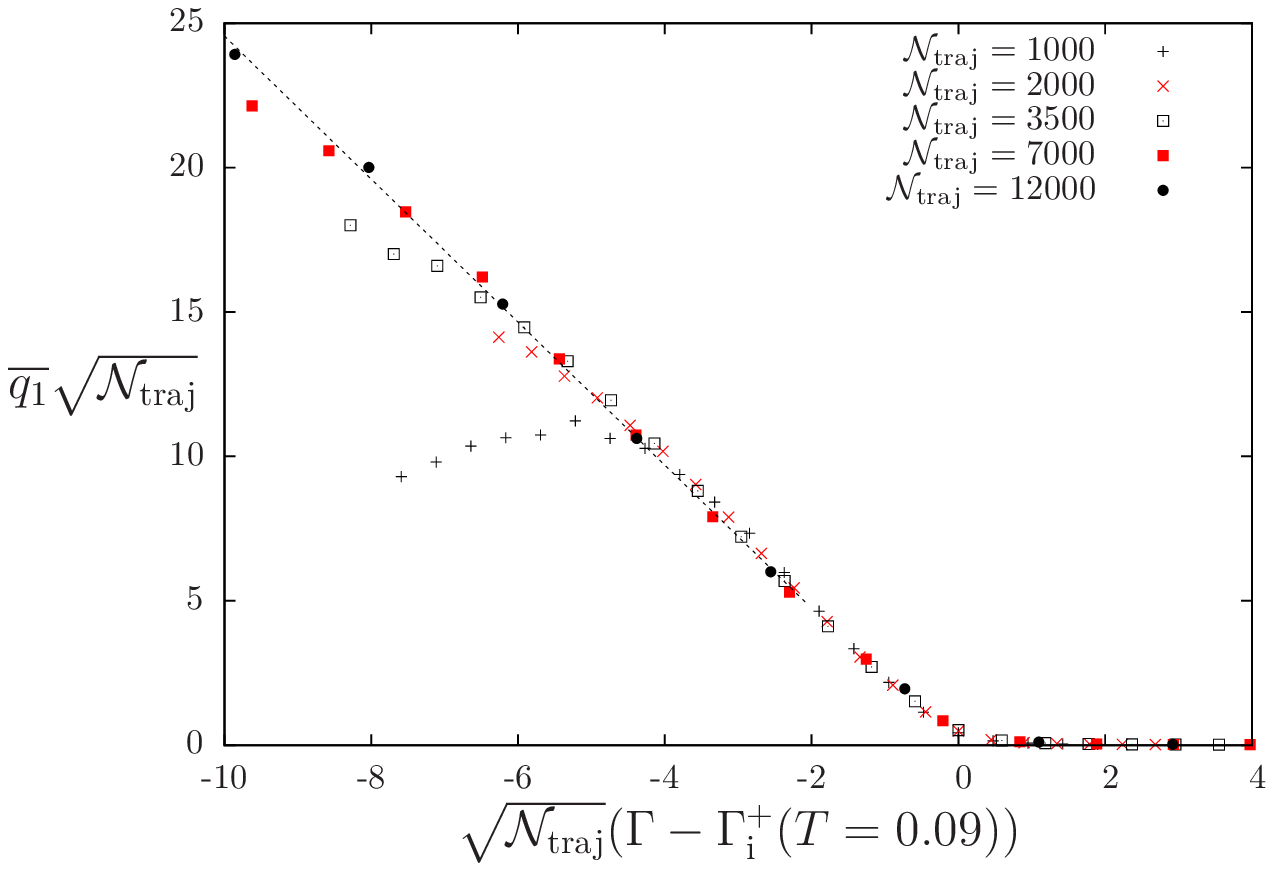}
\caption{Scaling form (\ref{eq_scaling_instability_N}) for $\overline{q_1} = q_1-1/\qcol$, for $c=9$, $T=0.09$, and the small field instability ($\Gi^-(T=0.09)=0.08$) on left panel, large field instability  ($\Gi^+(T=0.09)=0.44$). The dotted lines are the linear asymptotics for the scaling functions $\mathcal{F}$; $\Gi^{\pm}$ have been adjusted to obtain the best data collapse.}
\label{fig_scaling_instability_N}
\end{figure}

The simplest transition to find numerically is the static continuous 
transition at $\Ti(\G)$. In fact, it corresponds to the point where the replica-symmetric 
maximum becomes a saddle point within the larger 1RSB subspace; hence the 
RS solution, if slightly perturbed, will be unstable under iteration. 
Therefore, it is enough to initialize the 1RSB population dynamics 
equation (\ref{eq_cavity_classical_1RSB}) on the RS solution 
and to check whether this solution is stable under iteration, with respect to 
a small perturbation. This is detected by the growth of the intra-cluster overlap $q_1$, defined in (\ref{eq_def_q_quantum}) in the quantum case, from its RS value $1/\qcol$. Formally expanding in (\ref{eq_cavity_classical_1RSB})
$P$ around a delta-peaked distribution, one realizes that the condition of
local stability of the RS solution towards a non-trivial 1RSB solution
is independent of the Parisi replica-symmetry breaking parameter $m$. It is thus possible
to take $m=0$, which is very convenient in practice: the weights 
$z(\{ \eta_i \})^m$ are all equal to 1 (at non-zero temperature the factors
$z$ never vanish), the external population representation has thus
homogeneous weights and $\Npop$ can be reduced to a rather small value 
without too much negative effects on the numerical accuracy.
We can then concentrate on the finite population size effect of $\Ntraj$ and
build a scaling theory around the transition.
To simplify the discussion we focus on the case in which the temperature $T$ 
is fixed and $\Gamma$ is lowered from the quantum paramagnetic phase at large 
$\Gamma$ through the first continuous instability $\Gamma^+_{\rm{i}}(T)$. First of 
all, the RS solution has a very simple overlap structure: 
because of the symmetry between colors, one has 
$q_0 = q_1 = \frac{1}{\qcol}$ on the RS solution. 
Introducing $\overline{q_1} = q_1 - 1/\qcol$, one finds that within the RS 
phase $(\Gamma > \Gamma^+_{\rm{i}}(T))$, $\overline{q_1}$ has a finite population size 
behaviour as $1/\sqrt{\Ntraj}$. On the other hand, we expect from 
the discussion of Appendix~\ref{sec_spherical_pspin} that for 
$\Gamma < \Gamma^+_{\rm{i}}(T)$, $\overline{q_1} = - a \left 
( \Gamma- \Gamma^+_{\rm{i}}(T) \right )$. These two behaviours can be matched by 
introducing a scaling function $\mathcal{F}$ such that:
\beq 
\label{eq_scaling_instability_N} 
\overline{q_1}(\Gamma,T,\Ntraj) = \frac{1}{\sqrt{\Ntraj}} \mathcal{F} 
\left[ (\Gamma-\Gamma^+_{\rm{i}}(T)) \sqrt{\Ntraj} \right]  
\eeq
Moreover, $\mathcal{F}(x)$ admits the following asymptotic behaviours: 
$\lim_{x \rightarrow -\infty} \mathcal{F}(x)/x = -a$, 
$\lim_{x \rightarrow \infty} \mathcal{F}(x) = O(1)$. 
The value of $\Gamma-\Gamma^+_{\rm{i}}(T)$ is then determined in order to obtain the
best collapse of the numerical data for different values of $\Ntraj$; as
shown in Fig.~\ref{fig_scaling_instability_N} this scaling form gives
a very good collapse of the numerical curves. This procedure therefore allows one to determine
the line $\Ti(\G)$ with very good precision.

\subsection{The low $\G$ regime for $c=9$: the lines $\Tk(\G)$, $T_\star(\G)$, $\Td(\G)$}
\label{sec_c9_lowgamma}

Having discussed the determination of $\Ti(\G)$, let us now focus on the more complex determination of $\Tk(\G)$ and $T_\star(\G)$ that were defined in Sec.~\ref{sec_q4_c9_phase_diagram}
and appeared in Fig.~\ref{fig_phase_diagram_q4_c9_sp} and \ref{fig_phase_diagram_q4_c9}. We assume here that the reader is already familiar with the discussion 
of Sec.~\ref{sec_q4_c9_phase_diagram} but we will repeat some parts of the discussion for clarity.

\begin{figure}[t]
\includegraphics[width = 8 cm]{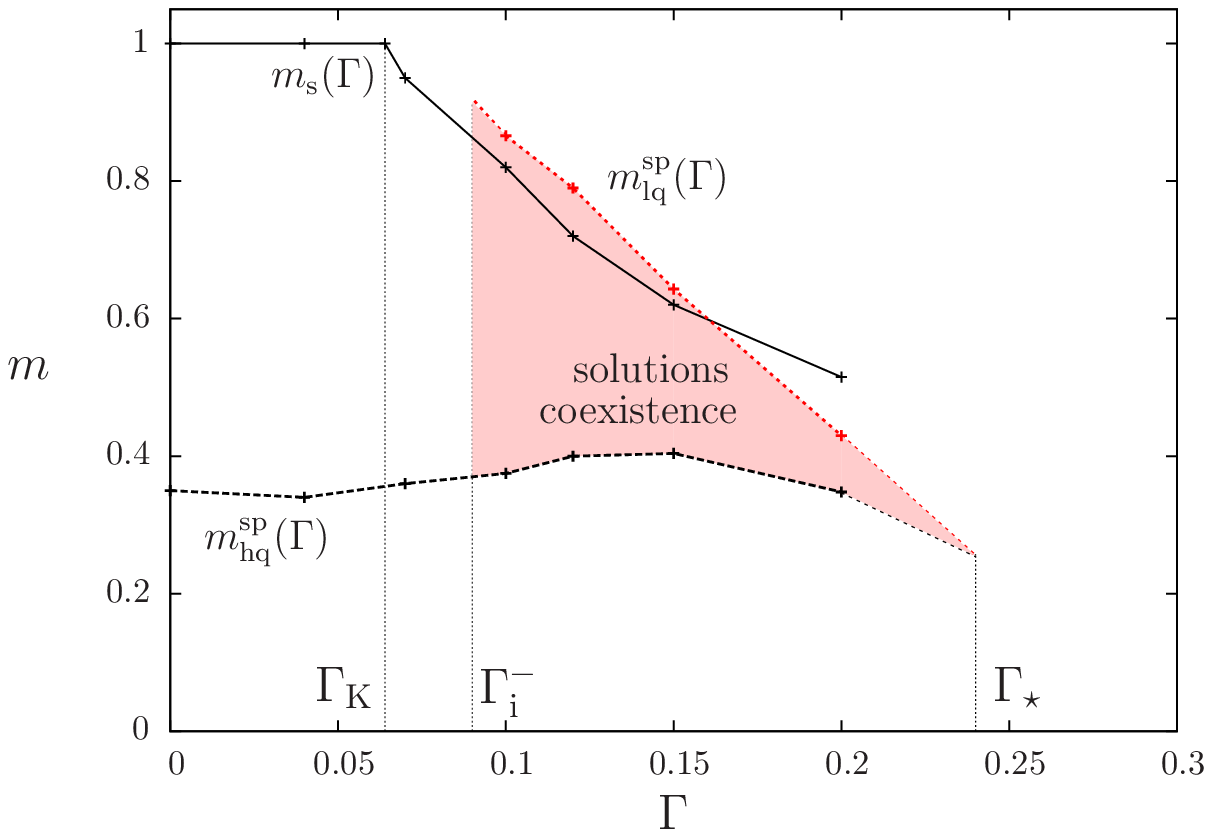}
\caption{Domains of existence of the solutions of the 1RSB equation for $c=9$, $T=0.09$, in the $(\Gamma,m)$ plane. The ${\rm 1RSB_{hq}}$ solution exists for $m \geq \msphq(\G)$, 
while the ${\rm 1RSB_{lq}}$ one exists for $\Gamma \geq \Gi^-$ and $m \leq \msplq(\G)$. 
The (extrapolated) intersection of $\msphq(\G)$ and $\msplq(\G)$ defines the point $\G_\star$ at which the two 1RSB solutions merge into a single one.
The region of coexistence is therefore delimited by $\Gi^- < \G < \Gamma_\star$ and $\msphq(\G) < m < \msplq(\G)$.
However, only the ${\rm 1RSB_{hq}}$ solution is thermodynamically relevant: the value of $m_{\rm{s}}(\Gamma)$ is equal to 1 for $\G<\Gk$, and it is smaller than 1 for $\G>\Gk$,
but it always correspond to the ${\rm 1RSB_{hq}}$ branch. 
 }
\label{fig_phase_diag_m}
\end{figure}

\begin{figure}[t]
\includegraphics[width = 6 cm]{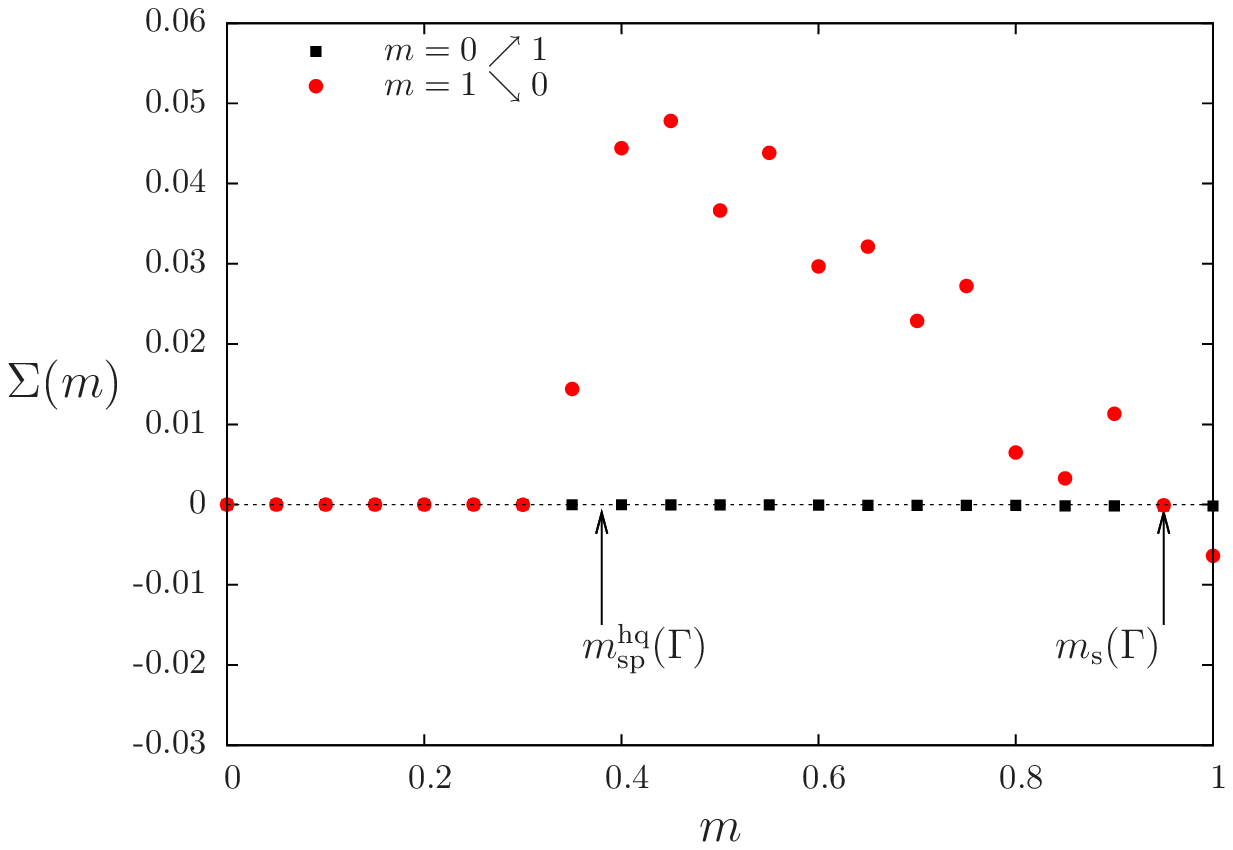}
\includegraphics[width = 5.55 cm]{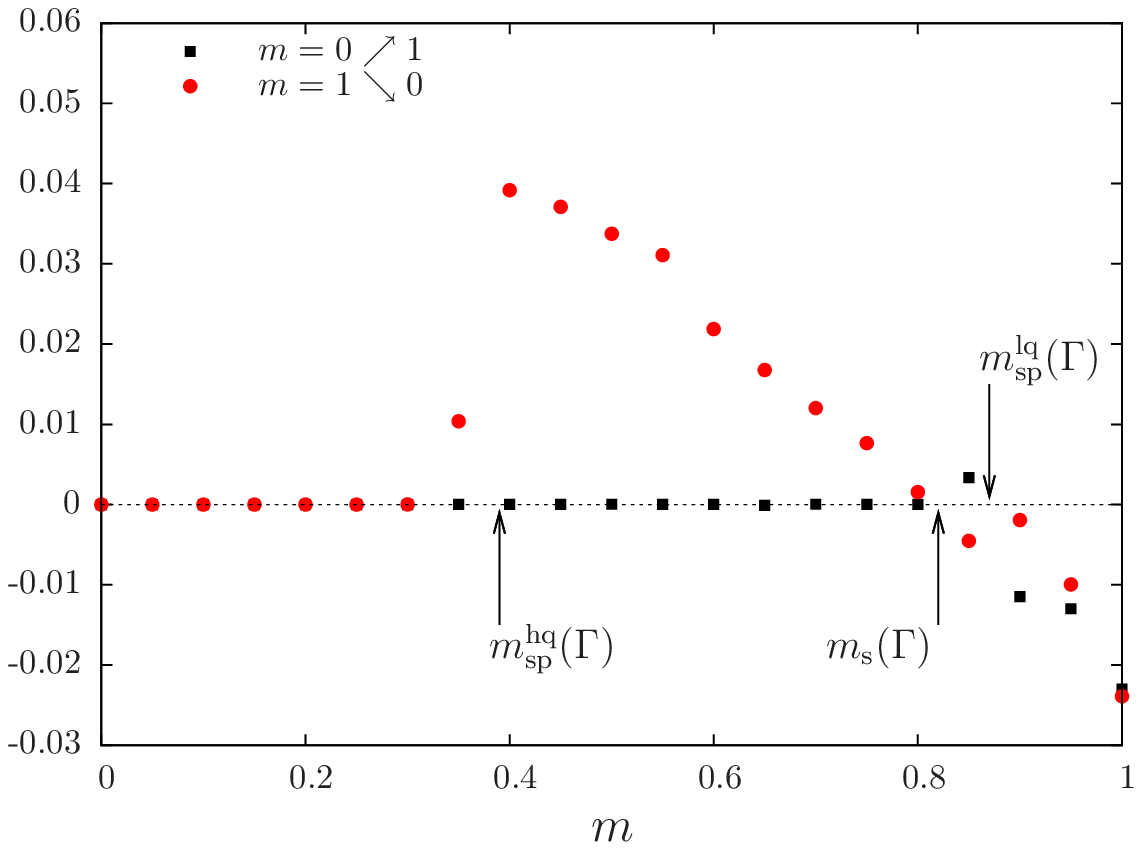}
\includegraphics[width = 5.5cm]{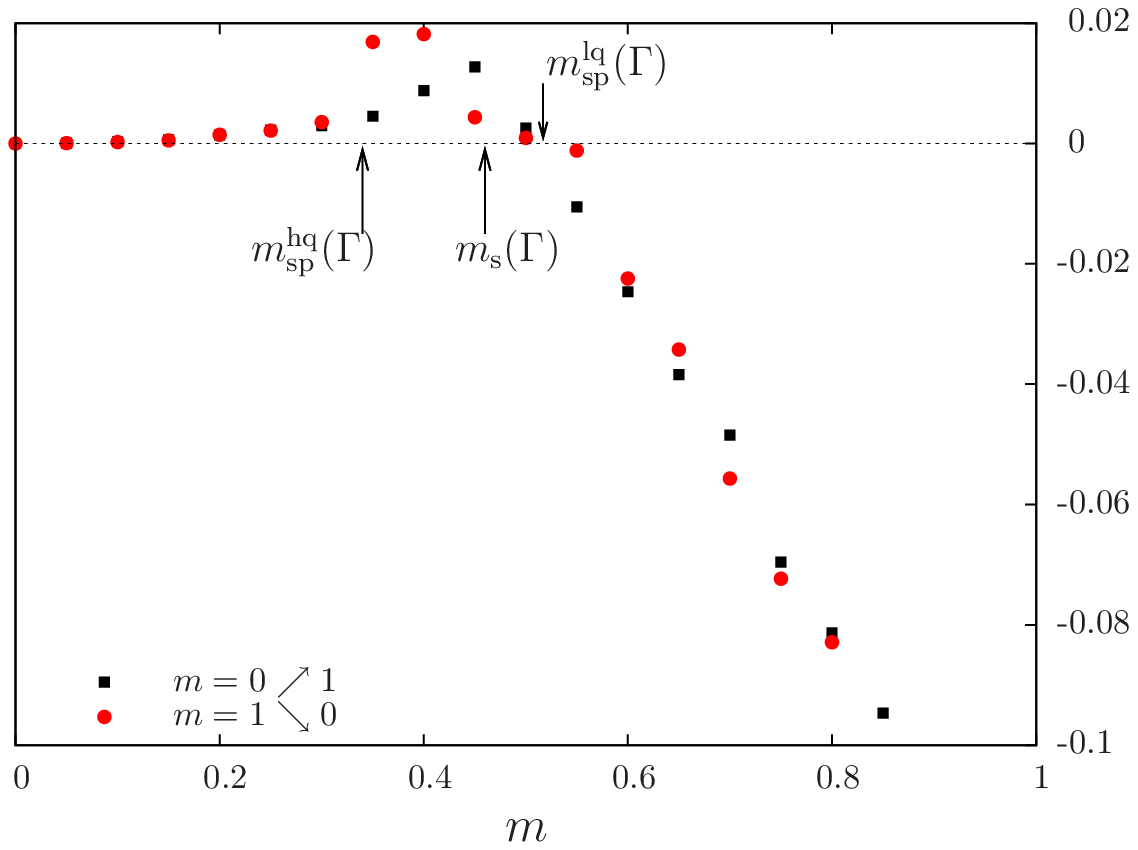}
\caption{Complexity $\Sigma(m)$ for $c=9$, $T=0.09$ and $\Gamma = 0.07$ (left), $\Gamma=0.1$ (middle), $\Gamma=0.2$ (right). Results obtained with $\Npop = 1200$ and $\Ntraj = 3500$. The red points are obtained by decreasing $m$ from 1 down to 0, thus selecting the ${\rm 1RSB_{hq}}$ solution, while the black points are obtained increasing $m$ from 0, thus 
selecting the ${\rm 1RSB_{hq}}$ solution (if any). Note the resemblance with Fig.~\ref{fig_coexistence}.
For this temperature, $\Gk \simeq 0.064$, $\Gi^- \simeq 0.08$, $\Gamma_\star \simeq 0.25$. The left plot is for $\Gk < \Gamma < \Gi^-$: in this region there exists only the ${\rm 1RSB_{hq}}$ 
solution with $\Sigma(m=1)<0$. 
For $\Gi^- < \Gamma < \Gamma_\star$ (middle and right panel) there exist two branches
${\rm 1RSB_{hq}}$ and ${\rm 1RSB_{lq}}$,
which can be seen to merge when $\Gamma$ approaches $\Gamma_\star$. 
To find the thermodynamic solution, one has to maximize the free energy of the system, 
which amounts to enforce the condition $\Sigma(m)=0$ (or $\Sigma(m=1)>0$): this condition always selects the ${\rm 1RSB_{hq}}$ solution.
}
\label{fig_coexist_sigma}
\end{figure}

\begin{figure}[t]
\includegraphics[width=8 cm]{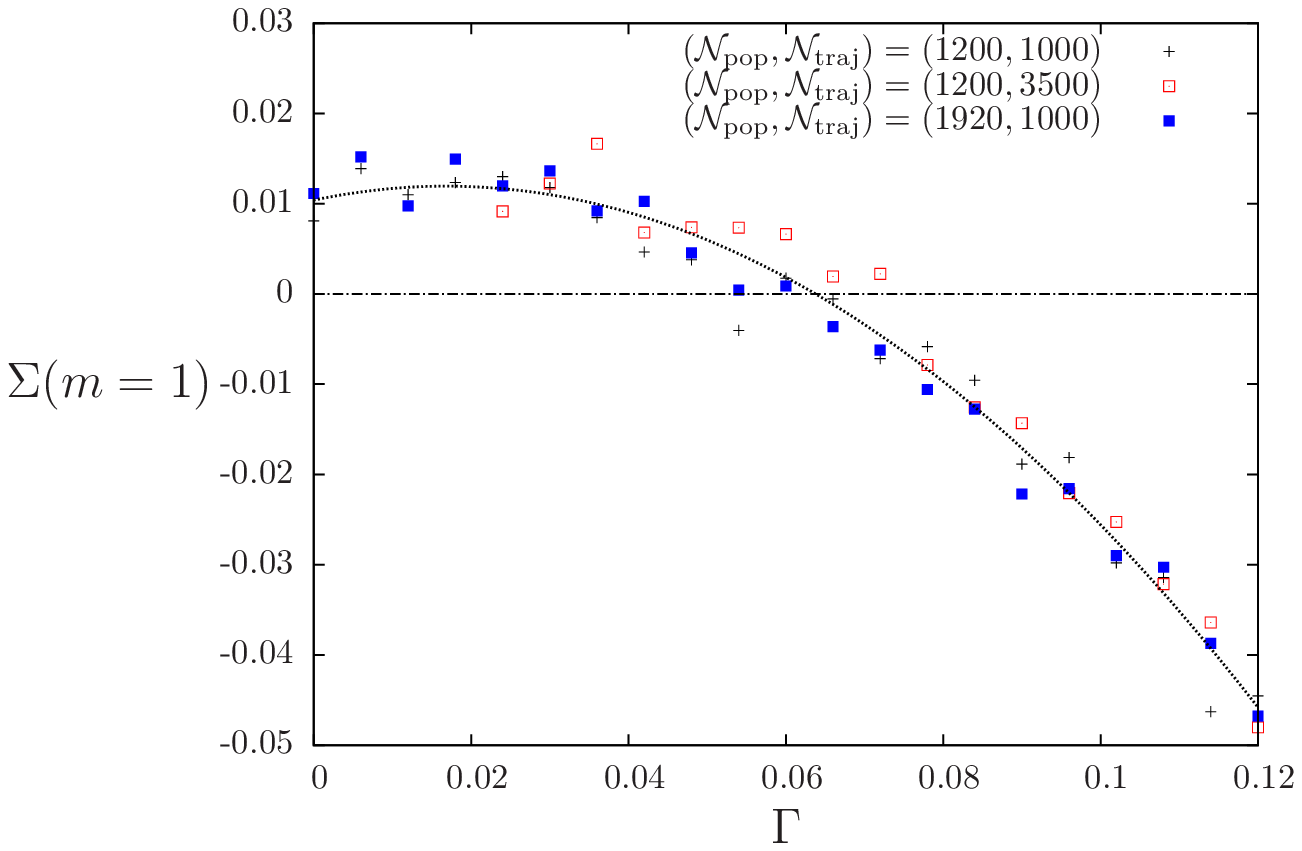} \hspace{0.5 cm}
\includegraphics[width=8 cm]{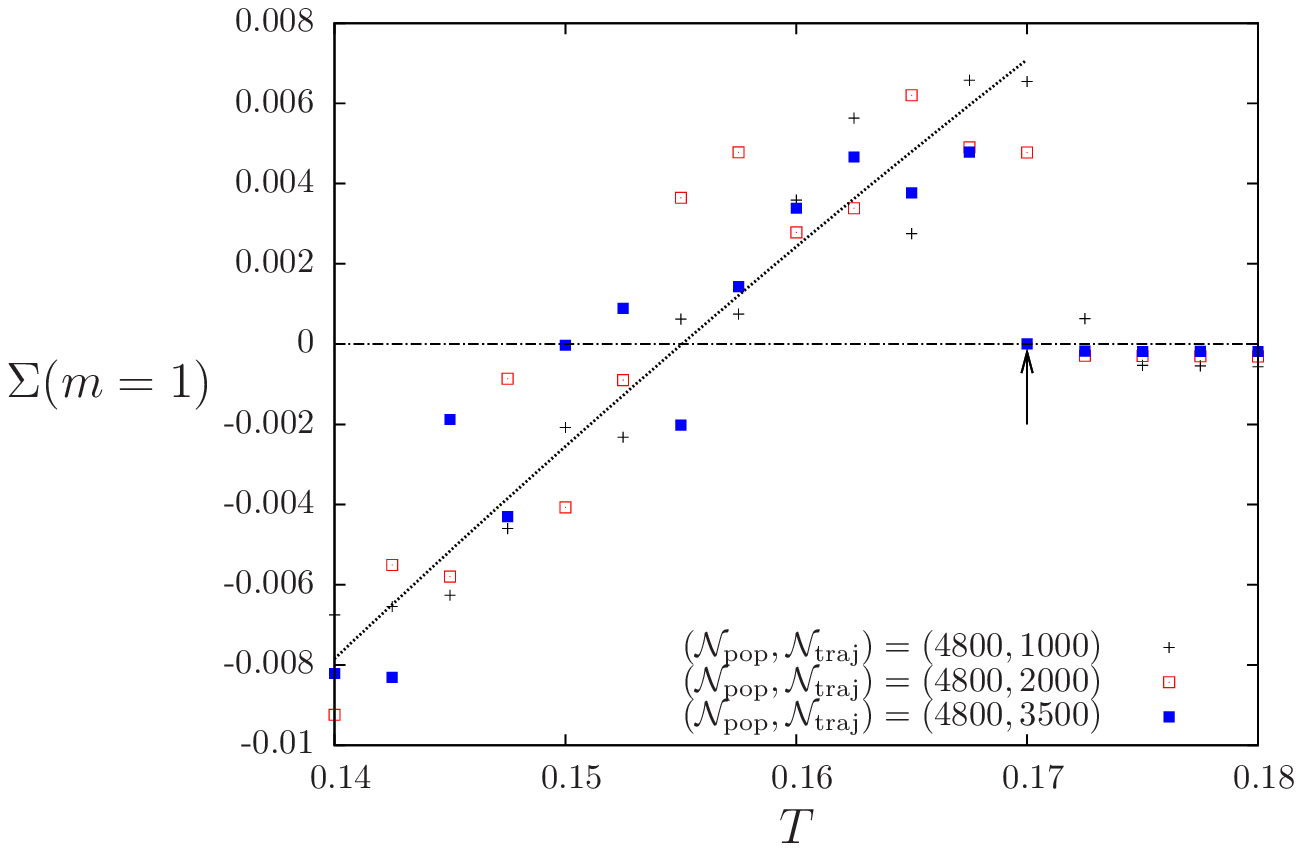}
\caption{Complexity of the 1RSB solution at $m=1$, for the coloring problem
with $c=9$. Left panel: as a function of $\G$ for $T=0.09$. Right panel: as a function of $T$ for $\Gamma = 0.13$. 
No clear finite size effect could be found so we report the data obtained with three large sizes. 
The black line is a quadratic fit on the range $\Gamma \in [0.02,0.12]$ (left panel), $T \in [0.14,0.17]$ (right panel), giving $\Gk(T=0.09)=0.063$ and $\Tk(\G=0.13)=0.155$.
The value of $\Td$ is marked by an arrow. 
}
\label{fig_q4_c9_Td_Tk}
\end{figure}

For simplicity, we will discuss the procedure at a fixed low temperature
$T=0.09$, and explain how $\Gk$ and $\G_\star$ are determined at this temperature.
At $\G=0$, there is a single 1RSB solution of the classical cavity equation, the ${\rm 1RSB_{hq}}$ one, that exists in an interval $m \in [ \msphq ,1]$.
This solution can easily be found by solving the classical 1RSB equations,
it is the relevant one and has positive complexity at $m=1$, because $\Td(\G=0)>0.09$ and $\Tk(\G=0)=0$.
The static Parisi parameter is therefore $m_{\rm{s}}(\G=0)=1$.
The evolution upon increasing $\G$ from $\G=0$ has already been sketched in Sec.~\ref{sec_q4_c9_phase_diagram} and in Fig.~\ref{fig_phase_diagram_q4_c9_sp}, 
but we give some additional details here (see a summary on 
Fig.~\ref{fig_phase_diag_m}):
\begin{itemize}
\item For $\Gamma < \Gk(T)$, in addition to the RS solution,
 there exists only the ${\rm 1RSB_{hq}}$ solution, which is defined for $m \in [ \msphq(\G), 1 ]$ and has a positive complexity at $m=1$. 
 It can be easily constructed by initializing the 1RSB population dynamics in this solution
 at $\G=0$ and following it at positive $\G$.
 The value of $\msphq(\G)$ can be determined numerically by following the evolution of the ${\rm 1RSB_{hq}}$ solution upon decreasing $m$ from 1 to 0 at constant $\G$, 
 as the value at which the solution
 disappears and the complexity jumps to zero (see Fig.~\ref{fig_coexist_sigma}, left panel).
\item For $\Gk(T) < \Gamma < \Gi^-(T)$ there still exists only the ${\rm 1RSB_{hq}}$ solution, but
its complexity is negative at $m=1$. 
The value of $\Gk(T)$ is therefore determined as the point where the complexity at $m=1$ in the ${\rm 1RSB_{hq}}$ goes continuously to zero, see Fig.~\ref{fig_q4_c9_Td_Tk}.
The static value of the Parisi parameter $m_{\rm{s}}(\G) \in (\msphq(\G),1)$ is determined as the point where the complexity vanishes
 (see again Fig.~\ref{fig_coexist_sigma}, left panel).
\item For $\Gi^-(T) < \Gamma < \Gamma_\star(T)$ the RS solution becomes unstable and it disappears in favor of the ${\rm 1RSB_{lq}}$ solution which exists in an interval
$m \in [0, \msplq(\G)]$. Therefore in this region the two 1RSB solutions coexist for $\msphq(\G) < m < \msplq(\G)$. The ${\rm 1RSB_{hq}}$ solution 
can still be obtained numerically
by following it in $\G$ at $m=1$, and then decreasing $m$ at fixed $\G$. On the contrary, the ${\rm 1RSB_{lq}}$ solution is obtained by starting at $m=0$ in the RS solution;
as the RS solution is unstable, the population evolves towards the ${\rm 1RSB_{lq}}$ fixed point which can then be followed by increasing $m$ at fixed $\G$. The point
$\msplq(\G)$ is defined as the point where the population jumps into the ${\rm 1RSB_{hq}}$ solution.
Again, in this region $m_{\rm{s}}(\G) \in (\msphq(\G),1)$ is the point where $\Si=0$ in the ${\rm 1RSB_{hq}}$ solution.
\item Upon increasing $\Gamma$, the coexistence region shrinks; 
the field $\Gamma_\star(T)$ at which $\msphq = \msplq$ marks the limit of coexistence of both solutions. However, when the coexistence region is too small it is hard
to determine the spinodals, so the point $\G_\star$ is obtained by extrapolating the two lines $\msphq(\G)$ and $\msplq(\G)$ from smaller $\G$ to determine their intersection
(see Fig.~\ref{fig_phase_diag_m}).
\item Finally above $\G_\star$ there is a single 1RSB solution, and again $m_{\rm{s}}(\G)$ is the point where its complexity vanishes.
\end{itemize}
The evolution of the complexity in Fig.~\ref{fig_coexist_sigma} is the best way to provide numerical support to the schematic picture for the free energy $\phi(m)$
outlined in the right panel of Fig.~\ref{fig_phase_diagram_q4_c9_sp}. Unfortunately, a direct computation of $\phi(m)$ is not possible because the data are too noisy.
The fact that the relevant solution for thermodynamics is the ${\rm 1RSB_{hq}}$ one is therefore not evident from Fig.~\ref{fig_coexist_sigma}. However, this can
be deduced from the consistency arguments presented in Sec.~\ref{sec_q4_c9_phase_diagram} and from the fact that the complexity of the ${\rm 1RSB_{hq}}$ is much
bigger, thus suggesting a more rapid rise of its free energy from the RS value.

\begin{figure}[t]
\includegraphics[width=8 cm]{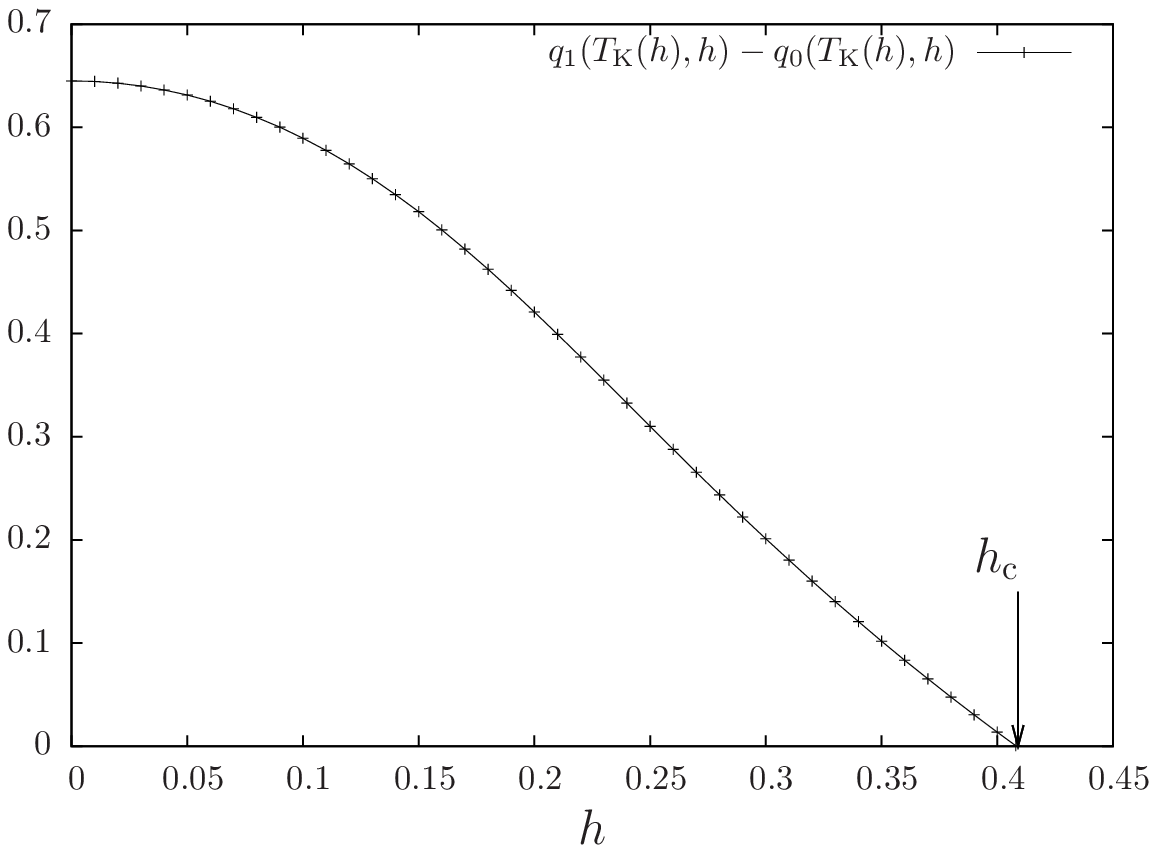} \hspace{0.5 cm}
\includegraphics[width=8 cm]{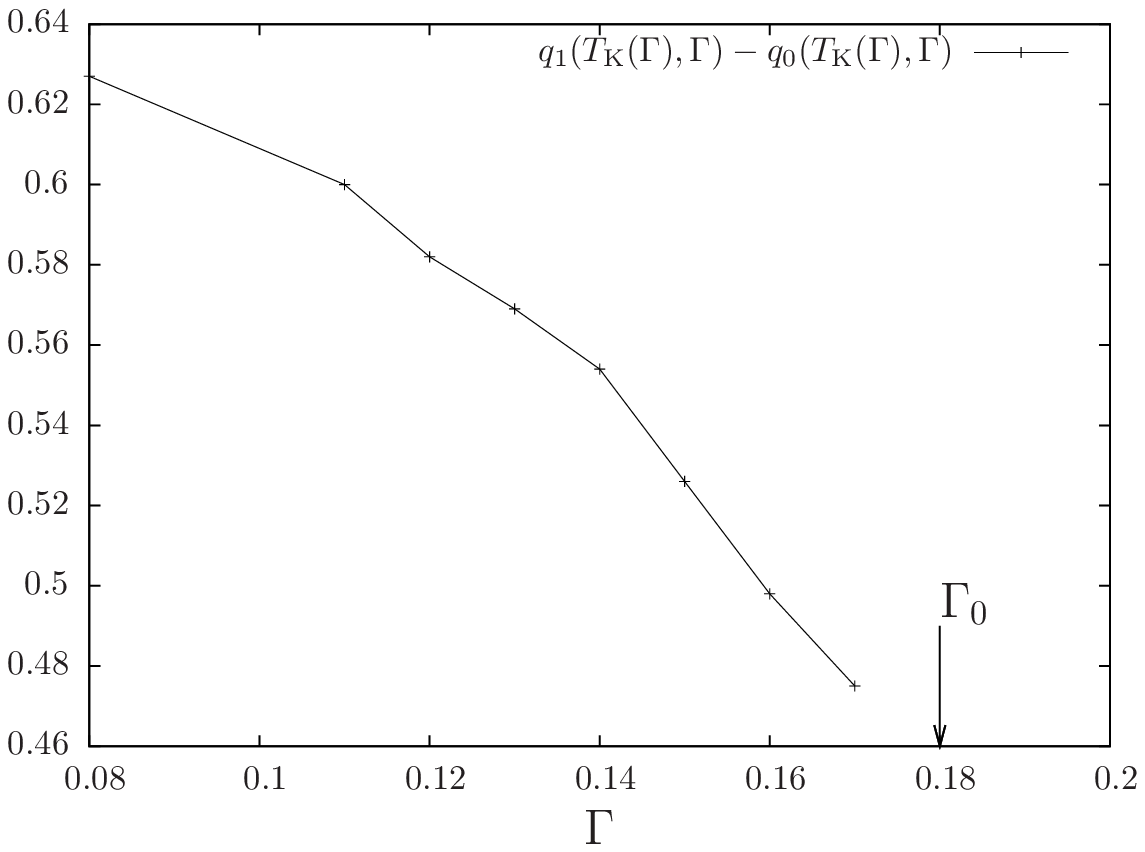}
\caption{Difference between the intra- and extra-cluster overlaps on the Kauzmann transition line for the spherical 3-spin model (left panel) and the coloring problem with $\qcol=4$ and $c=9$ (right panel). The triple points $h_{\rm c}$ and $\Gamma_0$ are defined in the text and do not have the same interpretation in the two cases. Note that the y-axis does not start from zero on the right panel.
}
\label{fig_qk}
\end{figure}

The dynamical transition temperature $\Td(\G)$ can be determined quite easily as the point where the ${\rm 1RSB_{hq}}$ solution at $m=1$ disappears on increasing
temperature at fixed $\G$, like in the classical case. An example of this procedure is shown on Fig.~\ref{fig_q4_c9_Td_Tk}.
It is important to stress that when the two lines $\Td(\G)$ and $\Tk(\G)$ merge, the overlap $q_1$ does not tend to its RS value, as shown in Fig.~\ref{fig_qk}.
This is an important difference with respect to the spherical $p$-spin model, where the two lines merge on a critical point where the transition becomes continuous.
In the coloring problem, the overlap $q_1$ remains non trivial, indicating that the lines $\Td$ and $\Tk$ do not merge on the instability line $\Ti$, consistently with
the phase diagram reported in Fig.~\ref{fig_phase_diagram_q4_c9}.

\subsection{The spurious RS first order transition for $c=13$: a study in the $(\G,m)$ plane}
\label{sec_c13_RS_spurious}

\begin{figure}[t]
\includegraphics[width=8cm]{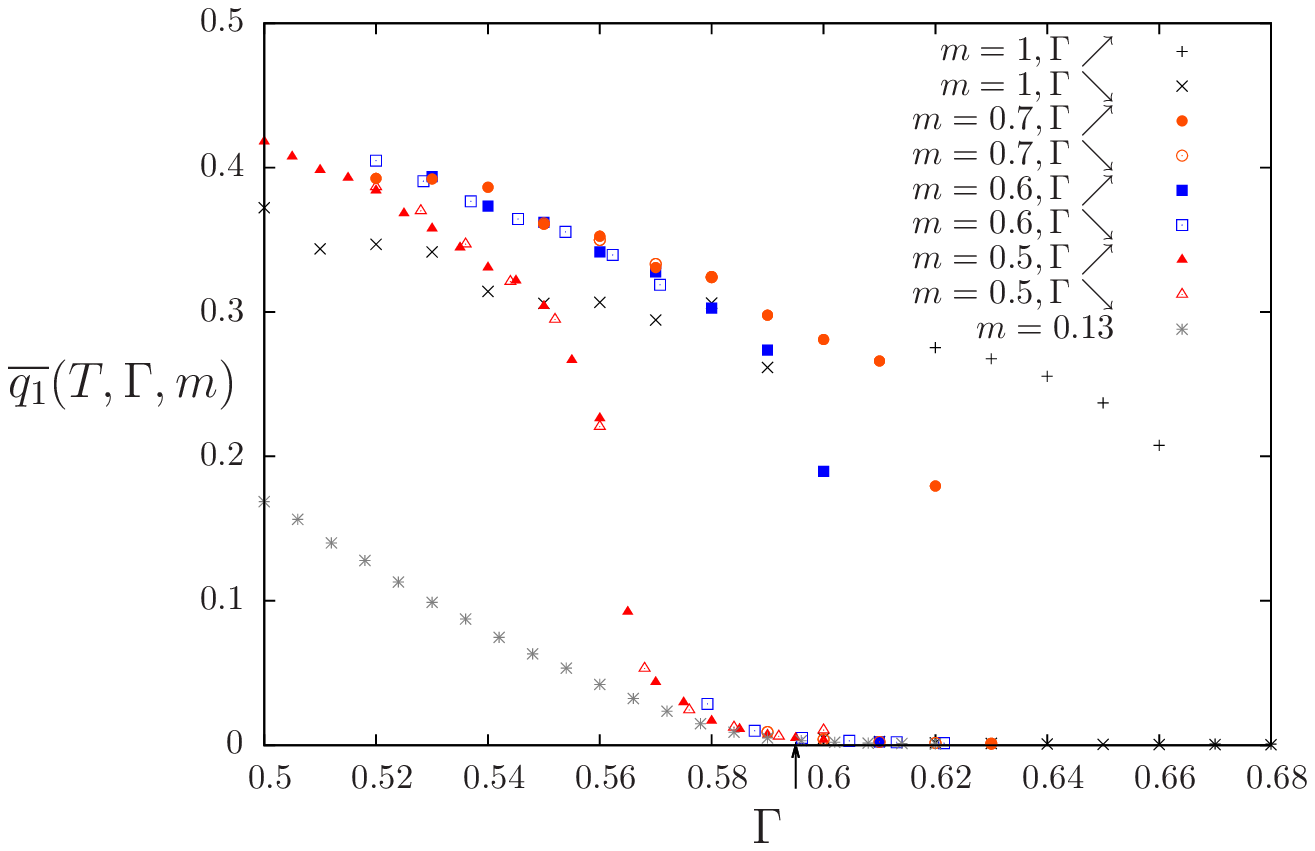}
\hspace{5mm}
\includegraphics[width=6.95 cm]{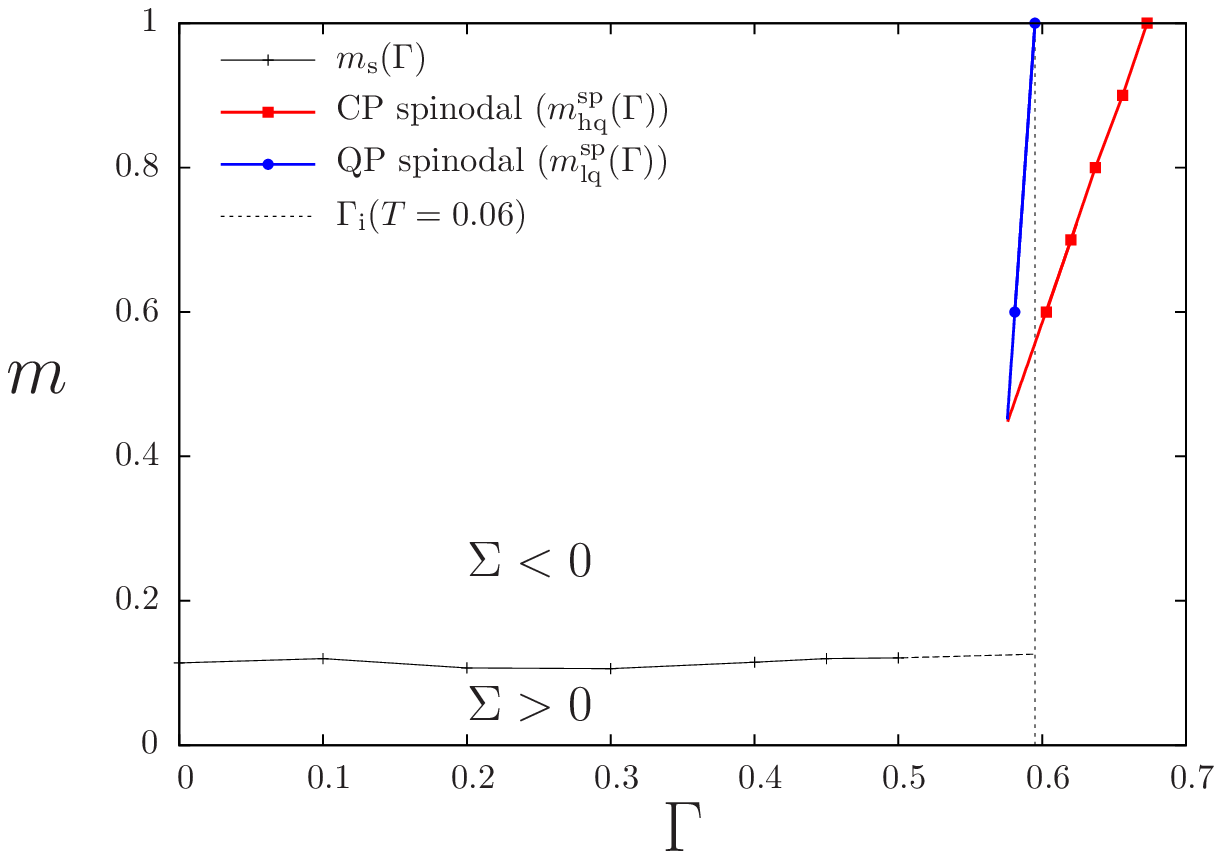} 
\caption{Coloring problem at $\qcol=4$, $c=13$ and $T=0.06$.
Left panel:
Internal overlap $\overline{q_1}(T,\G,m)$ for several $m$ as a function of $\G$. The value of the continuous instability $\Gamma_i(T)$ is indicated by an arrow. For $m \geq 0.6$ (including the RS case $m=1$), hysteresis can be seen between the curves coming from high $\G$ and those coming from small $\G$; but this hysteresis disapppears for $m \leq 0.5$, in particular for the static value $m_{\rm{s}}(T,\Gi(T)) \simeq 0.13$.
Right panel:
Phase diagram in the $(\G,m)$ plane. The replica-symmetry is broken on the left of the dashed black line, which represents the continuous instability of the RS solution. The high overlap solution exists above the red line with squares, and the low overlap one below the blue line with circles. These two lines meet at the value of $m$ below which no hysteresis remain. The relevant solution is always the low overlap one; it becomes RSB for $\G$ below the continuous instability shown as a vertical dashed black line. Finally, we also report in this region the static value of the Parisi parameter $m_{\rm{s}}(T,\G)$.}
\label{fig_q4_c12_m_gamma_1}
\label{fig_q4_c12_m_gamma_2}
\end{figure}

As mentioned above, at the replica-symmetric level and for $T \lesssim 0.12$, the $\qcol=4,c=13$ model exhibits a first order phase transition when $\G$ is varied, while the transition is only third order when the 1RSB computation at the static value $m_{\rm{s}}(T,\G)$ of the Parisi parameter is done. Altough the 1RSB equation (\ref{eq_cavity_classical_1RSB}) yields thermodynamic observables only for $m=m_{\rm{s}}(T,\G)$, the static value of the Parisi parameter, one can also solve it for any value of $m$. This gives a way to interpolate between the first order phase transition at $m=1$ and the continuous transition at $m=m_{\rm{s}}(T,\G)$. The results of cavity computations performed for various $m$ are shown on the left panel of Fig.~\ref{fig_q4_c12_m_gamma_1} for $T=0.06$. The value $m=1$ corresponds to the replica-symmetric calculation, in which the hysteresis between the solution coming from $\G=0$ and from large $\G$ is very well marked. Decreasing $m$, the hysteresis gets smaller and smaller, and cannot be seen anymore for $m=0.5$. In this case $m_{\rm{s}}(0.06,\Gi(0.06))\simeq 0.13$ and therefore the physical transition is continuous. As for the $c=9$ model at small field, it is possible to build a scenario to explain these numerical results and to understand in greater details the nature of the transition; we briefly sketch it herafter.

These results can be summarized by drawing a phase diagram in the $(\G,m)$ plane, for $T$ fixed (right panel of Fig.~\ref{fig_q4_c12_m_gamma_2}). At small $\G$ there exists only one 1RSB solution, defined for all values of $m$. 
However for larger values of $\G$ there exists a range of $m$ (between the blue and red lines of Fig.~\ref{fig_q4_c12_m_gamma_2}, right panel) where two
solutions of the 1RSB equation coexist, similarly to what we explained in Sec.~\ref{sec_spinodales_q4_c9} for $q=4,c=9$ model when $\G$ was decreased below $\G_\star$. In this case, one finds that the thermodynamic solution at $m_{\rm s}$ always lies on the ``low $q$'' branch, defined for $m \in [0, \msplq(T,\G)]$. Upon increasing the field, $\msplq$ reaches 1, and above the value of the linear instabilty $\G_{\rm{i}}(T)$, this low overlap branch becomes $m$-independent: this is the (quantum paramagnetic) replica-symmetric solution. On the other hand, the other ``high $q$'' branch still exists for $\G > \Gi$ not too large, and $m \in [\msphq(T,\G),1]$. It completely disappears only when $\msphq(T,\G)$ reaches 1, which corresponds to the limit of existence of the corresponding replica-symmetric solution. This branch is never relevant, because maximizing over $m$ leads to a solution of negative complexity, that has to be dismissed. However, it explains why a replica-symmetric computation gave the illusion of a first order phase transition.

\section{Perturbative expansions for small transverse field}
\label{app_small_field}
In this Appendix we explain how to derive a small transverse field 
perturbative expansion for observables in the quantum coloring problem. 
Let us first consider more generically an Hamiltonian 
$\hH(\G) = \hHp + \G \hHq$, where $\hHp$ is diagonal in the computational
basis, with diagonal elements denoted $E(\us)$, and where $\hHq$ is
purely off-diagonal. The expansion in $\G$ of the partition function reads
\beq \begin{split} Z(\beta,\Gamma) &
= \tr\left[ \e^{-\beta \hHp - \beta \G \hHq } \right]
\\ &= \tr \left[ e^{- \beta \hHp} -
\Gamma  \int_{0}^\beta \dd t \, e^{-(\beta-t)\hHp} \hHq e^{-t \hHp}   
+ \Gamma^2 \int_{0}^\beta \dd t_1 \int_0^{t_1} \dd t_2 \, e^{-(\beta -t_1)\hHp} 
\hHq e^{-(t_1-t_2)\hHp} \hHq e^{-t_2 \hHp} \right] + O(\G^3)
\\ & = Z(\beta, \G=0) + \G^2 \sum_{\us,\us'} | \la \us | \hHq | \us' \ra |^2
\int_{0}^\beta \dd t_1 \int_0^{t_1} \dd t_2 \,
e^{- \beta E(\us) } e^{-(t_1-t_2) (E(\us')-E(\us))}  + O(\G^3)
\\ &= Z(\beta,\G=0) 
+ \Gamma^2 \beta^2 \sum_{\us} e^{-\beta E(\us)}
\sum_{\us'}  | \la \us | \hHq | \us' \ra |^2 \left( 
\frac{1}{2} \delta_{E(\us),E(\us')} + 
\frac{1-\delta_{E(\us),E(\us')}}{\beta (E(\us')-E(\us))} \right) + O(\G^3) \ .
\end{split} 
\eeq
This expansion can of course be obtained from the usual formulas for the 
perturbation of eigenvalues, taking into account their possible degeneracies.
We now assume that $\hHq$ takes the form of an uniform ``transverse field''
$- \sum_i \hT_i$, with the flipping operators $\hT_i$ defined in 
Eq.~(\ref{eq_def_hTi}). For simplicity of notation we shall work with the 
free energy density instead of the partition function. The expansion for
this quantity reads
\beq
f(\beta,\G) = f(\beta,\G=0) - \beta \G^2 \sum_{\us} \mu(\us) 
\frac{1}{N} \sum_{i=1}^N \sum_{\s' \neq \s_i} \left( 
\frac{1}{2} \delta_{E(\us),E(\us^{(i,\s')})} + 
\frac{1-\delta_{E(\us),E(\us^{(i,\s')})}}{\beta (E(\us^{(i,\s')})-E(\us))}
\right) + O(\G^3) \ ,
\label{eq_dev_f}
\eeq
where $\us^{(i,\s')}$ denotes the configuration obtained from $\us$ by replacing
its $i$-th variable by $\s'$, and we introduced the classical Gibbs-Boltzmann
probability law $\mu(\us)=e^{-\beta E(\us)}/Z(\beta,\G=0)$. Deriving with
respect to $\G$ we obtain a similar equation for the transverse magnetization,
\beq
\la \hT_i \ra = \beta \G \sum_{\us} \mu(\us) 
\sum_{\s' \neq \s_i} \left( 
\delta_{E(\us),E(\us^{(i,\s')})} + 2
\frac{1-\delta_{E(\us),E(\us^{(i,\s')})}}{\beta (E(\us^{(i,\s')})-E(\us))}
\right) + O(\G^2) \ .
\label{eq_dev_T}
\eeq
We would like to
emphasize here that these first quantum corrections are expressed as an average
of simple, local, observables with respect to the classical Gibbs-Boltzmann
measure -- unfortunately, for higher order terms in $\Gamma$, the corresponding 
classical quantities involve correlation functions between variables at 
arbitrary distances. In particular, if the classical energy $E(\us)$ is such 
that $\mu$ is correctly described by the classical cavity method (be it in its RS or RSB 
version), then the first quantum correction can be obtained by solving the 
classical cavity equations, which are much simpler than the quantum ones.
As a matter of fact (\ref{eq_dev_f}) can be obtained from the quantum cavity 
method, either by expanding the quantum messages $\eta(\us)$ in powers of the 
number of discontinuities they contain (0 for the classical trajectories, 2 
at the lowest order in $\G$) and noting that their first quantum correction can
be expressed from the classical component of the incoming messages 
(see Eq.~\ref{eq_1rsb_b}) or by taking the derivative with respect to $\G^2$
in the variational expressions of the free energy provided by the cavity 
method in terms of its order parameter ($\eta$ at the RS level, $P(\eta)$
at the 1RSB one). In the 1RSB case the quantum corrections are thus the sum
over the postulated pure states of the correction inside each pure state; this
tells indirectly something about the classical cavity method, namely that
most pairs of configurations belonging to two different pure states are at
Hamming distance strictly larger than 1, as it should. 

As usual in perturbation theory, the presence or absence of close-by 
configurations with degenerate energies leads to qualitatively different
behaviors; this is particularly relevant at low temperatures, as the two
corresponding terms in Eqs.~(\ref{eq_dev_f},\ref{eq_dev_T}) 
have different scalings with $\beta$. In particular, for the coloring
problem, the transverse magnetization given in Eq.~(\ref{eq_dev_T})
satisfies
\beq
\lim_{\beta \to \infty} \lim_{\G \to 0}
\frac{\la \hT_i \ra}{\beta \G}= \la q_{\rm auth}(\us,i) - 1 \ra \ ,
\label{eq_dev_T_smallT}
\eeq
where in the r.h.s. the average is over the ground states $\us$ of the model,
and $q_{\rm auth}(\us,i)$ is the number of colors that $i$ is authorized to
take by its neighbors (in other words $q_{\rm col}$ minus the number of distinct
colors the neighbors of $i$ take in $\us$): the more ``floppy'' the spins are,
the stronger they respond to the quantum field. As mentioned above these
small $\G$ expansions can be obtained from the quantum cavity formalism, at
any level of replica-symmetry breaking. In particular at the 1RSB level the
formula (\ref{eq_dev_T_smallT}) can be interpreted as an expansion for
the transverse magnetization of a single cluster, provided the average in the
r.h.s. is restricted to this particular classical pure state. This provides
the justification for the interpretation of the Quantum Monte Carlo simulations
presented in Sec.~\ref{sec_q4_c9_clusters}, and in particular of Eq.~(\ref{eq_dev_mx}).

\bibliographystyle{h-physrev}

\bibliography{biblio}

\end{document}